\documentclass[reqno,12pt]{amsart}
\usepackage{graphics,amssymb,amsthm,amsmath}
\usepackage[english]{babel}
\usepackage{fullpage}
\usepackage{enumerate}
\usepackage{hyperref}
\usepackage{soul}

\usepackage{multirow}
\numberwithin{equation}{section}

\usepackage{graphics,amssymb,amsthm,amsmath}
\usepackage[english]{babel} %

\usepackage[normalem]{ulem}

\usepackage[hhmmss]{datetime}

\newcommand{\pof}{\noindent{\em Proof: }} \usepackage{mathtools}

\newcommand{\ad}{\operatorname{ad}}

\newcommand{\qdet}{{\operatorname{det}}_q}

\newtheorem{Thm}{Theorem}[section]
\newtheorem{Def/Prop}[Thm]{Definition/Proposition} \newtheorem{Rem}[Thm]{Remark}
\newtheorem{Def}[Thm]{Definition}
 
\newtheorem{Lem}[Thm]{Lemma}
\newtheorem{Cor}[Thm]{Corollary}
 
\newtheorem{Prop}[Thm]{Proposition} 

\newcommand{\qmidet}{{\operatorname{det}}_{a}}
 
\usepackage[usenames,dvipsnames]{xcolor}
\begin{document}

\fontsize{.49cm}{.49cm}\selectfont\sf

\quad

\vskip1cm

\setcounter{tocdepth}{3}

\bigskip \setcounter{section}{0} \bigskip

\title[Determinantal Ideals and the Canonical Commutation
relations]{Determinantal Ideals and the Canonical Commutation
Relations. Classically or Quantized}

\author{Hans Plesner Jakobsen}

\address{Department of Mathematical Sciences, University of
Copenhagen, Denmark}

\email{jakobsen@math.ku.dk}

\maketitle

\begin{abstract}We construct homomorphic images of
$su(n,n)^{\mathbb C}$ in Weyl Algebras ${\mathcal H}_{2nr}$.
More precisely, and using the Bernstein
filtration of  ${\mathcal H}_{2nr}$, $su(n,n)^{\mathbb C}$ is mapped into degree $2$ elements with the negative non-compact root spaces being mapped into second order creation operators. Using the Fock representation of ${\mathcal H}_{2nr}$, these homomorphisms give all
unitary highest weight representations of $su(n,n)^{\mathbb C}$
thus reconstructing the Kashiwara--Vergne List for the
Segal--Shale--Weil representation. Just as in the derivation of the their list, we construct a representation of $u(r)$ in the Fock space commuting with $su(n,n)^{\mathbb C}$, and this gives the multiplicities. The construction also gives an
easy proof that the ideals of $(r+1)\times (r+1)$ minors are
prime ($r\leq n-1)$. The quotients of all polynomials by such ideals
carry the more singular of the representations. As a consequence, these representations
can be realized in spaces of solutions to Maxwell type equations. We actually go one step further and determine exactly which representations from our list are missing some ${\mathfrak k}^{\mathbb C}$-types, thereby revealing exactly which covariant differential operators have unitary null spaces. 

We prove the analogous results for ${\mathcal
U}_q(su(n,n)^{\mathbb C})$. The Weyl Algebras are replaced by
the Hayashi--Weyl Algebras ${\mathcal H}{\mathcal W}_{2nr}$ and the Fock space by a $q$-Fock space. Further, determinants are replaced by $q$-determinants, and a commuting representation of ${\mathcal U}_q(u(r))$ in the $q$-Fock space is constructed. For this purpose a Drinfeld Double is used. We
mention one difference: The quantized negative non-compact root
spaces, while still of degree 2, are no longer given entirely by
second order creation operators.     \end{abstract} \medskip

\section{introduction} 

The most fundamental equation in analysis must be
\begin{equation}
[\frac{d}{dx}, x]=1.
\end{equation}
The algebraic structure corresponding to this is the Weyl Algebra. By inserting an $i$ in the equation it becomes the ``canonical commutation relation'' of mathematical physics. Another, equivalent way is to define the annihilator operator $A=\frac1{\sqrt{2}}(x + \frac{d}{dx})$. Its (formal) adjoint is $A^\dagger= \frac1{\sqrt{2}}(x - \frac{d}{dx})$ and is known as the creation operator. We have of course
\begin{equation}
[A,A^\dagger]=1.
\end{equation}

\medskip

The Metaplectic Group is the automorphism group of this equation.

\medskip

In the very first studies of the metaplectic  representation by L. van Hove, the representation was given on the Lie
algebra level. See (\cite{van-hove}), the summary
(\cite{van-hove}), as well as the review by I. E. Segal
(\cite{MR-Segal}). However, in subsequent studies by D. Shale
(\cite{shale}) (see also (\cite{segal})), K. Gross and R. Kunze
({\cite{gk1},\cite{gk2}), and the seminal article by M.
Kashiwara and M. Vergne (\cite{k-v}), the emphasis was shifted
to the group level. Here, the representation was called  the Segal-Shale-Weil (or
harmonic) representation.

\smallskip

This shift makes it easy to explain the basic observations
yielding a unitary representation of the
Metaplectic Group $Mp(n,{\mathbb R})$ (a $2$-fold cover of
$Sp(n,{\mathbb R})$): I) By the Stone--von Neumann Theorem
(\cite{stone}, \cite{vN}) there is an essentially unique
non-trivial irreducible representation of the Heisenberg Group
${\mathcal H}_n$; ${\mathcal H}_n\ni h\mapsto \pi(h)$. II) Any
$g\in Sp(n,{\mathbb R})$, appropriately interpreted, acts as an
automorphism of ${\mathcal H}_n$. Hence there must be a unitary
operator $T(g)$ such that $\pi(g(h))=T(g)\pi(h) T(g)^{-1}$. The
operator is not unique and therefore we do not immediately get a
representation of $Sp(n,{\mathbb R})$, but it may be seen that
this obstacle may be removed by passing to $Mp(n,{\mathbb R})$
(cf. \cite{k-v}).

\medskip

The infinitesimal version of this seems less well known though
occasionally touched upon in connection with the mathematical
treatment of the quantum mechanical harmonic operator. The
author learned about this representation from his masters
advisor who, among many other things,  studied the canonical commutation relations (see (\cite{nsp}). 

\smallskip

The infinitesimal level has the pleasing feature of building everything from the fundamental operators $x$ (as multiplication operator) and $\frac{\partial}{\partial x}$. Of course, one can extend to several, indeed many, variables.

\medskip

The Weyl algebra is defined as  ${\mathcal H}_{1}=Span_{\mathbb C}\{A^r,
(A^\dagger)^s\mid r,s\in{\mathbb N}_0 \}$. It is also the enveloping algebra of the Heisenberg group, hence the symbol. The Weyl
algebra ${\mathcal H}_{s}$ is just the product of $s$ copies of
${\mathcal H}_{1}$. Likewise, the so-called (by physicists)
Fock space ${\mathcal F}_s$ of ${\mathcal H}_{s}$ is generated by a cyclic
vector $v_s$ which is annihilated by all elements $A_i, i=1,\dots,s$. 
The algebra acts by the natural left action yielding the Stone--von Neumann representation $\pi^{SvN}_s$.  It is easy to see, and well known, that $\pi^{SvN}_s$ is injective on ${\mathcal H}_s$. These simple (in many ways)
representations  yield all unitarizable highest
weight representations of $su(n,n)^{\mathbb C}$.

\medskip

The idea is again to use the Stone--von Neumann theorem while
observing that the inclusion $su(n,n){\mathbb
C}\xhookrightarrow{} sp(2n,{\mathbb R})^{\mathbb C}$ yields
derivations of the Weyl algebra. These are known to be inner.

\medskip

In the end it turns out, however, that one can construct
the homomorphisms $su(n,n)^{\mathbb
C}\rightarrow{\mathcal H}_{2nr}$ directly.

\medskip

This is the path we take.

\medskip

The assignments \begin{equation}e=\frac{i}2A^2,\
f=\frac{i}2(A^\dagger)^2,\ h=[e,f] \end{equation} defines a
representation of $sl(2,{\mathbb C})=sp(1,{\mathbb
R})=su(1,1)^{\mathbb C}$.

\medskip

To explain some further details, let us consider ${\mathfrak
h}_{4}$. The following elements
\begin{eqnarray} W_{1,1}=A_1^\dagger A_3^\dagger &,&
W_{1,2}=A_1^\dagger A_4^\dagger,\\ W_{2,1}=A_2^\dagger
A_3^\dagger &,& W_{2,2}=A_2^\dagger A_4^\dagger \end{eqnarray}
generate a polynomial subalgebra ${\mathcal U}^-_{0}\subset
{\mathcal H}_{4}$. They can be seen to correspond to the negative non-compact roots of $su(2,2)$,  where e.g.   $W_{1,1}$ is the root vector
$f_\beta$ for the unique non-compact simple root $\beta$. The positive non-compact roots are given  by analogously quadratic expressions in  annihilation operators. Clearly, there is a non-trivial relation
\begin{equation} W_{1,1}W_{2,2}=W_{2,1}W_{1,2}, \end{equation}
and this is the equation for the vanishing of the determinant,
$\det W=0$ of the $2\times2$ matrix $W$ whose entries are the
above $W_{i,j};i,j=1,2$.

This implies that \begin{equation}{\mathcal U}^-_{0}\simeq
{\mathcal P}(W_{1,1}W_{1,2}W_{2,1}W_{2,2})/{\mathcal
I}_{2\times2} \end{equation} where ${\mathcal P}={\mathcal
P}(W_{1,1}W_{1,2}W_{2,1}W_{2,2})$ is the full polynomial algebra
in $4$ variables and ${\mathcal I}_{2\times2}$ is the ideal
generated by $\det W$. Since ${\mathcal U}^-_{0}$ is a
subalgebra of the algebra of polynomials in $A^\dagger_i,
i=1,2,3,4$, this easily implies that ${\mathcal I}_{2\times2}$
is a prime ideal.

\medskip

More generally, using the same method, we can prove 

\smallskip

\noindent{\bf Theorem~\ref{prime-th}} The ideal generated by all
$k\times k$ minors of an $r\times r$ matrix is prime.

\smallskip

This result is due to K. R. Mount (\cite{mo}), but  see also the
discussion in the end of Chapter 2 (\cite{det-ring}).

\medskip

Returning to the $2\times2$ case, one can easily decompose the
full Fock representation: Consider $\forall N,S\in{\mathbb N}_0$
the spaces \begin{eqnarray} V_N^{\uparrow}&=&Span_{\mathbb C}\{
(A_1^\dagger)^i(A_2^\dagger)^j\mid i+j=N\},\textrm{ and}\\
V_N^{\downarrow}&=&Span_{\mathbb C}\{
(A_3^\dagger)^i(A_4^\dagger)^j\mid i+j=S\}. \end{eqnarray} Then
the spaces ${\mathcal U}^-_{0}\otimes V_N^{\uparrow}$ and
${\mathcal U}^-_{0}\otimes V_S^{\downarrow}$ carry unitarizable
irreducible highest weight modules, and together they furnish a
decomposition of the Fock representation.

\medskip

In the analogous case for $su(n,n)^{\mathbb C}$ one has the 
Kashiwara--Vergne List of unitarizable highest weight modules. Recall that it was
conjectured by Kashiwara and Vergne and proved by the current author
in (\cite{jak-last}) that this list describes all irreducible,
unitarizable highest weight modules. A key point in the proof of this result was the Bernstein--Gelfand--Gelfand Theorem (\cite{bgg}) and {\em the last possible place of unitarity (\cite{jak-last})}. This was applied to the already constructed list. In the current study we bring this one step further and use these same tools to actually construct the list. Further ingredients are: {\em missing ${\mathfrak k}^{\mathbb C}$-types} ({\em covariant differential operators}), and {\em tensor products of generalized Verma modules}. 

In comparison to Kashiwara--Vergne, we actually go one step further and determine exactly which representations from our list are missing some ${\mathfrak k}^{\mathbb C}$-types, thereby revealing exactly which covariant differential operators have unitary null spaces. 

\medskip

We will refer to the above as the ``classical case'', the ``classical results'',  or sometimes even the `` $q=1$ case'' when discussing the ``quantized case''.

\medskip

In 1990, T. Hayashi (\cite{hi}) introduced a number of very
important quantum algebras and studied some basic representations of these. In his families are both quantized symplectic as well as
quantized spinor representations. His method and construction is
clearly the right way to quantize the harmonic representation,
and will form the foundation of our investigations in the case
of ${\mathcal U}_q(su(n,n)^{\mathbb C})$. This case was not covered by him.

\medskip

The case of the quantized enveloping algebra proceeds along
similar lines, though there occurs some additional
technicalities.

First of all the basic algebras are now the Hayashi-Weyl
Algebras ${\mathcal H}{\mathcal W}_{s}, s\in{\mathbb N}$: Let
$q$-be a non-zero complex number such that $q^2 \neq 1$. The
algebra ${\mathcal H}{\mathcal W}_{1}$ is defined as an
associative unital algebra with generators $B,^\dagger,L^{\pm
1}$ and relations \begin{align} & LL^{-1}=L^{-1} L=1, \\& LB
L^{-1}=q^{-1} B, \quad L B^\dagger L^{-1}=q B^\dagger, \\&
BB^\dagger-qB^\dagger B=L^{-1}, \quad BB^\dagger-q^{-1}B^\dagger
B=L .\end{align}

${\mathcal H}{\mathcal W}_{s}$ is the product of $s$ copies of
${\mathcal H}{\mathcal W}_{1}$.

These algebras are of fundamental importance. Early on, these
algebras were studied in (\cite{rosen}), Recently a renewed
interest has been observed (\cite{kilu}), see also
(\cite{jak-qdiff}).

\medskip

The definition of the $q$-Fock spaces ${\mathcal F}^{(q)}_{2nr}$ and $q$-Stone--von Neumann representation
$\pi^{qSvN}_{2nr}$, $r=1,\dots$,  are then straightforward.

\medskip

We obtain the analogues of the classical results for this
algebra. Specifically, we construct  homomorphisms ${\mathcal
U}_q(su(n,n)^{\mathbb C})\rightarrow {\mathcal H}{\mathcal
W}_{2nr}; r\in{\mathbb N}$  The homomorphisms are more complicated. While still of
degree $2$, one needs in general both creation and annihilation
operators to describe the (quantized) negative non-compact root
spaces. For the treatment here, we rely on representation theory as developed in (\cite{yam}) and  (\cite{rosso}). In (\cite{yam}) can also be found definitions and fundamental results relating to $q$-determinants.

We then obtain the following result which was
first proved in (\cite{qufd}).

\medskip

\noindent{\bf Theorem~\ref{q-prime}} The $q$-determinantal ideals are prime.

\medskip

As for the decomposition of the
quantized Fock spaces $F^{(q)}_{2nr}$, $r=1,2,\dots$, one has to construct a representation of ${\mathcal U}_q(u(r))$ on $F^{(q)}_{2nr}$  which commutes with the action of ${\mathcal U}_q(su(n,n))$ and even with  ${\mathcal U}_q(u(n,n))$. Here we rely on (\cite{yam}). The construction actually results in a Drinfeld Double. In the end, a result analogous to the classical case is obtained.  We call the result the  quantized Kashiwara--Vergne
List.  This result is more descriptive than a mere claim of ``similar representation theory'' because it constructs explicitly the highest weight vectors in terms of quantum minors.

We finish the last section by making a few remarks about ``quantized wave operators''.

\medskip

The topics we discuss here have been studied previously be many authors, and may easily be seen as special cases, pushed a but further, of many studies. We will not try to be comprehensive but will mention just two: (\cite{ew}) and (\cite{leclerc}).

\medskip

The results were first announced at the conference {\bf Quantum Groups, Quantum Symmetric Spaces, and Operator Algebras}, Kristineberg 2-8 June 2019, organized by L. Turovska,  A. Stolin, and G. Zhang to the memory of Leonid Vaksman. We were glad for the opportunity to pay homage.

\medskip

Besides this introduction, the material is arranged into the following five sections: \S2:
The (quantized) enveloping algebras, \S3:  The harmonic representation in the classical case,
\S4: The Kashiwara--Vergne list of unitary representations, \S5: The $q$-harmonic representation. \S6: The full decomposition of the $q$-Fock spaces.

\bigskip

\section{The (quantized) enveloping algebras}

We consider ${\mathfrak g}^{\mathbb C}=su(n,n)^{\mathbb C}$. The
simple roots are denoted
$\Pi=\{\mu_1,\dots,\mu_{n-1}\}\cup\{\beta\}\cup
\{\nu_1,\dots,\nu_{n-1}\}$, where $\beta$ is the unique
non-compact root. We have
\begin{equation}\label{fl}\begin{array}{ccccccccc}{\mathfrak
g}^{\mathbb C} &=&{\mathfrak k}^{\mathbb C}\oplus {\mathfrak
p}^{\mathbb C}&=&{\mathfrak p}^-\oplus {\mathfrak k}^{\mathbb C}\oplus
{\mathfrak p}^+, \\{\mathfrak k}^{\mathbb
C}&=&su(n)^{\mathbb C}\oplus {\mathbb C}\oplus su(n)^{\mathbb
C}&=&{\mathfrak k}_\mu^{\mathbb C}\oplus {\zeta} \oplus
{\mathfrak k}_\nu^{\mathbb C}\ .&&&& \end{array} \end{equation}
\begin{equation}\begin{array}{ccccccccc} \zeta \textrm{ is the
center, }&{\mathfrak p}^{\pm}\textrm{ are abelian ${\mathcal
U}({\mathfrak k}^{\mathbb C})$ modules,}& [{\mathfrak p}^+,{\mathfrak p}^-]\subseteq {\mathfrak k}^{\mathbb C}, \textrm{ and }\\
&{\mathcal U}({\mathfrak g}^{\mathbb C})={\mathcal P}({\mathfrak
p}^-)\cdot {\mathcal U}({\mathfrak k}^{\mathbb C})\cdot
{\mathcal P}({\mathfrak p}^+). \end{array} \end{equation} The
subalgebras ${\mathfrak k}_\mu^{\mathbb C},{\mathfrak
k}_\nu^{\mathbb C}$ above are generated by the simple roots
$\{\mu_1,\dots,\mu_{n-1}\}$, and $\{\nu_1,\dots,\nu_{n-1}\}$,
respectively. later we will also use the terminology
\begin{equation}
{\mathfrak k}_\mu^{\mathbb C}={\mathfrak k}_L^{\mathbb C}\textrm{ and }{\mathfrak k}_\nu^{\mathbb C}
={\mathfrak k}_R^{\mathbb C}.
\end{equation}

The generators are denoted by $E_\alpha,F_\alpha$, and we set
$h_\alpha=[E _\alpha,F_\alpha]$, for $\alpha\in\Pi$.

\smallskip

For later use, we fix the imbedding $su(n,n)\hookrightarrow gl(n,{\mathbb C})$ by the hermitian form $J=\left(\begin{array}{cc}I_n&0\\0&-I_n\end{array}\right)$. This means that $X\in su(n,n)\Leftrightarrow X^*J+JX=0$. As a consequence, we choose  the traceless diagonal matrices \label{page-hbeta}
 as our (compact) Cartan subalgebra. 
 We choose further $\mu_i=
 \varepsilon_{i}-\varepsilon_{i+1}$, $i=1,\dots,n-1$, 
 and $\nu_i=
 \varepsilon_{n+i}-\varepsilon_{n+i+1}$, $i=1,\dots,n-1$, and $\beta=
 \varepsilon_{n}-\varepsilon_{n+1}$. We write a  
weight $\Lambda =(\Lambda_\mu,\Lambda_\nu,\lambda_\beta)$ in
accordance with this. For specificity,
$\lambda_\beta:=\Lambda(h_\beta)$. Furthermore,
$\Lambda_\mu=(n_1,n_2,\dots,n_n)\in{\mathbb Z}^{n}$ and $\Lambda_\nu=(m_1,m_2,\dots,m_n)\in{\mathbb Z}^{n}$. We have that $n_1\geq n_2\geq\cdots\geq n_n$ and likewise $m_1\geq n_2\geq\cdots\geq m_n$. Then  $\Lambda_\mu(\mu_i)=n_i-n_{i+1}, i=1,\dots,n-1$ and $\Lambda_\nu(\nu_j)=m_j-m_{j+1}, j=1,\dots,n-1$. Notice that we do not insist that $n_n\geq 0$ or $m_n\geq0$, which turns out to be convenient. One can always find a representative in an equivalence class for which this holds. Here, two $n$ tuples in ${\mathbb Z}^n$ are equivalent if they differ by the addition of some $(p,p,\dots,p)$. Equivalent tuples define the same module. Notice: We make no coupling between these integers and $h_\beta$.

\smallskip

${\mathcal P}^\pm:={\mathcal P}({\mathfrak p}^\pm)={\mathcal
U}({\mathfrak p}^\pm)$ are polynomial algebras, given explicitly
as \begin{eqnarray} {\mathcal P}^-&=&{\mathbb C}[W_{i,j}\mid
i,j=1,\dots,n\},\label{try}\\ {\mathcal A}^+&=&{\mathbb C}[Z_{i,j}\mid
i,j=1,\dots,n\}, \end{eqnarray}
where
\begin{eqnarray}Z_{i,j}&=&\ad(E_{\nu_{j-1}}) \dots
\ad(E_{\nu_{0}})\cdot \ad(E_{\mu_{i-1}}) \dots
\ad(E_{\mu_{0}})(E_\beta),\\ W_{i,j}&=&\ad(F_{\nu_{j-1}}) \dots
\ad(F_{\nu_{0}})\cdot \ad(F_{\mu_{i-1}})\dots
\ad(F_{\mu_{0}})(F_\beta).\label{27}\end{eqnarray} Above, for
convenience, we set $\ad(F_{\mu_{0}})=\ad(F_{\nu_{0}})=I$.

\medskip In the quantum group ${\mathcal U}_q({\mathfrak
g}^{\mathbb C})$, we denote the generators by
$E_\alpha,F_\alpha,K_\alpha^{\pm1}$ for $\alpha\in\Pi$. There
are also decompositions \begin{eqnarray} {\mathcal
U}_q({\mathfrak g}^{\mathbb C})&=&{\mathcal A}_q^-\cdot
{\mathcal U}_q({\mathfrak k}^{\mathbb C})\cdot {\mathcal
A_q}^+,\\ {\mathcal U}_q({\mathfrak k}^{\mathbb C})&=&{\mathcal
U}_q({\mathfrak k}_L^{\mathbb C})\cdot {\mathbb
C}[K_\beta^{\pm1}] \cdot {\mathcal U}_q({\mathfrak k}_R^{\mathbb
C}). \end{eqnarray} Here, ${\mathcal A}_q^\pm$ are quadratic
algebras which are furthermore ${\mathcal U}_q({\mathfrak
k}^{\mathbb C})$ modules. Specifically, \begin{eqnarray} {\mathcal
A}_q^-&=&{\mathbb C}[W_{i,j}\mid i,j=1,\dots,n\},\textrm{ and}\\ {\mathcal
A}_q^+&=&{\mathbb C}[Z_{i,j}\mid i,j=1,\dots,n\}, \end{eqnarray}
with relations in ${\mathcal A}_q^-$ given by (these are the relations of the so-called ``standard'' quantized matrix algebra discovered in (\cite{frt}));
\begin{eqnarray}\label{frt}\textrm{ FRT}(q^{-1}):\qquad\qquad\\\label{a}\nonumber W_{ij}W_{ik} &=& q^{-1}W_{ik}W_{ij}
\textrm{ if }j < k,\\ \label{b}W_{ij}W_{kj} &=&\nonumber
q^{-1}W_{kj}W_{ij}\textrm{ if }i< k,\\\label{c} W_{ij}W_{st} &=&\nonumber
W_{st}W_{ij} \textrm{ if }i < s\textrm{ and }t <
j,\\\label{cross}\nonumber W_{ij}W_{st} &=&
W_{st}W_{ij}-(q-q^{-1})W_{it}W_{sj} = \textrm{ if }i < s\textrm{
and } j < t. \end{eqnarray} The algebra ${\mathcal A}_q^+$ has
the same relations, but the algebras ${\mathcal A}_q^\pm$ are
different as ${\mathcal U}_q({\mathfrak k}^{\mathbb C})$
modules. The elements $Z_{ij}$ and $W_{ij}$ are constructed by
means of the Lusztig operators. References \cite{l} and
\cite{jan} are general references of much of this. Using the
Serre relations one gets, setting $\mu_0=Id$, \begin{Lem}
\label{2.1}\begin{eqnarray}Z_{i,j}&=&T_{\nu_{j-1}}
T_{\nu_{j-2}}\dots T_{\nu_{0}}\cdot T_{\mu_{i-1}}
T_{\mu_{i-2}}\dots T_{\mu_{0}}(E_\beta),\textrm{ and}\\
W_{i,j}&=&(-q^{-1})^{i+j-2}T_{\nu_{j-1}}
T_{\nu_{j-2}}\dots T_{\nu_{0}}\cdot T_{\mu_{i-1}}
T_{\mu_{i-2}}\dots T_{\mu_{0}}(F_\beta).\label{26}\end{eqnarray}
\end{Lem}

\medskip

On simple roots $\alpha,\gamma$ one has \begin{eqnarray}
T_\alpha(E_\gamma)&=&E_\alpha E_\gamma-q^{-1}E_\gamma E_\alpha, \textrm{ and}\\
T_\alpha(F_\gamma)&=&F_\gamma F_\alpha-qF_\alpha F_\gamma.\label{t-si}
\end{eqnarray}

\begin{Rem}The factor $(-q^{-1})^{i+j-2}$ is inserted to make
the formulas have a nice limit for $q=1$. It leads to the same
relations as those for elements $\tilde
W_{i,j}=(-q^{-1})^{-i-j+2}W_{i,j}$ defined without the factor.
The latter choice has been preferred in our previous articles.
\end{Rem}

\medskip

We now recall the definition of a quantum minor  from \cite{yam} :

Suppose $I=\{i_1<1_2< \dots < i_m\}$ and $J=\{j_1<j_2< \dots <
j_m\}$ are subsets of $I=\{1,2,\dots, \min\{r,s\}\}$. Define
\begin{eqnarray}\label{2} {\qmidet}(I,J)&=&\Sigma_{\sigma\in
S_m}(-a)^{\ell(\sigma)}w_{i_1,j_{\sigma(1)}}w_{i_2,j_{\sigma(2)}}
\cdots w_{i_m,j_{\sigma(m)}}\\\label{3}&=&\Sigma_{\tau\in
S_m}(-a)^{\ell(\tau)} w_{i_{\tau(1)},j_1}w_{i_{\tau(2)},j_2} \cdots
w_{i_{\tau(r)},j_m}.\end{eqnarray} These elements are called quantum $m\times 
m$ minors. The parameter $a$ depends on the relations. The relations in  (\cite{yam}) are obtainable from the relations (\ref{frt}) by interchanging $q\leftrightarrow q^{-1}$. The $q$-determinants of ({\cite{yam}) are then those denoted by ${\det}_q$ whereas quantum minors in our setting will be taken as ${\det}_{q^{-1}}$.

We will also set

\begin{equation}
{\qmidet}(I,J)=\qmidet\left(\begin{array}{c}w_{i_1,j_1}\cdots
w_{i_1,j_m}\\\vdots\qquad\vdots\\w_{i_m,j_1}\cdots w_{i_m,j_m}\end{array}\right).
\end{equation}

In particular, we set 

\begin{equation}\label{triangle}
\triangle_{a,m}(w)=\qmidet\left(\begin{array}{c}w_{1,1}\cdots
w_{1,m}\\\vdots\qquad\vdots\\w_{m,1}\cdots w_{m,m}\end{array}\right).
\end{equation}

\bigskip

\section{The harmonic representation in the classical case}
\label{s3}
\subsection{The Weyl algebras}

The Weyl algebra ${\mathcal H}_{1}=Span_{\mathbb C}\{A^r,
(A^\dagger)^s\mid r,s\in{\mathbb N}_0 \}$ is defined in terms of
the Heisenberg canonical commutation relations: \begin{eqnarray}
\left[A, A^\dagger\right]&=&I, \end{eqnarray} and the Weyl
algebra ${\mathcal H}_{s}$ is just the product of $s$ copies of
${\mathcal H}_{1}$.

\medskip

\begin{Lem} \begin{eqnarray}
\left[AA^\dagger,A^\dagger\right]&=&A^\dagger\\
\left[AA^\dagger,A\right]&=&-A. \end{eqnarray} \end{Lem}

\medskip

\begin{Lem}
\begin{equation}
\forall s\in{\mathbb N}_0:\; AA^\dagger(A^\dagger)^s=(s+1)(A^\dagger)^s+(A^\dagger)^{s+1}A
\end{equation}
\end{Lem}

A simple but {key} result: \begin{Lem}[The Weyl-Serre Equations]
\begin{eqnarray} (A^\dagger)^2A-2(A^\dagger\label{ws}
AA^\dagger)+A(A^\dagger)^2=0\\
A^2A^\dagger-2(AA^\dagger\label{qs-rel} A)+A^\dagger A^2=0
\end{eqnarray} \end{Lem}

\medskip

\subsection{su(n,n)}

We now consider specifically the Weyl algebra
\begin{equation}{\mathcal H}_{2n}=Span_{\mathbb
C}\{A^r_i,(A^\dagger)^s_j\mid i,j=1,\dots,2n, r,s\in{\mathbb
N}_0\}.\end{equation} It is defined in terms of the Heisenberg
canonical commutation relations: \begin{eqnarray}\label{32}
\left[A_k, A_\ell\right]&=&0,\\ \left[A^\dagger_k,
A^\dagger_\ell\right]&=&0,\textrm{ and}\\ \left[A_k,
A^\dagger_\ell\right]&=&\delta_{k,\ell}.
\label{34}\end{eqnarray}

\medskip

Throughout \S\ref{s3} and \S\ref{4}, $su(n,n)$ is fixed, and $n$ will be suppressed whenever it is convenient.

\medskip

\begin{Def} $$\begin{array}{rclrcl}
	F^{(1)}_\beta&=&A^\dagger_1A^\dagger_{n+1},
&E^{(1)}_\beta&=&-A_1A_{n+1},\\
H^{(1)}_k&=&A^\dagger_{k}A_k,&H^{(1)}_\beta&=&-H^{(1)}_1-H^{(1)}_{n+1}-1,\\
F^{(1)}_{\mu_k}&=&A_{k+1}^\dagger
A_{k},&E^{(1)}_{\mu_k}&=&A_{k+1} A^\dagger_{k}\
(k=1,\dots,n-1),\\ F^{(1)}_{\nu_\ell}&=&A_{n+\ell}
A_{n+\ell+1}^\dagger,&E^{(1)}_{\nu_\ell}&=&A^\dagger_{n+\ell}
A_{n+\ell+1} \ (\ell=1,\dots,n-1),\\
H^{(1)}_{\mu_k}&=&H_k^{(1)}-H^{(1)}_{k+1},&H^{(1)}_{\nu_k}&=&H^{(1)}_{n+k}-H^{(1)}_{n+k+1}\
(k=1,\dots,n-1).\end{array}$$\label{s-def-1}
\end{Def}

\begin{Def}
\begin{equation}\Xi=\sum_{k=1}^{n}A_k^\dagger A_k-\sum_{k=n+1}^{2n}A_k^\dagger A_k.\label{def-z}
\end{equation}
  \end{Def}

\medskip

\begin{Prop} The elements in Definition~\ref{s-def-1}  satisfy the Serre relations for
$su(n,n)^{\mathbb C}$. The element $\Xi$ commutes with all of these.\end{Prop}

\proof This follows easily from (\ref{qs-rel}). \qed

\medskip

\begin{Rem}
The form of the elements $H_\star$ turns out to be convenient for what follows. The unusual form can be recast into a more familiar one using elements $h_i=-A_iA_i^\dagger; i=1,\dots, n$ and $h_i=A_i^\dagger A_i; i=n+1,\dots, 2n$.
\end{Rem}

\medskip

\begin{Cor}The relations above define a homomorphism 
\begin{equation}\psi_q^{(1)}: {\mathcal U}(su(n,n)^{\mathbb C})\rightarrow 
{\mathcal H}_{2n}.\end{equation}
\end{Cor}

\medskip

\begin{Def}
\begin{equation}W^{(1)}_{k,\ell}=A^\dagger_{k}A^\dagger_{n+\ell}\textrm{
and }Z^{(1)}_{k,\ell}=-A_{k}A_{n+\ell}.\end{equation} \end{Def}

\medskip

\begin{Lem}\label{L1}For $k,\ell<n$,$$W^{(1)}_{k+1,\ell}=
F_{\mu_k}W^{(1)}_{k,\ell}-W^{(1)}_{k,\ell}F^{(1)}_{\mu_k}
\textrm{ and
}W^{(1)}_{k,\ell+1}=F^{(1)}_{\nu_k}W^{(1)}_{k,\ell}-W^{(1)}_{k,\ell}F^{(1)}
_{\nu_k}.$$
\end{Lem}
	
\proof This follows from the definitions along with
(\ref{32}-\ref{34}). \qed

\medskip

\subsection{Unitarity; Stone--von Neumann}

The (infinitesimal) Stone--von Neumann representation
$\pi_1^{SvN}$ (by physicists often called the Fock
representation) is the irreducible unitary representation of
${\mathcal H}_1$ in a Hilbert space ${\mathcal F}_{1}$ defined in terms of a 
cyclic vector $v_1$ called the
vacuum vector which is annihilated by $A$. ($A$ is an
annihilation operator.)

	Specifically,

\begin{eqnarray}Av_1 &=&0, \textrm{ and}\\{\mathcal
F}_{1}&=&Span_{\mathbb C}\{(A^\dagger)^{r} v_1\mid
r\in{\mathbb N}_0\}.\end{eqnarray}

 \medskip
 
More generally, there is the analogous  unitary
representation $\pi_s^{SvN}$ of ${\mathcal H}_s$ in ${\mathcal
F}_{s}={\mathcal U}^-({\mathcal H}_s)$. This is just the (outer) tensor product
of $s$ copies of $\pi_1^{SvN}$. We denote by $v_s$ the $s$ fold tensor product of $v_1$.

Here,  ${\mathcal U}^-({\mathcal H}_s)$ denotes the subalgebra generated by all creation operators.  

\medskip

\begin{Prop}
$\pi_s^{SvN}$ is an irreducible representation of ${\mathcal U}({\mathcal H}_s)$.
\end{Prop}

\pof ${\mathcal H}_s$ acts as shift operators, omitting no positions. \qed

\medskip

\begin{Prop} Let $0\neq v\in {\mathcal
F}_{s}$. For $p\in {\mathcal U}^-({\mathcal H}_s)$ it holds that
\begin{equation}
 p\cdot v=0\Leftrightarrow p\cdot v_{s}=0\Leftrightarrow p=0.
\end{equation}
\end{Prop}

\pof This follows since ${\mathcal U}^-({\mathcal H}_s)$ is a polynomial (abelian) algebra. 

\medskip

Even more holds:

\begin{Prop}\label{new-prime}
Let $p\in {\mathcal U}^-({\mathcal H}_s)$ and $v\in {\mathcal
F}_{s}$. Then
\begin{equation}p\cdot v=0\Leftrightarrow [p=0\lor v=0] 
\end{equation}
\end{Prop}

\pof Using a standard monomial basis $\{X^{\mathbf a}\}$ of ${\mathcal U}^-({\mathcal H}_s)$ we can write
\begin{equation}
p\cdot v_s=\sum_{\mathbf b}\beta_{\mathbf b}(A^\dagger)^{\mathbf b}\cdot v_s\textrm{ and }v=\sum_{\mathbf a}\alpha_{\mathbf a}(A^\dagger)^{\mathbf a}\cdot v_s.
\end{equation}
Then
\begin{equation}
p\cdot v= \sum_{\mathbf a}\alpha_{\mathbf a}(A^\dagger)^{\mathbf a}\cdot p\cdot v_s= \sum_{{\mathbf a},{\mathbf b}}\alpha_{\mathbf a}\beta_{\mathbf b}(A^\dagger)^{{\mathbf a}+{\mathbf b}}\cdot v_s.
\end{equation}

By using some (lexicographic) order and focusing in highest order terms, it follows easily that either $\forall {\mathbf a}:\alpha_{\mathbf a}=0$ or  $\forall {\mathbf b}:\beta_{\mathbf b}=0$. \qed

\medskip

\subsection{Prime. $2\times2$ minors.}

\begin{Lem} Let ${\mathbb W}^{(1)}$ be the $n\times n$ matrix
whose $i,j$ entry is $W^{(1)}_{i,j}$. Then any $2\times2$ minor
vanishes. \end{Lem}

\proof This follows easily since the product of the diagonal
elements equals the product of the off diagonal elements.
Alternatively, set
\begin{equation}X^{(1)}=\left(\begin{array}{c}
A^\dagger_1\\A^\dagger_2\\\vdots\\A^\dagger_n\end{array}\right)
\textrm{ and
}Y^{(1)}=\left(\begin{array}{c}A^\dagger_{n+1}\\A^\dagger_{n+2}
\\\vdots\\A^\dagger_{2n}\end{array}\right).\end{equation}
Then ${\mathbb W}^{(1)}=X^{(1)}\cdot (Y^{(1)})^T$, so
${\mathbb W}^{(1)}$ is rank $1$.

\bigskip

\begin{Def} Let ${\mathcal P}$ denote the complex algebra
generated by $n^2$ commuting entries $W_{ij}$ of a complex
$n\times n$ matrix ${\mathbb W}$, and let ${\mathcal P}^{(1)}$
denote the complex algebra generated by the above $n^2$
commuting variables $W^{(1)}_{ij}$. Finally, let ${\mathcal
I}_{2\times 2}$ denote the ideal in ${\mathcal P}$ generated by
all $2\times 2$ minors of ${\mathbb W}$. \end{Def}

\medskip

The following is, for the first part obvious, and for the rest,
well-known: \begin{Lem} The spaces ${\mathcal P}^{(1)}$,
${\mathcal P}$, and ${\mathcal I}_{2\times2}$ are ${\mathfrak
k}_\mu\times {\mathfrak k}_\nu$ modules. \end{Lem}

\medskip

\begin{Prop} \label{3.11}\begin{equation} {\mathcal P}^{(1)}={\mathcal
P}/{\mathcal I}_{2\times2}. \end{equation} \end{Prop}

\proof We have a natural homomorphism $\Phi^{(1)}: {\mathcal
P}\rightarrow {\mathcal P}^{(1)}$ given on generators by
$W_{ij}\rightarrow W_{ij}^{(1)}$. The kernel of $\Phi^{(1)}$
contains ${\mathcal I}_{2\times2}$, but must in fact equal since
both spaces decompose in the same way under ${\mathfrak
k}_\mu\times {\mathfrak k}_\nu$. Specifically, the highest
weight vectors for ${\mathfrak k}_\mu\times {\mathfrak k}_\nu$
in ${\mathcal P}^{(1)}$ are the elements
$\left(W^{(1)}_{1,1}\right)^r; r=1,2,\cdots$ \qed

\medskip

The following theorem is then obtained:

\begin{Thm} \label{3.12}The ideal ${\mathcal I}_{2\times2}$ is prime.
\end{Thm}

\medskip

\proof Let $p^{(1)}_1, p^{(1)}_2\in {\mathcal P}^{(1)}$ satisfy
$p^{(1)}_1\cdot p^{(1)}_2=0$. We can assume that $p^{(1)}_i\cdot v_s\neq 0, \ i=1,2$, and $p^{(1)}_1\cdot p^{(1)}_2\cdot v_s=0$. This contradicts Proposition~\ref{new-prime}. \qed

\bigskip

\subsection{Rank $r$}

One can make an analogous construction based on ${\mathfrak
h}_{2nr}$: We label the elements as $A_i^{(s)},
(A_i^{(s)})^\dagger$, $i=1,\dots, 2n$ and $s=1,\dots, r$.

\begin{Def}\label{3.13}Set \begin{eqnarray}F_\beta{(r)}=\sum_{s=1}^r
(A_1{(s)})^\dagger(A_{n+1}{(s)})^\dagger &,&
E_\beta^{(r)}=-\sum_{s=1}^r (A_1{(s)})(A_{n+1}{(s)}),\\
F_{\mu_k}^{(r)}=\sum_{s=1}^r
(A_{k+1}{(s)})^\dagger(A_{k}{(s)}) &,&
E_{\mu_k}^{(r)}=\sum_{s=1}^r
(A_{k+1}{(s)})(A_{k}{(s)})^\dagger,\\
F_{\nu_\ell}^{(r)}=\sum_{s=1}^r
(A_{n+\ell}{(s)})(A_{n+\ell+1}{(s)})^\dagger &,&
E_{\nu_\ell}^{(r)}=
\sum_{s=1}^r
(A_{n+\ell}{(s)})^\dagger(A_{n+\ell+1}{(s)}).
\end{eqnarray}\begin{eqnarray}
H_{k}^{(r)}&=&\sum_{s=1}^r
(A^\dagger_{k}{(s)})(A_{k}{(s)}),\\
H_{\mu_k}^{(r)}&=&\sum_{i=1}^r A_k^\dagger(i)A_k(i)
 -\sum_{i=1}^r A_{k-1}^\dagger(i)A_{k-1}(i)  , k=1,\dots, n-1,\\
 H_{\nu_k}^{(r)}&=&\sum_{i=1}^r A_{n+k}^\dagger(i)A_{n+k}(i)
 -\sum_{i=1}^r A_{n+k+1}^\dagger(i)A_{n+k+1}(i)  , k=1,\dots, n-1,\\
H_\beta^{(r)}&=&r-\sum_{i=1}^rA_1(i)A_1^\dagger(i)-\sum_{i=1}^rA_{n+1}(i)A_{n+1}^\dagger(i)\\
&=&-r-H_1^{(r)}-H_{n+1}^{(r)}
\label{h-beta}.
\end{eqnarray}
\end{Def}

\begin{Prop} The above elements satisfy the Serre relations for
$su(n,n)^{\mathbb C}$. 
\end{Prop}

\pof As in the proof of the case $r=1$, this follows from the Weyl--Serre Equations. \qed

\medskip

Notice that the center in ${\mathfrak k}^{\mathbb C}$ in this version is generated by 
\begin{equation}\Omega^{(r)}_{{\mathfrak k}^{\mathbb C}}=-nr-\sum_{k=1}^{2n}\sum_{i=1}^r A_k^\dagger(i)A_k(i).
\end{equation}

\begin{Def}
\begin{equation}
\Xi^{(r)}=\sum_{k=1}^{n}A_k^\dagger(i)A_k(i)-
A_{n+k}^\dagger(i)A_k(i).
\end{equation}\label{z-r-def}
\end{Def}

One sees easily

\begin{Prop}The element $\Xi^{(r)}$ commutes with all the elements in Definition~\ref{3.13}. \end{Prop}

\smallskip

\begin{Cor}The elements above define a homomorphism 
\begin{equation}\psi^{(r)}: {\mathcal U}(su(n,n)^{\mathbb C})\rightarrow 
{\mathcal H}_{2nr}.\end{equation}
We will occasionally extend this map to ${\mathcal U}(u(n,n)^{\mathbb C})$ by including $\Xi^{(r)}$, and we will also use the name $\psi^{(r)}$ in these cases.
\end{Cor}

\medskip

\begin{Def}
The representation 
$${\mathcal R}^{harm}_r=\pi^{SvN}_{2nr}\circ\psi^{(r)}$$is the ($r$th) Harmonic Representation of ${\mathcal U}(u(n,n)^{\mathbb C})$ in ${\mathcal F}_{2nr}$.
\end{Def}

\medskip

Set

\begin{equation}X^{(k)}=\left(\begin{array}{cccc}A_1^{(1)}&A_1^{(2)}&\cdots&A_1^
{(k)}\\
A_2^{(1)}&A_2^{(2)}&\cdots&A_2^{(k)}
\\\vdots&\vdots&\ddots&\vdots
\\A_n^{(1)}&A_n^{(2)}&\cdots&A_n^{(k)}\end{array}\right)
\textrm{ and
}Y^{(k)}=\left(\begin{array}{cccc}A_{n+1}^{(1)}&A_{n+1}^{(2)}&\cdots&A_{n+1}^{
(k)}\\
A_{n+2}^{(1)}&A_{n+2}^{(2)}&\cdots&A_{n+2}^{(k)}
\\\vdots&\vdots&\ddots&\vdots
\\A_{2n}^{(1)}&A_{2n}^{(2)}&\cdots&A_{2n}^{(k)}
\end{array}\right).\end{equation}

Set \begin{equation}{\mathbb W}^{(k)}=(X^{(k)})(Y^{(k)})^T.\label{XYT}
\end{equation}

Analogously to Lemma~\ref{L1}, we have:

\begin{Lem}For
$k,\ell<n$,$$W^{(r)}_{k+1,\ell}=\ad(F^{r}_{\mu_k})W^{(r)}_{k,\ell}
\textrm{ and
}W^{(r)}_{k,\ell+1}=\ad(F^{(r)}_{\nu_\ell})W^{(1)}_{k,\ell}.$$
\end{Lem}

\medskip

Let ${\mathcal I}_{k\times k}$ denote the ideal in ${\mathcal
P}$ generated by all $k\times k$ minors of ${\mathbb W}$.

\medskip

In complete analogy with the $2\times2$ case we obtain the
following theorem and proposition. The result was first proved
in (\cite{qufd}).

\begin{Thm} \label{prime-th}The ideal ${\mathcal I}_{(k+1)\times
(k+1)}$ is prime in ${\mathcal P}$ for any $k=1,2,\dots$.
\end{Thm}

\medskip

\subsection{Three homomorphisms from ${\mathfrak u}^{\mathbb C}(r)$ to ${\mathcal H}_{2nr}$}

With an eye to applications in \S4, but also of its own interest, we now state:

\begin{Prop}\label{res33}The recipes
\begin{equation}\label{res3}\begin{array}{lcl}\forall i,j=1,\dots,r:\ E_{ij}&\stackrel{\Gamma^{(r)}_1}{\rightarrow}&\sum_{a=1}^nA^\dagger_{a}(i)A_{a}(j),\\
\forall i,j=1,\dots,r:\ E_{ij}&\stackrel{\Gamma^{(r)}_2}{\rightarrow}&\sum_{b=1}^nA^\dagger_{n+b}(i)A_{n+b}(j),\textrm{ and}\\
\forall i,j=1,\dots,r:\ E_{ij}&\stackrel{\Gamma^{(r)}_3}{\rightarrow}&\sum_{a=1}^nA^\dagger_{a}(i)A_{a}(j)-\sum_{b=1}^nA^\dagger_{n+b}(j)A_{n+b}(i)
\end{array}\end{equation} extend to homomorphisms ${\mathcal U}({\mathfrak u}^{\mathbb C}(r))\stackrel{\Gamma^{(r)}_i}{\rightarrow}{\mathcal H}_{2nr}, i=1,2,3$. 
\end{Prop}

\pof This again follows directly from the Serre-Weyl equations. The first two homomorphisms of course take values in two subalgebras isomorphic to ${\mathcal H}_{nr}$, and these subalgebras have trivial intersection. The third homomorphism can be seen as a classical analogue of a Drin'feld Double. See \S\ref{6} \qed

\medskip

\begin{Prop}\label{commut}
An element $h\in{\mathcal H}_{2nr}$ commutes with $\psi^{(r)}({\mathcal U}({\mathfrak u}(n,n)^{\mathbb C})$ if and only $h\in \Gamma^{(r)}_3({\mathcal U}({\mathfrak u}(r)^{\mathbb C})$.
\end{Prop}

We shall prove this result in the next section when more notation is available.

\bigskip

\section{The Kashiwara-Vergne list of unitary representations}

\label{4}

\subsection{The $U(n)\times U(n)\times U(r)\times U(r)$ module ${\mathcal F}_{2nr}$}
We first introduce some convenient notation which also facilitates a comparisons to  Kashiwara--Vergne (\cite{k-v}). Specifically, we consider variables

\begin{equation} \forall i=1,\dots,n; \forall j=1,\dots, r: \
x_{ij}=A^\dagger_i(j)\textrm{ and } y_{ij}=A^\dagger_{n+i}(j).
\end{equation}

We consider the space ${\mathcal P}_{2nr}$ of polynomials in these 
commuting variables. Let ${\mathcal P}_{nr}^{\bf X}$ and ${\mathcal 
P}_{nr}^{\bf Y}$ denote the spaces of polynomials in ${\bf x}$ and ${\bf y}$, 
respectively, so that ${\mathcal P}_{2nr}={\mathcal P}_{nr}^{\bf X}\cdot 
{\mathcal P}_{nr}^{\bf Y}$. Evidently there
is a monomial basis \begin{equation} {\bf x}^{\bf a}{\bf y}^{\bf
b}=\prod_{(i,k)=(1,1)}^{(n,r)}x_{i,k}^{a_{i,k}}
\prod_{(i,k)=(1,1)}^{(n,r)}y_{i,j}^{b_{i,k}} \end{equation}
of ${\mathcal P}_{2nr}$, where ${\mathbf a},{\mathbf b}$ are elements in ${\mathbb N}_0^{nr}$.

\medskip

We will think of the variables $x_{ij}$ and $y_{ij}$ as the entries of 
$n\times
r$ matrices ${\mathbf x}$ and ${\mathbf y}$, respectively. 

\medskip

There is an action of $Gl(n,{\mathbb C})\times Gl(n,{\mathbb C})$
from the left; \begin{equation} \forall a,b\in Gl(n,{\mathbb
C})\times Gl(n,{\mathbb C}): ((a,b)\star p)({\mathbf x},{\mathbf
y})=p(a^{T}\cdot{\mathbf x},b^{T}\cdot{\mathbf y}).\label{lefttot}
\end{equation}

\medskip

We will equip the groups $GL(n,{\mathbb C})$ with a superscript $X$ or $Y$ according to which type of variables they act on. We will also add a subscript $L$ for this left action. Similarly, there is an action on the Lie algebra level. Later there will be a right action of a matrix group which will be labeled with a subscript $R$, and appropriate superscripts. The Lie algebras get labels $X,Y,L$, and $R$, similarly. We maintain the notation ${\mathcal P}_{2nr}$, now as a $gl^X_L(n,{\mathbb C})\times gl^Y_L(n,{\mathbb C})$ module as described above.

\medskip

It is clear that there is a vector space isomorphism \begin{equation}{\mathcal 
I}_{2nr}: {\mathcal P}_{2nr}\rightarrow {\mathcal F}_{2nr}.
\end{equation}
We have that ${\mathcal F}_{2nr}$ is a ${\mathcal U}(u(n,n)^{\mathbb C})$ 
module via $\psi^{(r)}\circ \pi_{2nr}^{SvN}$, and thus ${\mathcal P}_{2nr}$
becomes a $u(n,n)^\mathbb C$ module. We denote this module by ${\mathcal P}^{\mathcal F}_{2nr}$. An easy computation gives that as a $gl_R^X(n,{\mathbb C})\times gl_L^Y(n,{\mathbb C})$ module, it is twisted in comparison to ${\mathcal P}_{2nr}$:\begin{equation}{\mathcal P}^{\mathcal F}_{2nr}={\mathcal P}_{2nr}\otimes_{\mathbb C}{\mathbb C}_{\frac12 r},\end{equation} where ${\mathbb C}_{\frac12 r}$ is the $1$-dimensional module in which each $H_k$ acts as ${\frac12 r}$. Notice in this connection that we work with an ``abstract'' realization of $u(n,n)$ as $\infty\times \infty$ matrices. In our present setting, there is no underlying faithful finite dimensional representation. 

One easily obtains the following expressions for the generators when expressed 
on ${\mathcal P}_{2nr}^{\mathcal F}$:

\begin{Lem}
\begin{eqnarray}\label{e-b-no}
e_\beta^{(r)}&=&-\sum_{i=1}^r\frac{\partial}{\partial
x_{1,i}}\frac{\partial}{\partial y_{1,i}},\\
e_j^{(r)}&=&\sum_{i=1}^r x_{j,i}\frac{\partial}{\partial
x_{j+1,i}}\quad (j=1,\dots,n-1), \\ e_j^{(r)}&=&\sum_{i=1}^r
y_{j,i}\frac{\partial}{\partial y_{j+1,i}}\quad
(j=n-1,\dots,2n-1),\\ f_\beta^{(r)}&=&\sum_{i=1}^r
x_{1,i}y_{1,i},\\ f_j^{(r)}&=&\sum_{i=1}^r
x_{j+1,i}\frac{\partial}{\partial x_{j,i}}\quad
(j=1,\dots,n-1),\\ f_j^{(r)}&=&\sum_{i=1}^r
y_{j+1,i}\frac{\partial}{\partial y_{j,i}}\quad
(j=n+1,\dots,2n-1).\end{eqnarray}
\begin{eqnarray}
H_k^{(r)}&=&\frac{r}2+\sum_{i=1}^r
x_{k,i}\frac{\partial}{\partial x_{k,i}}\quad
(k=1,\dots,n),\\
H_{k+n}^{(r)}&=&\frac{r}2+\sum_{i=1}^r
y_{k,i}\frac{\partial}{\partial y_{k,i}}\quad
(k=1,\dots,n),\\
H_\beta^{(r)}&=&-r-\sum_{i=1}^r
x_{1,i}\frac{\partial}{\partial x_{1,i}}-\sum_{i=1}^r
y_{1,i}\frac{\partial}{\partial y_{1,i}},
\\\Omega^{(r)}_{{\mathfrak k}^{\mathbb C}}&=&-nr-\sum_{k=1}^n\sum_{i=1}^r
x_{k,i}\frac{\partial}{\partial x_{k,i}}-\sum_{k=1}^n\sum_{i=1}^r
y_{k,i}\frac{\partial}{\partial y_{k,i}},
\\
\Xi^{(r)}&=&\sum_{k=1}^n\sum_{i=1}^r
x_{k,i}\frac{\partial}{\partial x_{k,i}}-\sum_{k=1}^n\sum_{i=1}^r
y_{k,i}\frac{\partial}{\partial y_{k,i}}.
\end{eqnarray} \end{Lem}

\medskip

There is an additional group acting on ${\mathcal P}_{2nr}$:

\begin{equation}\label{addi}\forall p\in {\mathcal P}_{2nr}, \forall u\in
U(r): (p\star u)({\mathbf x},{\mathbf y})=p({\mathbf x}\cdot
u^{-1},{\mathbf y}\cdot  u^T). \end{equation}

We may even consider an action of $Gl^X_R(r,{\mathbb C})\times Gl^Y_R(r,{\mathbb C})$:
\begin{equation}\forall p\in {\mathcal P}_{2nr}, \forall c,d\in Gl^X_R(r,{\mathbb C})\times Gl^Y_R(r,{\mathbb C})
: (p\star (c,d))({\mathbf x},{\mathbf y})=p({\mathbf x}\cdot
c^{-1},{\mathbf y}\cdot (d^T)).\label{417}
\end{equation}
We have labeled the groups by $R$ for this right action.
The previous action of $U(r)$ is then the restriction to the diagonal of this, more general, action. Corresponding to this, we let \begin{equation}\label{418}Gl^D_R(r,{\mathbb C})=\{(u,u)\in Gl^X_R(r,{\mathbb C})\times Gl^Y_R(r,{\mathbb C})\mid u\in Gl(r,{\mathbb C})\}\end{equation} denote the diagonal subgroup of the product.

\medskip

These actions can be carried over to ${\mathcal F}_{2nr}$ by letting the two copies of $U(r)$ act trivially (i.e. as $1$) on $v_{2nr}$.

This action clearly commutes with the action of $Gl^X_L(n,{\mathbb C})\times Gl^Y_L(n,{\mathbb C})$ from the left. 

The action of the diagonal sub-algebra is given as follows (c.f. the third formula in (\ref{res3})) 
\begin{equation}\label{Oij}\forall i,j=1,\dots,r: p\star O_{ij}:=p\star (E_{i,j},-E_{j,i})=\left(-\sum_{k=1}^n x_{k,i}\frac{\partial}{\partial x_{kj}}+\sum_{k=1}^n y_{k,j}\frac{\partial}{\partial y_{ki}}\right)p.
\end{equation} 
\bigskip

\noindent{\bf Proof of Proposition~\ref{commut}:} It follows easily from the relations (\ref{32}-\ref{34}) that \begin{equation}\label{commu}\Gamma^{(r)}_3({\mathcal U}({\mathfrak u}(r)^{\mathbb C})\subseteq \left(\psi^{(r)}({\mathcal U}({\mathfrak u}(n,n)^{\mathbb C})\right)'.\end{equation} (The right hand side is the commutant in ${\mathcal H}_{2nr}$).   Let then  $h\in \left(\psi^{(r)}({\mathcal U}({\mathfrak u}(n,n)^{\mathbb C})\right)'$. We may analyze $h$ through its action $\pi^{SvN}(h)$  in ${\mathcal F}_{2nr}$, or directly.

\begin{equation}
h=\sum_{i,j,k,\ell} p_i^{(1)}p_j^{(2)}p_k^{(3)}p_\ell^{(4)},\end{equation}
where $p_i^{(1)}$ is a polynomial on the variables $A^\dagger_a(b), 1\leq a\leq n, 1\leq b\leq r$, and $p_j^{(2)},p_k^{(3)}$, and $p_\ell^{(4)}$ are polynomials on the variables $A_a(b), A^\dagger_{n+a}(b),1\leq a\leq n, 1\leq b\leq r$, and $A_{n+a}(b),1\leq a\leq n, 1\leq b\leq r$, respectively. Consider
\begin{equation}\sum_{i,j}p_i^{(1)}p_j^{(2)}p_k^{(3)}p_
\ell^{(4)},\end{equation}
$p_k^{(3)}p_\ell^{(4)}$ fixed, but arbitrary. This must be invariant under the $u(n)$ defined from the the elements $F_{\mu_k}^{(r)},E_{\mu_k}^{(r)}, k=1,\dots,n-1$. The elements $p_i^{(1)}$ transform under $u(n)$ according to a sum of irreducible representations, and so do elements $p_j^{(2)}$. To pick up the trivial representation, we need the representations to be pairwise duals of each other. It follows that the degree of  homogeneity of $p^{(1)}$ must be the same as that of $p^{(2)}$, though of course in different variables, Indeed, taken together they are  products of first order differential operators. These are {\bf left} invariant. This means that they come from the right action. By similar reckoning for the other variables, it follows that in the representation, $h$ can be represented by an element in ${\mathcal U}(gl(r,{\mathbb C})\times gl(r,{\mathbb C}))$ (acting from the {\bf right}).
Let ${\mathcal U}_D(r)$ be the subalgebra of this algebra generated by the elements $(C,-C^T)$ and let  ${\mathcal U}_X(r)$ be the subalgebra generated by the elements $(C,0)$. Clearly,

\begin{equation}
 {\mathcal U}(gl(r,{\mathbb C})\times gl(r,{\mathbb C}))={\mathcal U}_X(r)\cdot {\mathcal U}_D(r).
\end{equation}

Let 
\begin{equation}
{\mathcal M}=\textrm{Ideal}\left(\{\sum_{s=1}^r
(A_i{(s)})(A_{n+j}{(s)});\ i,j=1,\dots n\}\right), 
\end{equation} 
and let 
\begin{equation}
{\mathcal M}^\dagger= \textrm{Ideal}\left(\{\sum_{s=1}^r
(A_i{(s)})^\dagger(A_{n+j}{(s)})^\dagger;\ i,j=1,\dots n\}\right).
\end{equation}
The generators of  these are easily seen to belong to $\left(\psi^{(r)}({\mathcal U}({\mathfrak u}(n,n)^{\mathbb C})\right)$ and hence $h$ must commute with both ideals. 

We have already established (though left it to the reader) that ${\mathcal U}_D(r)$ commutes with the ideals. If $h$ does not belong to ${\mathcal U}_D(r)$, e.g.
\begin{equation}
h=\sum_i u_X^{(i)}u_D^{(i)},
\end{equation}
with at least one $u_X^{(i)}$ different from a complex constant,
then it follows that there must be a non-trivial element of ${\mathcal U}_X(r)$ that commutes with both ideals. But observe that the elements $A_{n+i}^\dagger(j)$ and $A_{n+k}(\ell)$ are invariant placeholders in the commutators. It follows that there must be a non trivial element in 
 ${\mathcal U}_X(r)$ that commutes with all elements in ${\mathcal H}_{nr}$ (the first $r$ places) and hence even with ${\mathcal H}_{2nr}$. It is well known that the center of the Weyl algebra is ${\mathbb C}$, so this is a contradiction. \qed

\bigskip

We first consider the spaces ${\mathcal P}_{nr}^{\bf X}$ and ${\mathcal P}_{nr}^{\bf Y}$ separately. The respective left-right actions correspond to (\ref{lefttot}) and (\ref{addi}).

\smallskip

\begin{Def}\label{asDef}
\medskip $\forall j=1,\dots,r$:
\begin{equation} \triangle_j({\mathbf
x})=\det\left(\begin{array}{c}x_{1,1}\cdots
x_{1,j}\\x_{j,1}\cdots x_{j,j}\end{array}\right) \textrm{ and }
\tilde\triangle_j({\mathbf
y})=\det\left(\begin{array}{c}y_{1,k-j+1}\cdots
y_{1,k}\\y_{j,k-j+1}\cdots y_{j,k}\end{array}\right).
\end{equation}
\end{Def}

\medskip

The following is a well-known result of De Concini, Eisenbud, and Procesi (\cite{EPdC}):

\begin{Prop}[\cite{EPdC}] \label{asDef2}Let $c_1,\dots,c_i,\cdots, c_m$, where ${\mathfrak m}=\min\{n,r\}$, be non-negative integers. Then
\begin{equation}\triangle({\mathbf x})^{\mathbf
c}= \prod_{k=1}^{\mathfrak m}
\triangle^{c_k}_k({\mathbf x})\label{element} \end{equation} is a
highest weight vector for the  combined action of $Gl_L^X(n,{\mathbb C})\times Gl^X_R(r,{\mathbb C})$ on ${\mathcal P}_{nr}^X$, and any highest weight vector is of this form. There are no multiplicities. The corresponding highest weight
for  $Gl^X_R(r,{\mathbb C})$, given as a linear functional on the diagonal of 
$gl^X_R(r,{\mathbb C})$, is \begin{eqnarray}
&\nonumber\Lambda_{gl^X_R(r)}=\\&(c_1+\dots+
c_i,c_2+\dots+c_{\mathfrak m},\dots,c_{\mathfrak m},\underbrace{0,\dots,0}_{r-{\mathfrak m}}), 
\end{eqnarray} \label{p4.5} 
and, analogously,  
\begin{eqnarray}
&\nonumber\Lambda_{gl_L^X(n)}=\\&(c_1+\dots+
c_i,c_2+\dots+c_{\mathfrak m},\dots,c_{\mathfrak m},\underbrace{0,\dots,0}_{n-{\mathfrak m}}). 
\end{eqnarray}
(The pair $(\Lambda_{gl_L^X(n)},\Lambda_{gl^X_R(r)})$ is often called a  ``double tableaux''.) Let $V^X_{LR;{\mathfrak m}}$ denote the corresponding representation.

Similarly, let $d_1,\dots,d_{\mathfrak m}$ be non-negative integers. Then
\begin{equation}\tilde\triangle({\mathbf y})^{\mathbf d}=  \prod_{\ell=1}^{\mathfrak m}
\tilde\triangle^{d_\ell}_\ell({\mathbf y})\end{equation} is a
highest weight vector for the combined action of 
$Gl_L^Y(n,{\mathbb C})\times Gl^Y_R(r,{\mathbb C})$ on ${\mathcal P}_{nr}^Y$. Again, this is the most general form and there are no multiplicities.  The highest weight
for the $Gl^Y_R(r,{\mathbb C})$, given as a linear functional on the diagonal of 
$gl^Y_R(r,{\mathbb C})$, is \begin{eqnarray}
&\nonumber\Lambda_{gl^Y_R(r)}=\\&(\underbrace{0,\dots,0}_{r-{\mathfrak m}},-d_{\mathfrak m},\dots,-d_{\mathfrak m}-d_{{\mathfrak m}-1}\dots
-d_{1}).\quad
\end{eqnarray}
Here we get
\begin{eqnarray}
&\nonumber\Lambda_{gl^Y(n)}=\\&(d_1+d_2+\cdots+d_{\mathfrak m},d_2+\cdots+d_{\mathfrak m} ,\cdots, d_{\mathfrak m},\underbrace{0,\dots,0}_{n-{\mathfrak m}}),\quad
\end{eqnarray}
and we let $V^Y_{LR;{\mathfrak m}}$ denote the corresponding representation.
    \end{Prop}

\medskip

If ${\mathcal N}_{\mathfrak m}$ denotes the set of such ${\mathfrak m}$ tuples, then 
\begin{equation}
{\mathcal P}_{nr}^X=\oplus_{{\mathbf a}\in {\mathcal N}_{\mathfrak m}}V^X_{LR;{\mathbf a}}.
\end{equation}
 Likewise, 
\begin{equation}
{\mathcal P}_{nr}^Y=\oplus_{{\mathbf b}\in {\mathcal N}_{\mathfrak m}}V^Y_{LT;{\mathbf b}}.
\end{equation}

\smallskip

We then have the following easy corollary, where we still use the restriction of the action (\ref{addi}) to the relevant spaces:
\begin{Cor}\label{right-form}If a $Gl_L^X(n,{\mathbb C})$ highest weight vector $p\in {\mathcal P}_{nr}^X$ of highest weight ${\mathbf a}=(a_1,\cdots,a_{\mathfrak m},\underbrace{0,\cdots,0}_{n-{\mathfrak m}})$ is given, then there exists $u\in{\mathcal U}^+(gl^X_R(r,{\mathbb C}))$ such that \begin{equation}p\star u=\triangle({\mathbf x})^{\mathbf
a}.\end{equation}
If a $Gl_L^Y(n,{\mathbb C})$ highest weight vector $\hat p\in {\mathcal P}_{nr}^Y$ of highest weight $(b_1,\cdots, b_{\mathfrak m},\underbrace{0,\cdots,0}_{n-{\mathfrak m}}))$ is given, then there exists $u\in{\mathcal U}^+(gl^Y_R(r,{\mathbb C}))$ such that \begin{equation}\tilde p\star  u=\tilde\triangle({\mathbf y})^{\mathbf b}.\end{equation}
\end{Cor}

\medskip

\subsection{Highest weighs in two different bases}
\label{sub4.4}
Set ${\mathbf e}=(\underbrace{1,1,\cdots,1}_n)$. Any highest weight vector $p^X_{\mathbf a}p^Y_{\mathbf b}$ in ${\mathcal P}_{2nr}$ of highest weight $({\mathbf a},{\mathbf b})$ defines a highest weight vector $p^X_{\mathbf a}p^Y_{\mathbf a}\cdot v_{2nr}$ for $u(n)\times u(n)$ in ${\mathcal F}_{2nr}$  whose weight $\Lambda_{{\mathbf a},{\mathbf b}}=({\mathbf a}+\frac{r}2{\mathbf e}, {\mathbf b}+\frac{r}2{\mathbf e})$, we want to write according to our other parametrization; \begin{equation}\label{a,b}\Lambda_{{\mathbf a},{\mathbf b}}=((i_1,\cdots,i_{n-1},0),(j_1,\cdots,  j_{n-1},0),\lambda_\beta,\zeta),\end{equation} where $(i_1,\cdots,i_{n-1},0)$ and $(j_1,\cdots,  j_{n-1},0)$ determine the same weights of ${\mathfrak k}_L^{\mathbb C}$ and ${\mathfrak k}_R^{\mathbb C}$, respectively, as does $({\mathbf a},{\mathbf b})$, and where
\begin{equation}
H^{(r)}_\beta p^X_{\mathbf a}p^Y_{\mathbf b}\cdot v_{2nr}=\lambda_\beta \cdot p^X_{\mathbf a}p^Y_{\mathbf b}\cdot v_{2nr}\textrm{ and }\ \Xi^{(r)} p^X_{\mathbf a}p^Y_{\mathbf b}\cdot v_{2nr}=\zeta \cdot p^X_{\mathbf a}p^Y_{\mathbf b}\cdot v_{2nr}.\end{equation}
Notice that e.g. $x_{11}\frac{\partial}{\partial x_{11}}$ applied to an element as in (\ref{element}) gives $(c_1+c_2+\dots+ c_{\mathfrak m})$ times the element, it follows that
\begin{eqnarray}
i_k&=&a_k-a_n,\ k=1,\dots, n-1,\\
j_k&=&b_k-b_n,\ k=1,\dots, n-1,\\
\lambda_\beta&=&-r-a_1-b_1=-r-i_1-j_1-(a_n+b_n),\textrm{ and}\label{l1-b1}\\
\zeta&=&\sum_i a_i-\sum_j b_j=n(a_n-b_n)+\sum_k(i_k-j_k).
\end{eqnarray}Notice that if we set\begin{equation}\Omega_{{\mathfrak k}^{\mathbb C}}^{(r)}p^X_{\mathbf a}p^Y_{\mathbf b}\cdot v_{2nr}=\lambda_c\cdot p^X_{\mathbf a}p^Y_{\mathbf b}\cdot v_{2nr},
\end{equation}then $\lambda_c=-nr- \sum_i a_i-\sum_j b_j$. A small computation reveals that the two parameters $\lambda_\beta,\lambda_c$ are connected by the equation
\begin{equation}
\lambda_\beta = -i_1-j_1+\frac1{n}(\lambda_c+\sum_k(i_k+j_k)).
\end{equation}

\medskip

We further notice that
\begin{equation}\label{l-b}
\lambda_\beta\leq -r-i_1-j_1.
\end{equation}
Hence, $b:=-\lambda_\beta-r-i_1-j_1\geq 0$,
It follows by linear algebra that to a given set $((i_1,\cdots,i_{n-1},0), (j_1,\cdots,j_{n-1},0),\lambda_\beta)$ with $\lambda_\beta$ as in (\ref{l1-b1}), the following set of $({\mathbf a},{\mathbf b})$ correspond to this;
\begin{equation}
{\mathbf a}=\left(\begin{array}{c}i_1+a\\i_2+a\\\hdots\\i_{n-1}+a\\a\end{array}\right)\;\ {\mathbf b}=\left(\begin{array}{c}j_1+b-a\\j_2+b-a\\\hdots\\j_{n-1}+b-a\\b-a\end{array}\right), \ 0\leq a\leq b.
\end{equation}
The value $\zeta(a)$ corresponding to a solution determined by the parameter $a$ is \begin{equation}\zeta(a)=n(2a-b)+\sum_ki_k-\sum_kj_k.\label{440}
\end{equation}

\medskip

The value $\zeta$ above is important later when we discuss multiplicities, but for the next several sections it will be omitted, since we here discuss ${\mathcal U}(su(n,n)^{\mathbb C})$ modules in which $\Xi$ is a constant, $\zeta$, times the identity.

\bigskip

\subsection{Top pluri-harmonic elements}

Recall (see e.g. \cite[\S1.5]{r-v}) that unitarity on a highest weight module $M$ for
$su(n,n)^{\mathbb C}$ is defined in terms of the Hermitian form $\langle\cdot,\cdot\rangle$ uniquely
given by the anti-linear anti-involution $\omega$,
\begin{eqnarray}\omega(E_\alpha)&= &F_\alpha\textrm{ if $\alpha$
is compact, and }\\ \omega(E_\beta)&=&-F_\beta. \end{eqnarray}
Then, $su(n,n)$ is the $-1$ eigenspace of $\omega$ in $su(n,n)^{\mathbb C}$.

The defining property is
\begin{equation}
\forall v_1,v_2\in M,\ \forall X\in su(n,n)^{\mathbb C}: \langle Xv_2,v_1\rangle=\langle v_2,\omega(X)v_1\rangle.
\end{equation}

\medskip

It is easy to see that
\begin{equation}
\forall X\in su(n,n)^{\mathbb C}: ({\mathcal R}^{harm}_r(X))^*={\mathcal 
R}^{harm}_r(\omega(X)).
\end{equation}

\medskip

The representation ${\mathcal R}^{harm}_r$ of $su(n,n)^{\mathbb C}$ in the 
(pre-)Hilbert space ${\mathcal F}_{2nr}$ is obviously unitarizable.

\medskip

If we take the cyclic span of the vector $v_{2nr}$ we get in particular:

\begin{Prop} \label{4.1} In the space \begin{equation}{\mathcal
H}_{(0,0,-1)}=Span_{\mathbf
C}\{W^{(1)}_{i_1,j_1}W^{(1)}_{i_2,j_2}\cdots
W^{(1)}_{i_r,j_r}\cdot v_0\mid r\geq0, i_1,\dots, j_r\in
\{1,\dots,n\}\} \end{equation} there is a unitarizable
representation of $su(n,n)$ of highest weight $(0,0,-1)$.
\end{Prop}

\bigskip

We now wish to decompose ${\mathcal F}_{2nr}$ completely under ${\mathcal R}_r^{harm}$.
Clearly, we then need to  look for vectors $v_\Lambda\in {\mathcal F}_{2nr}$ that are weight vectors for the Cartan subalgebra of weight $\Lambda$ and
satisfy

\begin{eqnarray}\label{1} E_{\beta}^{(r)}v_\Lambda&=&0,\\
E_{\mu_k}^{(r)}v_\Lambda&=&0,\textrm{ and}\\ E_{\nu_k}^{(r)}v_\Lambda&=&0,\label{33}
\end{eqnarray}

or, equivalently, weight vectors $v_\Lambda$ of weight $\Lambda$ that satisfy

\begin{eqnarray} \sum_{s=1}^r
(A_1^{(s)})(A_{n+1}^{(s)})v_\Lambda&=&0\label{top-beta},\\ \sum_{s=1}^r
(A_{k+1}^{(s)})(A_{k}^{(s)})^\dagger v _\Lambda&=&0,\textrm{ and}\\
\sum_{s=1}^r
(A_{n+\ell}^{(s)})^\dagger(A_{n+\ell+1}^{(s)})v_\Lambda&=&0.
\end{eqnarray}

\begin{Def}
We call a non-zero vector $v_\Lambda$ {\bf top pluri-harmonic}  of weight
 $\Lambda$ if it 
satisfies (\ref{1}-\ref{33}) and is a weight vector of $su(n,n)$  of weight $\Lambda$. Equivalently, if  ${\mathcal U}^+v_\Lambda=0$. If it furthermore is an eigenvector of $\Xi^{(r)}$ of eigenvalue $\zeta$, we say that it has weight $(\Lambda,\zeta)$.
\end{Def}

As previously, $v_{2nr}$ denotes the vacuum vector in ${\mathcal
F}_{2nr}$. Of course, it is top pluri-harmonic.

The origin of this terminology can be found in (\cite{k-v}).

\smallskip

It is clear that any top pluri-harmonic vector will determine an invariant 
sub-module. Furthermore, a maximal set of orthogonal top 
pluri-harmonic vectors will yield the full decomposition. 

\medskip

Let us first look at the case $r=1$, i.e. ${\mathcal F}_{2n}$: It is easy to see that this space transforms under $u(n)\times u(n)$ as sums of symmetrized tensor products of the defining representation and that one indeed has:
\begin{Prop}As a $u(n)\times u(n)$ module,\label{lo-le}
\begin{equation}
{\mathcal F}_{2n}=\oplus_{N,S\in {\mathbb N}_0}V_{N,S},
\end{equation}
where $V_{N,S}$ is the irreducible module with highest weight vector $(A_{1}^\dagger)^N
(A_{n+1}^\dagger)^Sv_{2n}$.

The top pluri-harmonic elements are exactly the elements
\begin{equation}(A_{1}^\dagger)^Nv_{2n}\textrm{ and }(A_{n+1}^\dagger)^Sv_{2n};\ N,S\in{\mathbb N}_0.\end{equation}
We set
\begin{equation}
V^L_{N}=V_{((N,0,\cdots,0),{\mathbf 0},-1-N)}\textrm{ and }V^R_{S}=V_{({\mathbf 0},(S,0,\cdots,0),-1-S)}
\end{equation}
for the corresponding ${\mathfrak k}^{\mathbb C}$ modules, written in accordance with (\ref{a,b}) but ignoring the value $\zeta$.

Let $\hat M(V^L_N)$ and $\hat M(V^R_S)$ denote the unitary  representations of $su(n,n)$ in ${\mathcal F}_{2n}$ generated by the above top pluri-harmonic elements.

The $su(n,n)$ highest weights are, respectively, written according to (\ref{a,b}) but ignoring the value $\zeta$,
\begin{equation}
V^L_{N}=V_{((N,0,\cdots,0),{\mathbf 0},-1-N)}\textrm{ and }V^R_{S}=V_{({\mathbf 0},(S,0,\cdots,0),-1-S)}.
\end{equation}

Then, as a $su(n,n)$ module,
\begin{equation}
{\mathcal F}_{2n}=\oplus_{N=0}^\infty \hat M(V^L_{N})\bigoplus \oplus_{S=1}^\infty \hat M(V^R_{S}).
\end{equation}

These representations are irreducible. They are indeed {\it the} irreducible quotients of the  corresponding generalized Verma modules $M(V^L_{N})$ and $M(V^R_{S})$, respectively.

The ${\mathfrak k}^{\mathbb C}$-types are as follows:
\begin{eqnarray}\label{k-type}
\hat M(V^L_{N}):&& \textrm{ ${\mathfrak k}^{\mathbb C}$-types: } \oplus_{i\in{\mathbb N}_0}V_{((N+i,0,\cdots),(i,0,\cdots), 
-N-2i-1)},\textrm{ and}\\
\hat M(V^R_{S}):&& \textrm{ ${\mathfrak k}^{\mathbb C}$-types: } \oplus_{j\in{\mathbb N}_0} V_{((j,0,\cdots),(S+j,0,\cdots), 
-S-2j-1)},
 \end{eqnarray}
corresponding to the highest weight vectors $(A_{1}^\dagger)^{N+i}(A_{n+1}^\dagger)^{i}v_{2n}$ and $(A_{1}^\dagger)^{j}(A_{n+1}^\dagger)^{S+j}v_{2n}$, respectively.
\end{Prop} 
 
 \pof The first statement about the ${\mathfrak k}^{\mathbb C}$-types is obvious, and so is the statement about the highest weight vectors. The irreducibility and the content of ${\mathfrak k}^{\mathbb C}$-types of the individual representations go hand in hand: The spectrum of ${\mathfrak k}^{\mathbb C}$-types is in a very obvious way $1$-dimensional. If some $\hat M(V^L_{N})$ or $\hat M(V^R_{S})$ contained fewer ${\mathfrak k}^{\mathbb C}$-types than the stated (i.e. that some types were ``missing''), there would be a gap in the spectrum which would force the unitarizable $su(n,n)$ representation to be finite-dimensional. The only such is the trivial representation, and this is clearly not one of ours since the vacuum vector is not annihilated by $F_\beta$.     \qed
 
 \medskip
 
\begin{Rem} We here encounter for the first time the notion of ``missing'' ${\mathfrak k}^{\mathbb C}$-types.   We return to these notions below and their relation to homomorphisms between Verma modules.  In this context another proof of the above will be furnished as a special case of the general analysis.

The representations above are generalizations of the so-called ladder representations (see e.g.\cite{mack}). Even though they contain all the mentioned ${\mathfrak k}^{\mathbb C}$-types, they have very thin sets of ${\mathfrak k}^{\mathbb C}$-types and in a very specific sense, many other ${\mathfrak k}^{\mathbb C}$-types are ``missing'' (see below). They are the spaces of the mass $0$, spin $N/2$, or spin $S/2$ particles, though it might be better to talk about helicity instead of spin.
 \end{Rem}
 \medskip

 More generally,  in ${\mathcal F}_{2nr}$ there are some obvious top pluri-harmonic vectors: Let us denote by ${\mathcal P}_u^{(r)}$ the span of vectors of the form
\begin{equation}
p_u((A_1(1)^\dagger,\cdots, (A_n(r)^\dagger)v_{2nr},
\end{equation}
and, likewise, by 
${\mathcal P}_d^{(r)}$ the span of vectors of the form
\begin{equation}
p_d((A_{n+1}(1)^\dagger,\cdots, (A_{2n}(r)^\dagger)v_{2nr}.
\end{equation}
Both  spaces are clearly invariant under the action of ${\mathfrak k}^{\mathbb C}$, an action that also preserves degrees.

${\mathcal P}_u^{(r)}$ has a trivial action of ${\mathfrak k}_R^{\mathbb C}$ and decomposes under ${\mathfrak k}_L^{\mathbb C}$ into the $r$th fold tensor products of the symmetric tensor representations $(a,0,\cdots)$, $a\in{\mathbb N}$. For $r<n$, any highest weight representation of the form
\begin{equation}((a_1,\cdots,a_r,0,\cdots),{\mathbf 0},-a_1-r)\end{equation} is obtained. For $r\geq n$  any highest weight representation of the form \begin{equation}((a_1,\cdots,a_{n-1},0),{\mathbf 0},-a_1-\tilde r),\tilde r= r, r+1,r+2,\cdots\end{equation} is obtained. 

Similarly, ${\mathcal P}_d^{(r)}$ has a trivial action of ${\mathfrak k}_L^{\mathbb C}$ and decomposes under ${\mathfrak k}_R^{\mathbb C}$ into the $r$th fold tensor products of the symmetric tensor representations $(b,0,\cdots)$, $b\in{\mathbb N}$. For $r<n$, any highest weight representation of the form \begin{equation}({\mathbf 0},(b_1,\cdots,b_r,0,\cdots),-b_1-r)\end{equation} is obtained. For $r\geq n$  any highest weight representation of the form \begin{equation}({\mathbf 0},(b_1,\cdots,b_{n-1},0),-b_1-\tilde r),\tilde r= r, r+1,r+2,\cdots\end{equation} is obtained. 

\medskip

\begin{Lem}\label{left-right} Let $p_{u}v_{2nr}$ be a highest weight vector in ${\mathcal P}_u^{(r)}v_{2nr}$ of highest weight as mentioned right above.  Then it is top pluri-harmonic.

Likewise, Let $p_{d}v_{2nr}$ be a highest weight vector in ${\mathcal P}_d^{(r)}v_{2nr}$ of highest weight as mentioned right above. Then it is top pluri-harmonic. 
   \end{Lem}

\proof We need only examine the operators (\ref{top-beta}). Since the summands contain both $A_1(s)$ and $A_{n+1}(s)$, the stated vectors will be annihilated. \qed

\medskip

Another important observation, which follows from the above, or directly from (\cite{EPdC}), is

\begin{Prop}\label{most-gen}If $r<n$, the most general ${\mathfrak k}^{\mathbb C}$-type in ${\mathcal F}_{2nr}$ is of the form
\begin{equation}
V_{(a_1,\cdots,a_r,0,\cdots),(b_1,\cdots,b_r,0,\cdots),-a_1-b_1-r)}.
\end{equation}
If $r\geq n$, the most general ${\mathfrak k}^{\mathbb C}$-type in ${\mathcal F}_{2nr}$ is of the form
\begin{equation}
V_{(a_1,\cdots,a_{n-1},0),(b_1,\cdots,b_{n-1},0,),-a_1-b_1-\tilde r)},\textrm{ with }a_{n-1},b_{n-1}\geq0,\textrm{ and } \tilde r=r, r+1,r+2,\dots.
\end{equation}
The multiplicities of these types have their origin in the $u(r)$ module structures:

\smallskip

  \underline{If $r<n$} the multiplicity of $V_{(a_1,\cdots,a_r,0,\cdots),(b_1,\cdots,b_r,0,\cdots),-a_1-b_1-r)}$ is given as the product of the dimensions of the  irreducible representations of $u(r)$ of highest weights $(a_1,\cdots,a_r)$ and $(b_1,\cdots,b_r)$, respectively. 
  
 \medskip

  \underline{If $r\geq n$} the multiplicity of the ${\mathfrak k}^{\mathbb C}$-type 
 $V_{((n_1,\cdots,n_{n-1},0),(m_1,\cdots,m_{n-1},0,\cdots),\lambda_\beta)}$ in ${\mathcal F}_{2nr}$ is determined as follows:  Set $\Gamma=-\lambda_\beta-r-n_1-m_1$. Then the multiplicity is zero unless   $\Gamma$ is a non-negative integer. Set ${\mathbf e}=(1,1,\cdots,1)$ then then pair ${\mathbf a},{\mathbf  b}$, where 
 \begin{eqnarray}
 {\mathbf a}&=(a_1,a_2,\cdots,a_n)=&(n_1,\cdots,n_{n-1},0)+a_n\cdot {\mathbf e}\\
 {\mathbf b}&=(b_1,b_2,\cdots,b_n)=&(m_1,\cdots,m_{n-1},0)+n_n\cdot {\mathbf e}
 \end{eqnarray}
 corresponds to $V_{((n_1,\cdots,n_{n-1},0),(m_1,\cdots,m_{n-1},0,\cdots),\lambda_\beta)}$ precisely when $a_n+b_n=\Gamma$ and $a_b,b_n\in{\mathbb N}_0$. Let $A_n$ denote the dimension of the irreducible finite-dimensional representation of $u(r)$ of highest weight $(a_1,a_2,\cdots,a_n,\underbrace{0,0,\cdots,0}_{r-n})$ determined as above,  and let $B_n$ denote the dimension of the irreducible finite-dimensional representation of $u(r)$ of highest weight $(b_1,b_2,\cdots,b_n,\underbrace{0,0,\cdots,0}_{r-n})$ also determined as above.
 
 \smallskip
 
 The multiplicity of ${\mathfrak k}^{\mathbb C}$-type 
 $V_{((n_1,\cdots,n_{n-1},0),(m_1,\cdots,m_{n-1},0,\cdots),\lambda_\beta)}$ in ${\mathcal F}_{2nr}$ is
  then given as 
  \begin{equation}
  \sum_{\stackrel{a_n+b_n=\Gamma}{a_n,b_n\in {\mathbb N}_0 }} A_n\cdot B_n.
  \end{equation}
\end{Prop}

\medskip

\begin{Cor}\label{mis-cor}If $r<n$, ${\mathcal  F}_{2nr}$ is missing all ${\mathfrak k}^{\mathbb C}$-types 
\begin{equation}
((a_1,\cdots,a_x,0,\cdots),(b_1,\cdots,b_{y},0,\cdots ),-a_1-b_1-r)\textrm{ with }x>r\textrm{ or }y>r.
\end{equation}
\end{Cor}

\medskip

In this sense, the general case $r\in{\mathbb N}$ is similar to the case $r=1$. However, the general case is much more complicated; there are other representations to take into account, and the group $u(r)$ becomes much more important.

\bigskip

\subsection{The Littlewood-Richardson Rule}

For a statement and proof of this celebrated rule, see \cite[p. 60-65]{l-w}.

\begin{Lem}[The Littlewood-Richardson Rule, very special case]                    
\begin{equation}
V_{(k_1,\cdots,k_r,0,\cdots)}\otimes V_{(a,0,\cdots)}=\oplus_{(j_1,j_2,\cdots,j_{r+1},0,\cdots)\in T}V_{(j_1,j_2,\cdots,j_{r+1},0,\cdots)}.
\end{equation}The set $T$ is defined as follows:  $(j_1,j_2,\cdots,j_{r+1},0,\cdots)\in T$ if and only if it satisfies:
\begin{equation}j_1\geq k_1, \sum_\alpha j_\alpha=a+\sum_\beta i_\beta\textrm{ and }
j_x>k_x\Rightarrow j _x\leq k_{x-1} (x=2,\dots,r+1).
\end{equation}      \label{L-R}               	              	                 
\end{Lem}

\bigskip

\subsection{On ``missing'' ${\mathfrak k}^{\mathbb C}$-types}

\label{sub-missing}

\medskip

We now, briefly, describe ``missing ${\mathfrak k}^{\mathbb C}$-types'':

 Until further notice, we consider a general Lie algebra ${\mathfrak g}^{\mathbb C}$ associated to a Hermitian symmetric space and we let ${\mathfrak g}^{\mathbb C}={\mathfrak p}^{-}\oplus {\mathfrak k}^{\mathbb C}\oplus {\mathfrak p}^{+}$ be a standard decomposition (\cite[p. 589]{H-C-2}).

\medskip

Recall that the Generalized Verma Module $M(V_\tau)$ defined by a 
finite dimensional (irreducible) ${\mathfrak k}^{\mathbb C}$-module $V_\tau$ is
\begin{equation}
M(V_\tau)={\mathcal U}\otimes_{{\mathcal U}({\mathfrak p}^+\oplus{\mathfrak 
k})}V_\tau,
\end{equation}
where we let ${\mathcal U}({\mathfrak p}^+)$ act trivially on $V_\tau$. The 
action is then from the left. The action of ${\mathfrak k}^{\mathbb C}$ on  ${\mathcal U}({\mathfrak p}^-)$ is well understood. It preserves degree. If 
\begin{equation}
 {\mathcal U}({\mathfrak p}^-)_d=\oplus_i V_{0,d,i}
\end{equation}
is the  decomposition of the degree $d$ elements into irreducible 
representations, we shall refer to the representations  $V_{0,d,i}$ as the $d$th 
order representations. Clearly, the ${\mathfrak k}^{\mathbb C}$-types of $M(V_\tau)$ are
\begin{equation}
 M(V_\tau)=\tau \otimes \left(\oplus_{d=0}^\infty\oplus_i V_{0,d,i}\right).
\end{equation}

Let $v_\tau$ be an essentially unique highest weight vector of $M(V_\tau)$ 
and let $\tilde M(V_\tau)$ be a quotient 
module of $M(V_\tau)$ in which the image $\tilde v_\tau$ of $v_\tau$ is 
non-zero. We will quite generally call such modules  \underline{top 
quotients.} There is no demand of irreducibility, so a generalized Verma module 
itself is also considered a top quotient.

\begin{Def}
 An irreducible representation $\mu_0\in \tau \otimes \left(\oplus_i 
\mu_{d,i}\right)$ is said to be  
missing from $\tilde M(V_\tau)$  if its multiplicity in $\tilde M(V_\tau)$ is smaller than its multiplicity in $M(V_\tau)$. 
\end{Def}

\medskip

\begin{Rem}\label{cov-dif}In this context, if there are missing ${\mathfrak k}^{\mathbb C}$-types, there is also a covariant differential operator. Specifically, a missing ${\mathfrak k}^{\mathbb C}$-type $V_{\tau_s}$ of smallest possible degree defines a homomorphism
\begin{equation}
M(V_{\tau_s})\stackrel{\psi}{\rightarrow} M(V_{\tau}).
\end{equation}
The covariant differential operator $D_{\mathcal\psi}$ is then the dual
of this map. And vice versa. 

Furthermore, the null-space of this covariant differential operator is a highest weight module for the opposite algebra ${\mathcal U}({\mathfrak p}^-\oplus {\mathfrak k}^{\mathbb C})$.

\smallskip

The map $\psi$ is completely determined by the map $\psi_0:V_{\tau_s}\rightarrow M(V_\tau)$ which we can view as an element \begin{equation}\label{cov-dif-0}
{\psi_0}\in{\mathcal U}({\mathfrak p}^-)\otimes Hom_{\mathbf C}(V_{\tau_s},V_\tau).
\end{equation}

By duality this directly translates into 
\begin{equation}
D_{\psi_0}\in{\mathcal U}({\mathfrak p}^+)\otimes Hom_{\mathbf C}(V'_{\tau},V'_{\tau_s}),
\end{equation}
where the elements in ${\mathcal U}({\mathfrak p}^+)$ are easily read off and may be interpreted as differential operators on ${\mathfrak p}^-$. 

\smallskip

See (\cite{harrish} and \cite{basic} for details about this duality.\end{Rem}

\medskip

Suppose now that a first order ${\mathfrak k}^{\mathbb C}$-type 
  $V_{\tau_1}$ is missing from $\tilde M(V_\tau)$. 
 The assumption is equivalent to the existence of a ${\mathfrak k}^{\mathbb C}$-type $V_{\tau_1}$ in ${\mathfrak p}^-\otimes V_\tau$ such that there is a non-trivial homomorphism
\begin{equation}
\tilde M(V_{\tau_1})\rightarrow \tilde M(V_\tau).
\end{equation}
 The highest weights are related by
 \begin{equation}
 \Lambda_{\tau_1}=\Lambda_{\tau}-\gamma,
 \end{equation}
where $\gamma$ is a weight of ${\mathfrak p}^+$.
\begin{Lem}
Under the above assumptions,
\begin{equation}
\langle \Lambda_\tau+\rho,\gamma\rangle=1.
\end{equation}
\end{Lem}
\pof This follows from Harish-Chandra \cite[p. 41]{h-c}, cf. the proof by Verma (\cite{Verma}). \qed

\bigskip

The following is straightforward:

\begin{Lem}
The inner product on first order polynomials is given as
\begin{equation}
\langle W_{-\gamma_1}v_1, W_{-\gamma_2}v_2\rangle=-\langle v_1, Z_{\gamma_1}W_{-\gamma_2}v_2\rangle=-\lambda \cdot
\delta_{\gamma_1,\gamma_2}\langle v_1, v_2\rangle +\ l.o.t.
\end{equation}
More generally, the inner product between polynomials $p_1v_1,p_2v_2$ of degree $d_1,d_2$, respectively, is given as 
\begin{equation}
(-\lambda)^{d-1}\delta_{d_1,d_2}	\langle p_1,p_2\rangle\langle v_1,v_2\rangle + l.o.t.
\end{equation}
\end{Lem}

\medskip

Recall that we have unitarity for $\lambda\in{\mathbb R}$ sufficiently small. The lemma leads directly to the last possible place of unitarity:

\begin{Cor}
Let $\lambda_{crit}$ be the smallest $\lambda$ for which there exists a non-compact positive root $\gamma$ such that $\Lambda-\gamma$ is a ${\mathfrak k}^{\mathbb C}$ highest weight in ${\mathfrak p}^-\otimes V_\Lambda$ and such that $\langle \lambda+\rho,\gamma\rangle=1$. Then there can be no unitarity for $\lambda>\lambda_{crit}$.
\end{Cor}

\pof It follows that the inner product on any ${\mathfrak k}^{\mathbb C}$-type in ${\mathfrak p}^-\otimes V_\Lambda$ is of the form $\lambda\cdot I + \ l.o.t.$. There is then a unique value where a such type vanishes. After that the inner product becomes negative as $\lambda$ increases. It is easy to see that a vanishing first order polynomial $p_\gamma$ of some ${\mathfrak k}^{\mathbb C}$ highest weight $\Lambda_\gamma$ leads to a homomorphism. Specifically, 
\begin{equation}
\forall v\in V_\Lambda:\ \langle p_\gamma,W_{-\delta}v\rangle=0,
\end{equation}
hence $Z^+_{\delta}p_\gamma=0$. 
\qed

\medskip

Consider $\Lambda=\left((i_1,\dots,i_x,0,\dots),(j_1,\dots,j_y,0,\dots),\lambda)\right)$ with $x,y\leq n-1$. (recall \S\ref{sub4.4}.) It is easy to see that $\lambda_{crit}$ is attained at the first order polynomial corresponding to the representation  $\Lambda_2=\left((i_1,\dots,i_x,1,0,\dots),(j_1,\dots,j_y,1,0,\dots),\lambda\right)$. The corresponding critical value then satisfies the equation
\begin{equation}\label{4.35}
\lambda_{crit}+1+i_1+j_1+x+y=1\textrm{ that is, }\lambda_{crit}=-i_1-j_1-x-y.
\end{equation}

\medskip

\begin{Def}
The value $\lambda_{crit}$ is called the last possible place of unitarity.
\end{Def}

\begin{Rem}
The word ``last'' above has its origins in the history of the subject. One imagines that one starts at very negative values of $\lambda$ where the representations are square integrable and then proceeds towards bigger values. At very small  values of $\lambda$ one has unitarity for all $\lambda$ if one passes to the universal covering group. Increasing $\lambda$  towards the last possible place one eventually passes through a finite discrete set; the so-called Wallach set (\cite{wallach}, \cite{r-v}).
\end{Rem}

\medskip

Along the same lines, we note:

\begin{Prop}\label{uni-prop}
If a highest weight module $\tilde M(V_\tau)$ is unitarizable, then it is irreducible.
\end{Prop}

\pof If not, there will be an element $p$ of weight different from $v_\tau$ which is annihilated by all $Z+$. The norm of this element will then be zero, hence $p=0$. \qed

\bigskip

\subsection{The irreducible representations occurring in the ${\mathcal U}(u(n,n)^{\mathbb C})$ module ${\mathcal F}_{2nr}$.}

We first determine the irreducible summands. Later, we will determine the multiplicities. It turns out to be convenient to   extend all ${\mathcal U}(su(n,n)^{\mathbb C})$ modules to ${\mathcal U}(u(n,n)^{\mathbb C})$ modules by including, from Definition~\ref{z-r-def}, the central element $\Xi^{(r)}$. Up to this point it has been irrelevant and hence dormant. In all representations we consider, it will  act by multiplication by some scalar $\zeta$.

We will denote specific  ones by e.g. $\check M((i_1,\cdots,i_x,0,\cdots),(j_1,\cdots,j_y,0,\cdots),\lambda,\zeta)$. The value of $\zeta$ will always be given in terms of data from the entering representations. We will also use the  
notation $M_{(\Lambda,\zeta)}$ where $(\Lambda,\zeta)=
{((i_1,\dots,i_n),(j_1,\cdots,j_n),\lambda,\zeta)}$ with $\lambda$ as on p. \pageref{page-hbeta}, and with $\zeta\cdot {\mathbf 1}$ the action of $\Xi^{(r)}$ on $V_\tau$. 

\medskip

It is clear that we also may view ${(\Lambda,\zeta)}\in \widehat{u(n)\times u(n)}$ and that any element in the latter is determined uniquely by such a pair.

\bigskip

We have already described the lowest level $r=1$ in Proposition~\ref{lo-le}. From the list, it is clear, in our new terminology, that the following first order ${\mathfrak k}^{\mathbb C}$-types are missing (assume for simplicity that $n>1$):

\begin{eqnarray*}
&{\mathbf r=1}:\\
&V_{((i,1,0,\dots),(1,0,\dots),-i-2)}\textrm{ is missing from }\tilde M(V_{((i,0,\dots),{\mathbf 0},-i-1)}), \textrm{ and}\\
&V_{((1,0,\dots)(i,1,0,\dots),-i-2)}\textrm{ is missing from }\tilde M(V_{({\mathbf 0},(i,0,\dots),-i-1)}).
\end{eqnarray*}

\medskip

The following definition extends terminology from $r=1$:

\begin{Def}
If a highest weight module $\tilde M(V_{(\Lambda,\zeta)})$ appears in the decomposition of some ${\mathcal F}_{2nr}$ we denote it by $\hat M(V_{(\Lambda,\zeta)})$.
\end{Def}

\medskip

Already, from Lemma~\ref{left-right} and Proposition~\ref{most-gen}, we have: \begin{Lem}\label{special-lem}Suppose $r<n$. Then ${\mathcal F}_{2nr}$ contains all modules
\begin{equation}
\hat M(V_{(a_1,\cdots,a_r,0,\cdots),{\mathbf 0},-a_1-r)})\textrm{ and }\hat M(V_{{\mathbf 0},(b_1,\cdots,b_r,0,\cdots),-b_1-r)}).
\end{equation}
The ${\mathfrak k}^{\mathbb C}$-type $V_{(a_1,\cdots,a_r,1,\cdots),(1,0,\cdots),-a_1-1-r)}$  is missing from the first mentioned module,  and the ${\mathfrak k}^{\mathbb C}$-type $V_{(1,0,\cdots),(b_1,\cdots,b_r,0,\cdots),-1-b_1-r)}$ is missing from the other.
\end{Lem}
(The last part also follows directly from (\ref{4.35}).

\medskip

\begin{Rem}
Of course, if a ${\mathfrak k}^{\mathbb C}$-type is missing, many more ${\mathfrak k}^{\mathbb C}$-types are also missing. The stated ones are in an obvious sense the smallest amongst the missing. They become highest weight spaces in generalized Verma modules in which all the ${\mathfrak k}^{\mathbb C}$-types are missing from the initial  module.
\end{Rem}

\bigskip

We will now consider a module \begin{equation}\hat M(V_{((i_1,\dots,i_x,0,\dots),(j_1,\dots,j_y,0,\dots),\lambda_\beta,\zeta)})\label{mod}\end{equation}
with, by convention, $x<n, y<n$. Preparing for cases where $r\geq 2n$, we need to introduce two extra integers $x\leq\tilde x\leq n$ and $y\leq\tilde y\leq n$ to fully accommodate this situation. We utilize the structures defined in (\ref{a,b}):
\begin{eqnarray}
\tilde x&:=&\max\{i=1,\dots,n\mid a_i\neq0\},\textrm{ and}\\
\tilde y&:=&\max\{j=1,\dots,n\mid b_j\neq0\}.
\end{eqnarray}

\bigskip

\begin{Thm}\label{mainth}
If a module \begin{equation}\hat M(V_{((i_1,\dots,i_x,0,\dots),(j_1,\dots,j_y,0,\dots),\lambda_\beta,\zeta)})\label{mod}\end{equation}
with $x<n, y<n$ appears in ${\mathcal F}_{2nr}$,       
then there exist integers $a_1\geq\cdots a_{\tilde x}> 0$ and integers $b_1\geq\cdots b j_{\tilde y}> 0$ such that \begin{equation}\tilde x +\tilde y\leq r\textrm{ and }\tilde x,\tilde y\leq n,\end{equation}  and such that
\begin{eqnarray}\underline{\textrm{If }\tilde x\leq n-1}:&&
x=\tilde x \textrm{ and } i_k=a_k,k=1,\cdots, x.\\
\underline{\textrm{If }\tilde y\leq n-1}:&&
y=\tilde y \textrm{ and } j_k=b_k,k=1,\cdots, y.\\  
\underline{\textrm{If }\tilde x=n}:&&\nonumber
\exists x\in\{1,\cdots,n-1\}:a_1>\cdots>a_x>a_{x+1}=a_{x+2}=\cdots=a_n;\\&&i_1=a_1-a_{n},\cdots,i_{n-1}=a_{n-1}-a_n. \\\underline{\textrm{If }\tilde y=n}:&&\nonumber 
\exists y\in\{1,\cdots,n-1\}:b_1>\cdots>b_y>b_{y+1}=b_{y+2}=\cdots=b_n;\\&&j_1=b_1-b_{n},\cdots,j_{n-1}=b_{n-1}-b_n.
\\\textrm{ In all cases}:
&&\lambda_\beta= -a_1-b_1-r\textrm{ and }\zeta=a_1+\cdots +a_{\tilde x}-b_1-\cdots -b_{\tilde y}.
\end{eqnarray}
Observe that $-a_1-b_1-r=-i_1-j_1-r-(a_1-a_n)-(b_1-b_n)\leq -i_1-j_1-r$.

\smallskip 

The representations $\hat M(V_{((i_1,\dots,i_x,0,\dots),(j_1,\dots,j_y,0,\dots),\lambda_\beta,\zeta)})$ are irreducible and unitarizable.

All modules satisfying these constraints are obtained. (The multiplicities are determined later in Theorem~\ref{k-v-thm}).

\medskip

If $\tilde x=n$ or $\tilde y=n$ we have that \begin{equation}\hat M(V_{((i_1,\dots,i_x,0,\dots),(j_1,\dots,j_y,0,\dots),\lambda_\beta)})= M(V_{((i_1,\dots,i_x,0,\dots),(j_1,\dots,j_y,0,\dots),\lambda_\beta)})\end{equation} and no ${\mathfrak k}^{\mathbb C}$-types are missing. 

Suppose from here on that $\tilde x<n$ and $\tilde y<n$. Then $\tilde x=x$ and $\tilde y=y$ and $x+y\leq r$.

  If $x+y=r$ the first order ${\mathfrak k}^{\mathbb C}$-type $((i_1,\dots,i_x,1,0,\dots),(j_1,\dots,j_y,1,0,\dots),-i_1-j_1-x-y)$ is missing\footnote{If $x=n-1$ or $y=n-1$ this is then not of the form we usually insist on. We refrain from changing it since it is utterly clear how to do it.}. 
  
  In this case,
  \begin{eqnarray}&
  \hat M(V_{((i_1,\dots,i_x,0,\dots),(j_1,\dots,j_y,0,\dots),-i_1-j_1-x-y})=\\&M(V_{((i_1,\dots,i_x,0,\dots),(j_1,\dots,j_y,0,\dots),-i_1-j_1-x-y)})/ M(V_{((i_1,\dots,i_x,1,0,\dots),(j_1,\dots,j_y,1,0,\dots),-i_1-j_1-x-y)}).\nonumber
  \end{eqnarray}
   
   More generally, define $r_0$ by $x+y=r_0\leq r$. Assume for clarity that $x\geq y$; the case $y>x$ follows analogously.

      In this case, the representation 
\begin{equation}\hat M(V_{((i_1,\dots,i_x,0,\dots),(j_1,\dots,j_y,0,\dots),-r_0-i_1-j_1-i)})\label{mod1}, i=0,1,\dots,n-x-1\end{equation}
is missing the ($i+1$)th order ${\mathfrak k}^{\mathbb C}$-type 
\begin{equation}V_{((i_1,\dots,i_x,{\tiny\underbrace{1,\dots,1}_{i+1}},0,\dots,0),(j_1,\dots,j_y,{\tiny\underbrace{1,\dots,1}_{i+1}},0,\dots,0),-r_0-i-i_1-j_1)}.\label{mod2}\end{equation}  
  
In this case,
  \begin{eqnarray}&
  \hat M(V_{((i_1,..,i_x,0,\dots),(j_1,..,j_y,0,\dots),-i_1-j_1-x-y})=\\&M(V_{((i_1,..,i_x,0,\dots),(j_1,..,j_y,0,..),-i_1-j_1-r_0-i)})/ M(V_{((i_1,..,i_x,{\tiny\underbrace{1,..,1}_{i+1}},0,..,0),(j_1,..,j_y,{\tiny\underbrace{1,..,1}_{i+1}},0,..,0),-r_0-i-i_1-j_1)}).\nonumber
  \end{eqnarray}  
  
  The representations 
  \begin{equation}M(V_{((i_1,\dots,i_x,0,\dots),(j_1,\dots,j_y,0,\dots),\lambda_\beta)})\label{mod3}, \ \lambda_\beta<-n-y+1-i_1-j_1 \end{equation}
  are not missing any ${\mathfrak k}^{\mathbb C}$-types.
   \end{Thm}

\pof

First assume that $r<n$.

The $Gl(n,{\mathbb C})^X_L$ representations come from the space of polynomials in the variables $A_i^\dagger(j), i=1,\dots,n; j=1,\cdots, r$. We here take tensor products of $U(n)$ representations corresponding to the various columns $j=1,\dots,r$;
\begin{equation}\label{subrep}
\underbrace{V_{(a_1,0,\cdots,0)}\otimes V_{(a_2,0,\cdots,0)} \otimes\cdots\otimes V_{(a_r,0,\cdots,0)}}_r.
\end{equation}
Any such tensor product is a sub-representation of the tensor product of  $a=a_1+a_2+\cdots a_r$ copies of the defining representation. The latter may be represented by any $A_1^\dagger(j), j=1,\dots, r$, or a Young diagram with just one box. Any irreducible summand of (\ref{subrep}) is then represented by a Young diagram with $a$ boxes. Suppose an irreducible summand has highest weight $(i_1,\cdots,i_x,0,\cdots)$. Then, evidently, $x\leq r$, and its highest weight vector is of the form $(A_1^\dagger(1))^{\alpha_1}\cdots (A_1^\dagger(r))^{\alpha_r} \cdot p$, where $p$ contains no elements of the form $(A_1^\dagger(i))$ and where, obviously, $\alpha_1+\cdots+\alpha_r=i_1$. Actually, we must allow for sums of such elements, but this does not affect the conclusion. 
   A similar statement can be made for the $Gl(n,{\mathbb C})^Y_L$ representation, whose highest weight vector is a sum of terms of the form $(A_{n+1}^\dagger(1))^{\beta_1}\cdots (A_{n+1}^\dagger(r))^{\beta_r} \cdot \hat p$, where $\hat p$ contains no elements of the form $(A_{n+1}^\dagger(i))$, and where $\beta_1+\cdots+\beta_r=j_1$. The tensor product of a  $Gl(n,{\mathbb C})^X_L$ representation and a $Gl(n,{\mathbb C})^Y_L$ representation is  similarly describable and we get a module $V_{((i_1,\dots,i_x,0,\dots),(j_1,\dots,j_y,0,\dots),\lambda_\beta,\zeta)}$. Let us say that the  $Gl(n,{\mathbb C})^Y_L$ representation  occurs inside the tensor product of  $b=b_1+b_2+\cdots b_r$ copies of the defining representation so that it is represented by a Young diagram with $b$ boxes. We now assume that our $Gl(n,{\mathbb C})^X_L\times Gl(n,{\mathbb C})^Y_L$ module  belongs to a top pluri-harmonic vector $v_{(\Lambda,\zeta)}$.  From the above,  it follows that $H_\beta^{(r)}p\cdot v_{2nr}=(-i_1-j_1-r)p\cdot v_{2nr}$, and $\zeta= i_1+\cdots + i_{x}- j_1-\cdots - j_{y}$. Here, $i_1+\cdots +i_x=a$,  $j_1+\cdots +j_y=b$, $i_1=\alpha_1+\cdots + \alpha_r$ and $j_1=\beta_1+\cdots + \beta_r$, where $(A_{n+1}^\dagger(1))^{\alpha_1}\cdots (A_{n+1}^\dagger(r))^{\alpha_r}$ is the total contribution of $A_{n+1}^\dagger(k)$s to the highest weight vector.

    Since the representation defined by $v_{(\Lambda,\zeta)}$ is unitarity, we have that $ -i_1-j_1-r\leq \lambda_{crit}$. From (\ref{4.35}) we then get \begin{eqnarray} -i_1-j_1-r&\leq&-i_1-j_1-x-y,\textrm{ hence}\\ x+y&\leq& r.
\end{eqnarray}
Also, by the assumptions on $p$, 
\begin{equation}
\lambda_\beta= - i_1- j_1-r\textrm{ and }\zeta= i_1+\cdots + i_{x}- j_1-\cdots - j_{y}.
\end{equation}

\medskip

 If $r\geq n$, using an idea from (\cite{k-v}), we choose $N>r$ and embed $su(n,n)\hookrightarrow su(N,N)$ as the top left corner. If  $p\cdot v_{2nr}$  is a highest weight vector for $su(n,n)$, where $p$ is a polynomial in   
the variables $A_i^\dagger(j)$, $i=1,\cdots,n$ and $j=1,\cdots,r$, then $p\cdot v_{2nr}$ is a highest weight vector for any $su(N,N)$ with $N>n$ which follows by looking at the form of the elements $E_\beta, E_{\mu_k}$, and $E_{\nu_k}$ in Definition~\ref{3.13}.

\medskip

Using the proof of the case $r<n$ we get a representation of $U(N)\times u(N)$
\begin{equation}\label{tilde-form}
V_{((\underbrace{a_1,\dots,a_{\tilde x},0,\dots}_N),(\underbrace{b_1,\dots,b_{\tilde y},0,\dots}_N),\lambda_\beta,\zeta)},\end{equation}
where $\tilde x+\tilde y\leq N$, but where also, by construction, we have $\tilde x,\tilde y\leq n$.
Similarly we have 
\begin{equation}
\lambda_\beta= -a_1-b_1-r\textrm{ and }\zeta=a_1+\cdots +a_{\tilde x}-b_1-\cdots -b_{\tilde y}.
\end{equation}

If $\tilde x=n$ or $\tilde y=n$, or both, there remains the task of bringing the representation in (\ref{tilde-form}) into a form as in (\ref{mod}), and this is straightforward.

\medskip

Irreducibility of the modules $\hat M$ comes from Proposition~\ref{uni-prop}.

\medskip

That the representations all occur follows by explicitly exhibiting highest weight vectors. See Corollary~\ref{missing} later.

\medskip

Let us now examine if there are missing ${\mathfrak k}^{\mathbb C}$-types in the cases where either $\tilde x=n$, $\tilde y=n$, or both. The strategy is similar, so we only do $\tilde x=n$, $\tilde y<n$. Then $r\geq n+y$. 

\medskip

Missing ${\mathfrak k}^{\mathbb C}$-types would easily lead  to a homomorphism between generalized Verma modules
\begin{equation}
M(V_{\Lambda_2})\rightarrow M(V_{\Lambda}).
\end{equation}

Now we recall the famous BGG Theorem; see (\cite{bgg}, \cite{boe}), and also (\cite[Proposition~3.6]{herm}). 

\begin{Prop}[BGG]\label{bgg}
If there is a homomorphism $M(V_{\Lambda_1})\rightarrow M(V_{\Lambda})$ then the following condition (A) is satisfied:  

\medskip

\noindent{\bf (A)}There exists a sequence of positive non-compact roots $\gamma_1,\cdots,\gamma_s$ such that 
\begin{equation}
\Lambda_1+\rho=\sigma_{\gamma_s}\circ\cdots\circ \sigma_{\gamma_1}(\Lambda+\rho)
\end{equation}
and such that 
(set $\sigma_{\gamma_0}=Id$)
$\sigma_{\gamma_i}\circ\cdots\circ \sigma_{\gamma_1}(\Lambda+\rho)-\sigma_{\gamma_{i-1}}\circ\cdots\circ \sigma_{\gamma_1}(\Lambda+\rho)=-n_i\gamma_i$ with $n_i\in{\mathbb N}$ for $i=1,\dots,s$. 

\medskip
In particular,
$\Lambda_1=\Lambda-n_1\gamma_1-\cdots-n_s\gamma_s$. Furthermore, since inner products between positive non-compact roots are positive,
$(\Lambda+\rho)(H_{\gamma_i})>0$ for all $i=1,\dots,s$.
\end{Prop}

\smallskip

Consider now a positive non-compact root
$\gamma_I=\beta+\mu_1+\cdots+\mu_{a_i-1}+\nu_1
+\cdots+\nu_{b_i-1}$ and assume it is the first in a sequence as in condition (A) above. We get from the last condition in the above proposition;
\begin{equation}
i_1-i_{a_i}+j_1-j_{b_i}+(a_i-1)+(b_i-1)+\lambda+1=n_1\in{\mathbb N}.
\end{equation}
Since $\lambda=-i_1-j_1-r-c$ for some $c\in{\mathbb N}$
and since $r\geq n+y$, this leads to
\begin{equation}
a_i+b_i=n_i+i_{a_i}+j_{b_i}+1+c+r\geq n_i+1+c+n+y.
\end{equation}
Furthermore, $a_i+b_i\leq n+b_i$, and hence 
\begin{equation}
\forall i: b_i\geq 3+y. 
\end{equation}

But then $\Lambda_1(H_{\nu_{y+1}}+\cdots+ H_{\nu_{n-1}})<0$ contradicting that $\Lambda_1$ is a  highest weight for a finite dimensional representation. Hence there can be no missing ${\mathfrak k}^{\mathbb C}$-types in this domain.

\medskip

Let us finally return to the case $\tilde x=x, \tilde y=y$. Of course, when $x+y=r$ (then $\lambda=\lambda_{crit}$), a first order polynomial is missing. 

\medskip

It follows that this ${\mathfrak k}^{\mathbb C}$ type is in the radical of the Hermitian form. Now we invoke the results from (\cite{DES}), (\cite{EJ}), or (\cite{jak-compo}) that give that this radical is a generalized Verma module. 

\medskip

Finally observe that 
\begin{eqnarray}&\hat M(V_{((i_1,\dots,i_x,0,\dots),(j_1,\dots,j_y,0,\dots),-r_0-i_1-j_1-i)})\subset\\&\hat M(V_{((i_1,\dots,i_x,0,\dots),(j_1,\dots,j_y,0,\dots),-r_0-i_1-j_1)})\otimes \hat M(V_{{\mathbf 0},{\mathbf 0},-i})\ i=1,\dots,n-x-1.\end{eqnarray}
Since $\hat M(V_{{\mathbf 0},{\mathbf 0},-i})$ is missing all $(i+1)\times (i+1)$ minors, the claim follows easily along the same line as for the case $x+y=r$.

This completes the proof of Theorem~\ref{mainth}. \qed

\medskip

\begin{Cor}
The representations with missing ${\mathfrak k}^{\mathbb C}$-types in Theprem~\ref{mainth} are exactly those that, via duality, define covariant differential operators whose null spaces carry unitary representations.
\end{Cor}

\medskip

\begin{Def}If $\hat M(V_{((i_1,\dots,i_x,0,\dots),(j_1,\dots,j_y,0,\dots),\lambda,\zeta)})$ as above appears in ${\mathcal F}_{2nr}$, in the same notation as above, we associate a highest weight representation of $u(r)$ of highest weight 
\begin{equation}
(\Lambda,\zeta)_{u(r)}:=(a_1,\cdots,a_{\tilde x},\underbrace{0,\cdots,0}_{r-\tilde x-\tilde y},-b_{\tilde y},\cdots,b_1).
\end{equation}
\end{Def}
The reason for this will become clear later. We may also refer to $(\Lambda,\zeta)_{u(r)}$ as the irreducible representation of $u(r)$ determined by $(\Lambda,\zeta)$.

\bigskip

Recall that the ${\mathfrak k}^{\mathbb C}$-types always can be assumed to be of the form  
\begin{equation}\Lambda=((i_1,\dots,i_x,0,\dots),(j_1,\dots,j_y,0,\dots),\lambda_\beta,\zeta)\textrm{ with }x,y<n.\end{equation}
As an aside, notice that this convention is different from that of (\cite{k-v}).

\bigskip

\subsection{Generalized Verma modules and tensor products thereof.} 

To deal with the question of multiplicities, we first go through some general theory. To begin we return to the general set-up as in Subsection~\ref{sub-missing}.    

Let $\{W_\alpha\}_{\alpha\in A}$ and $\{Z_\alpha\}_{\alpha\in A}$ be bases of ${\mathfrak p}^-$ and ${\mathfrak p}^+$, respectively, and assume that 
$\forall \alpha: Z_\alpha=-\omega(W_\alpha)$. In our later  application, ${\mathfrak p}^\pm$ are representable by the space of $n\times n$ complex matrices. 

\medskip

The following method was discovered by S. Martens (\cite{mar}). It was further developed, independently, of Jakobsen--Vergne. Here we go one step further and describe the dual version:

\smallskip

We assume given two highest weight modules $\tilde M(V_{\tau_1}), \tilde 
M(V_{\tau_2})$ that are top quotients in their corresponding generalized 
Verma Modules. 
We wish to decompose $\tilde M(V_{\tau_1})\otimes \tilde M(V_{\tau_2})$.

As a ${\mathfrak k}^{\mathbb C}$-module we have a decomposition into irreducible submodules
\begin{equation}
V_{\tau_1}\otimes V_{\tau_2}=\oplus^N_{i=1}V_{\tilde \tau_i}.
\end{equation}

Let \begin{equation}\label{Y}W^L_{\alpha}=W_{\alpha}\otimes1,\ W^R_{\alpha}= 1\otimes W_{\alpha},\textrm{ and }Y_{\alpha}=W^L_{\alpha}-W^R_{\alpha}\ (\alpha\in A),\end{equation} and,  for each $\ell$,  let ${\mathcal P}_\ell(Y,V_{\tilde\tau_i})$ denote the span of all expressions 
\begin{equation}
Y_{\alpha_1}Y_{\alpha_2}\cdots Y_{\alpha_k}v_{\tilde\tau_i} \in \tilde 
M(V_{\tau_1})\otimes \tilde M(V_{\tau_2})
\end{equation}
	 with $k\leq \ell$  and $v_{\tilde\tau_i}\in  V_{\tilde\tau_i}$. 
	 
It is clear that they transform under ${\mathfrak k}^{\mathbb C}$ according to a sub-representation of 
$		V_{\tilde\tau_1}\otimes V_{\tilde\tau_2}\otimes {\mathcal U}({\mathfrak p}^-)$. As we shall 
see, it may happen that this, due to cancellations, is a proper 
sub-representation.

 One  has a filtration of ${\mathfrak k}^{\mathbb C}$ modules

$$
\emptyset\subseteq {\mathcal P}_0(Y,V_{\tilde\tau_i})\subseteq {\mathcal P}_1(Y,V_{\tilde\tau_i})\subseteq \cdots \subseteq {\mathcal P}_\ell(Y,V_{\tilde\tau_i})\subseteq\cdots .
$$

It is straightforward to see that the elements $X_{\alpha}=(Z_{\alpha}\otimes 1 +1\otimes Z_{\alpha})$ have the following property:
\begin{equation}
\forall j,\forall\alpha\in A; X_{\alpha}: {\mathcal P}_\ell(Y,V_{\tilde\tau_i})\rightarrow {\mathcal P}_{\ell-1}(Y,V_{\tilde\tau_i}).
\end{equation}

This easily has the following consequence: 

\smallskip

\begin{Prop}\label{filt}
For each $\ell,\tilde\tau_i$, let $R_\ell(V_{\tilde\tau_i})$ denote the ${\mathcal U}$ module obtained by letting elements $u\in{\mathcal U}$ act through the co-product $\triangle(u)$ on  ${\mathcal P}_\ell(Y,V_{\tilde\tau_i})$ from the left. 
Then we have a ${\mathcal U}$ filtration of $\tilde M(V_{\tau_1})\otimes\tilde M(V_{\tau_2})$:
\begin{equation}
\emptyset\subseteq R_0(V_{\tilde\tau_i})\subseteq R_1(V_{\tilde\tau_i})\subseteq \cdots \subseteq R_\ell(V_{\tilde\tau_i})\subseteq\cdots\subseteq \tilde M(V_{\tau_1})\otimes \tilde M(V_{\tau_2}).
\end{equation}
The quotient modules $R_{\ell+1}(V_{\tilde\tau_i})/R_\ell(V_{\tilde\tau_i})$ are 
finite sums of top quotients $\tilde M(V_\xi)$ where
\begin{equation}
\xi\textrm{ occurs in }V_{\tau_1}\otimes V_{\tau_2}\otimes({\mathfrak p}^-)^{\otimes_s^{\ell+1}}.
\end{equation}
Here ${\mathfrak p}^-$ is spanned by the elements $Y_{\alpha}$ (\ref{Y}). However, we may also choose elements from just ${\mathfrak p}_L^-$ or from ${\mathfrak p}_R^-$, the spaces generated by the elements $W^L_{\alpha}$ and  $W^R_{\alpha}$, respectively.  
\end{Prop}

\smallskip

The last statement in the proposition is true since we are working modulo expressions generated by the elements $W^L_{\alpha}+W^R_{\alpha}$, so $Y_{\alpha}$ is equivalent to the left or right expressions.

\medskip

It is clear that 
\begin{equation}
R_0(V_{\tilde\tau_i})=\tilde M(V_{\tilde\tau_i})
\end{equation}
for some top quotient $\tilde M(V_{\tilde\tau_i})$.

\medskip

Analogous statements hold at the higher levels except that there is a major new feature which can be seen already at degree $1$: Each of the modules $R_1(V_{\tilde\tau_i})$ is sum of modules of the form $\tilde M(V_{\tau_{i,1}})$ where $V_{\tau_{i,1}}$ occurs in $V_{\tilde\tau_i}\otimes{\mathfrak p}^-$, but it may happen, as we shall see below, that there may be an irreducible summand $V_{\tau_{i,1}}$ from the latter for which no top quotient  $\tilde M(V_{\tau_{i,1}})$ occurs in the decomposition of  $R_1(V_{\tilde\tau_i})/
R_0(V_{\tilde\tau_i})$. In this case we say that the highest weight module of highest weight $\Lambda_{\tilde\tau_i}$  is {\bf missing } from $\tilde M(V_{\tau_1})\otimes\tilde M(V_{\tau_2})$. A related problem which we shall only partially discuss is that of determining which ${\mathcal U}({\mathfrak k}^{\mathbb C})$  representations that actually {\bf do} occur in the tensor product. 

\medskip

\begin{Def} A ${\mathfrak k}^{\mathbb C}$-type $V_\xi$ {\bf appears at level $d$} in $\tilde M(V_{\tau_1})\otimes\tilde M(V_{\tau_2})$ if it is a non-zero ${\mathfrak k}^{\mathbb C}$-type in the quotient of ${\mathcal P}_d(Y)\otimes V_{\tau_1}\otimes V_{\tau_2}$ modulo the U-modules generated by all (non-zero) elements in all ${\mathcal P}_{d'}(Y)\otimes V_{\tau_1}\otimes V_{\tau_2}$ with $d'<d$. \end{Def}

\bigskip

\subsection{Multiplicities}

To determine multiplicities, we now invoke the commuting algebra ${\mathcal U}(u(r))$. Specifically, the action of $u(r)$ given in (\ref{Oij}). This method is inspired by (\cite{k-v}).

The span ${\mathcal T}{\mathcal P}(\Lambda)$ of the set of top pluri-harmonic vectors $v_\Lambda$ of a given weight $\Lambda$ is invariant ${\mathcal U}(u(r))$, but when $r\geq 2n$ this space needs not be irreducible. The dimension $\dim_{u(r)}(\Lambda)=\dim  {\mathcal T}{\mathcal P}(\Lambda)$ gives  the multiplicity of the $su(n,n)^{\mathbf C}$ module of highest weight $\Lambda$. To handle this situation, we need to be more precise and turn to the space ${\mathcal T}{\mathcal P}((\Lambda,\zeta))$  of top pluri-harmonic vectors $v_{(\Lambda,\zeta)}$ of a given weight $(\Lambda,\zeta)$.  This space will always define an irreducible representation $(\Lambda,\zeta)_{u(r)}$ of $u(r)$, but, as we shall see,  if $r>n$ these representations will not exhaust $\widehat{u(r)}$.  

We have

\begin{equation}
{\mathcal T}{\mathcal P}(\Lambda)=\oplus_\zeta {\mathcal T}{\mathcal P}((\Lambda,\zeta)).
\end{equation}

The sum over $\zeta$ above is finite. $\zeta$ is essentially determined as in (\ref{440}). See also (\ref{4411}) and further below.

\medskip

We now embark on the details of this:

\smallskip

\begin{Prop}\label{irr-mult}Let $p$ be top pluri-harmonic polynomial of weight $(\Lambda,\zeta)$ and let $\tilde x, \tilde y$ be as in Theorem~\ref{mainth}. Then $\tilde x + \tilde y\leq r$.  Set \begin{equation}{\mathbf n}=(n_1,\cdots, n_{\tilde x},\underbrace{0,\cdots,0}_{r-\tilde x})\textrm{ and }{\mathbf m}=(m_1,\cdots, m_{\tilde y},\underbrace{0,\cdots,0}_{r-\tilde y}).\end{equation}In the cyclic $Gl^D_R(r,{\mathbb C})$  module generated by $p$ there is a polynomial $p^T= \triangle^{\mathbf n}({\mathbf x})\tilde\triangle({\mathbf y})^{\mathbf m}$ such that $p^T\cdot v_{2nr}$ is a highest weight vector of the same weight  $(\Lambda,\zeta)$ as $p$.
The $gl^D_R(r,{\mathbb C})$ module of top pluri-harmonic vectors of weight $(\Lambda,\zeta)$  is irreducible of highest weight 
\begin{equation}\label{4411}
(\Lambda,\zeta)_{u(r)}=(n_1,\cdots,n_{\tilde x},0,\dots,0,-m_{\tilde y},\dots,-m_1).
\end{equation}
\end{Prop}

\proof We prove this by induction. The case $r=1$ is clear. Consider then an irreducible representation $\tilde M(V_{(\Lambda_{r+1},\zeta_{r+1})})$ in ${\mathcal F}_{2n(r+1)}$ with highest weight vector $p_{(\Lambda_{r+1},\zeta_{r+1})}$. We know that $\tilde M(V_{(\Lambda_{r+1},\zeta_{r+1})})\subseteq \tilde M(V_{(\Lambda_{r},\zeta_{r})})\otimes
\tilde M(V_{(\Lambda_{1},\zeta_{1})})$ (by unitarity it is a summand) where  $\tilde M(V_{(\Lambda_{r},\zeta_{r})})$ comes from ${\mathcal F}_{2nr}$ and $\tilde M(V_{(\Lambda_{1},\zeta_{1})})$ comes from ${\mathcal F}_{2n}$. Without loss of generality we may assume that $V_{(\Lambda_{1},\zeta_{1})}=V_{((a,0,\dots),{\mathbf 0},\lambda=-a-1,\zeta=a)}$ for some integer $a\geq 0$. 

We view $Gl^D_R(r,{\mathbb C})$   embedded  in $Gl^D_R(r+1,{\mathbb C})$ in the top left $r\times r$ positions with the auxiliary  entries set equal to $0$ except the $(r+1),(r+1)$ entry which is set to $1$. 

We may assume, by the induction assumption, that $\tilde M(V_{(\Lambda_{r},\zeta_{r})})$ is represented by a highest weight vector \begin{equation}\label{u-eq}p_{(\Lambda_{r},\zeta_{r})}\cdot v_{2nr}=u\cdot p^T_{(\Lambda_{r},\zeta_{r})}\cdot v_{(\Lambda_{r},\zeta_{r})}\end{equation} for some $u\in{\mathcal U}(gl^D_R(r,{\mathbb C}))$. Using Proposition~\ref{filt} we know that it suffices, for $p_{(\Lambda_{r+1},\zeta_{r+1})}\cdot v_{2n(r+1)}$, to look at the tensor products of vectors from the ${\mathfrak k}^{\mathbb C}={\mathcal U}(u(n)^{\mathbb C})\times {\mathcal U}(u(n)^{\mathbb C})$ type corresponding to $p_{(\Lambda_{r},\zeta_{r})}$    with vectors coming from $\tilde M(V_{(\Lambda_{1},\zeta_{1})})$. 

The element $u$ in (\ref{u-eq}) only depends on the variables in the aforementioned upper left $r\times r$ corner (when identifying $gl^D_R(r+1,{\mathbb C})$ with $gl(r+1,{\mathbb C})$). By assumption we see that we may write

 {\small\begin{eqnarray} &p_{(\Lambda_{r+1},\zeta_{r+1})}=\\&u\cdot\left(\sum_i\hat p_{r,i,{\mathbf x}}(x_{11},\dots, x_{n\tilde x})\hat p_{r,i,{\mathbf y}}(y_{1,r-\tilde y+1},\dots, y_{n,r})\hat p_{r+1,i}(x_{1,r+1},\cdots, x_{n,r+1}, y_{1,r+1},\cdots, y_{n,r+1})\right):=\quad\nonumber\\&
 u\cdot \tilde p_{(\Lambda_{r+1},\zeta_{r+1})}
\end{eqnarray}}
with $\tilde x+\tilde y\leq r$. The $u$ is in  $gl^D_R(r,{\mathbb C})\subseteq gl^D_R(r+1,{\mathbb C})$.

\medskip

We analyze the expression $\tilde p_{(\Lambda_{r+1},\zeta_{r+1})}$ in the bracket in the above equation.

\medskip

We do induction in the $y$ degrees of the second term. The degree $0$ is clear by the Littlewood-Richardson Rule (\ref{L-R}). $\tilde p_{(\Lambda_{r+1},\zeta_{r+1})}$ now is pluri-harmonic modulo the (repeated) action of expressions  $W_\alpha^L+W_\alpha^R$ on lower order degrees $p_{lower}$. 
By induction, $p_{lower}$ can be brought to the desired form  $\triangle^{\tilde{\mathbf n}}({\mathbf x})\triangle({\mathbf y})^{\tilde{\mathbf m}}$ for some $\tilde{\mathbf n},\tilde{\mathbf m}$   by $gl^D_R(r+1,{\mathbb C})$. Moreover,
 $gl^D_R(r+1,{\mathbb C})$ commutes with the elements $W_\alpha^L+W_\alpha^R$.
 
We now show that $\tilde p_{(\Lambda_{r+1},\zeta_{r+1})}$ can be brought into a similar form $\triangle^{\hat{\mathbf n}}({\mathbf x})\triangle({\mathbf y})^{\hat{\mathbf m}}$ by the action of $gl^D_R(r+1,{\mathbb C})$. 

\medskip

 Suppose first that $\tilde x+\tilde y=r$. Also assume $\tilde x\neq0$. Use first the permutation $\sigma_{\tilde x+1,r+1}$ of columns $\tilde x+1$ and $r+1$. Clearly, $\sigma_{\tilde x+1,r+1}\in Gl^D_R(r+1)$. If $\tilde x=0$, this step is not needed. After that we have polynomials in $X$ which depend on the first $x+1$ columns multiplied with polynomials in $Y$ which depend on the columns $x+1, \cdots.r+1$. 
Observe that the operators $E_\mu$ and $O_{ij}$ (of (\ref{Oij})), $j>i$  ``pushes'' towards $(A_1(1))^\dagger$ while the operators $E_\nu$ and $O_{ij}$, $j>i$
``pushes'' towards $(A_{n+1}(r))^\dagger$. 

Indeed, the elements in $gl^D_R(r+1,{\mathbb C})$ we utilize in bringing the polynomials in $X$ to the  desired corner position (using Corollary~\ref{right-form}), are of the form 
\begin{equation} 
\sum_{a=1}^n\left(x_{a,i}\frac{\partial}{\partial x_{a,j}}-
y_{a,j}\frac{\partial}{\partial y_{a,i}}\right)\textrm{ with }x+1\geq j>i\geq1,
\end{equation} 
hence they will give zero when applied to polynomials in $Y$ which are zero in the first $x$ columns. This means that we can bring these polynomials into the corner position in the $X$ variables without affecting the $Y$ variables and vice versa.  So, we can bring the $X$ variables into corner form covering the first $x+1$ columns and the $y$ variables into corner form covering the last $y+1$ columns. We then arrive at a vector $\triangle^{\hat{\mathbf n}}({\mathbf x})\tilde\triangle({\mathbf y})^{\hat{\mathbf m}}\cdot v_{2nr}$ which is a highest weight vector for ${\mathfrak k}^{\mathbb C}$ as well as $GL^D_R(r+1,{\mathbb C})$. The assumption is that, additionally, this is top pluri-harmonic modulo lower orders. Especially, the data $\hat{\mathbf n},\hat{\mathbf m}$ corresponds to $(\Lambda_{r+1},\zeta_{r+1})$, and falls under Theorem~\ref{mainth} to allow us to conclude that either the $X$ variables actually are bound only to the $x$ first columns (then the $Y$ variables will  be supported on the last $y+1$ columns) or the $Y$ variables are tied only to the last $y$ columns (with the $X$ variables  supported on the first $x+1$ columns).

The case where $\tilde x+\tilde y<r$ is clear since then there is a position $\ell$ with $\tilde x<\ell<r+1-\tilde y$ and we can use a permutation from $GL^D_R(r+1,{\mathbb C})$ that interchanges $\ell$ and $r+1$. After that, follow the same proof.

It follows easily that also $ p_{(\Lambda_{r+1},\zeta_{r+1})}$ can be brought into the form $\triangle^{\hat{\mathbf n}}({\mathbf x})\triangle({\mathbf y})^{\hat{\mathbf m}}$.

Finally, by easy inspection, the top term $\triangle^{\hat{\mathbf n}}({\mathbf x})\tilde\triangle({\mathbf y})^{\hat{\mathbf m}}$ is pluri-harmonic. Hence the lower order terms are actually $0$.

We finally need to address the irreducibility of the representation of $GL^D_R(r+1,{\mathbb C})$: The element $\triangle^{\mathbf n}({\mathbf x})\tilde\triangle({\mathbf y})^{\mathbf m}\cdot v_{2nr}$ in Proposition~\ref{irr-mult} is evidently a highest weight vector for $gL^D_R(r+1,{\mathbb C})$ of weight $(n_1,\cdots,n_{\tilde x},0,\dots,0,-m_{\tilde y},\dots,-m_1)$, hence defines an irreducible finite-dimensional representation of $gL^D_R(r+1,{\mathbb C})$. Notice that  $\triangle^{\mathbf n}({\mathbf x})\cdot v_{2nr}$ is a highest weight vector for $GL(n,{\mathbb C})^X_L\times GL(r,{\mathbb C})^X_R$ and $\triangle({\mathbf y})^{\mathbf m}\cdot v_{2n(r+1)}$  is a  highest weight vector for $GL(n,{\mathbb C})^Y_L\times GL(r,{\mathbb C})^Y_R $ (The latter acting by $(u^T)^{-1}$, c.f. \ref{addi})). The  representation of $gl^D_R(r+1,{\mathbb C})$ thus has a highest weight vector, and this is uniquely, and explicitly, determined by the ${\mathfrak k}^{\mathbb C}$ type involved, i.e. by the highest weight $(\Lambda,\zeta)$. It follows that the representation of $GL^D_R(r+1,{\mathbb C})$ in the space of top pluri-harmonics of type $(\Lambda,\zeta)$ is irreducible. \qed

\medskip

Quite generally, any element $\triangle^{\mathbf n}({\mathbf x})\tilde\triangle({\mathbf y})^{\mathbf m}\cdot v_{2nr}$ in ${\mathcal F}_{2nr}$ defines an irreducible $GL^D_R(r,{\mathbb C})$ module and, if the variables ${\mathbf x},{\mathbf y}$ do not overlap, it is also  a top pluri-harmonic vector of a weight $(\Lambda,\zeta)$ directly given $(n_1,\cdots,n_{\tilde x},0,\dots,0,-m_{\tilde y},\dots,-m_1)$. 

\medskip

\begin{Cor}\label{missing}
All the representations of Theorem~\ref{mainth}  appear in the decomposition.
\end{Cor}
                                     
\pof  This follows from Lemma~\ref{special-lem}, by forming tensor products between representations in which $x=0$ with representations in which $y=0$. \qed 

\medskip

To finally settle the question of multiplicities we need one more step in our  analysis: This is because of the appearance of $\zeta$.

\medskip

From now on, we identify $gl^D_R(r,{\mathbb C})$ with $gl(r,{\mathbb C})$.  Observe that any highest weight $\xi_{gl(r,{\mathbb C})}$ of $gl(r,{\mathbb C})$ can be written uniquely as 
\begin{equation}\label{a-b-n}
\xi_{gl(r,{\mathbb C})}=(a_1,n_2, \cdots,a_{\tilde x},\underbrace{0,\cdots,0}_{r-\tilde x-\tilde y},-b_{\tilde y},-b_{\tilde y-1}, \cdots,-b_1)
\end{equation}
with $a_1\geq a_2\geq\cdots\geq a_{\tilde x}>0$ and $b_1\geq b_2\geq\cdots\geq b_{\tilde y}>0$. The cases $\tilde x=0$ or $\tilde y=0$ are allowed and mean that the string  $a_1,\cdots$, or $b_1,\cdots$,   is not present.

\medskip

We must then demand $\tilde x\leq n$ and $\tilde y\leq n$ for $\Lambda_{gl(r,{\mathbb C})}$ to correspond to a top pluri-harmonic representation. This is necessary when $r>n$. We set ${\mathbf a}=(a_1,\cdots,a_n)$ where, if $\tilde x<n$, we set $a_k=0$ for $\tilde x<k\leq n$. The symbol ${\mathbf b}=(b_1,\cdots,b_n)$ is defined analogously.  Further, we set $\vert {\mathbf a}\vert=a_1+\cdots+a_n$, and $\vert {\mathbf b}\vert=b_1+\cdots+b_n$.

\begin{Def}\label{extra}
$\Sigma^{(r)}_n$ denotes those irreducible representations of $gl(r,{\mathbb C})$ for which, in (\ref{a-b-n}), $\tilde x,\tilde y\leq n$.

Given  $\xi=\xi_{u(r)}\in \Sigma^{(r)}_n$, written as in (\ref{a-b-n}), we set 
\begin{eqnarray}&(\Lambda,\zeta)^{\xi}=((i_1,\dots,i_x,0,\dots),(j_1,\dots,j_y,0,\dots),\lambda_\beta,\zeta)\textrm{ where}=\\&(i_1,\dots,i_x,0,\dots)=(a_1-a_n, \cdots,a_{n-1}-a_n,0),\\ &(j_1,\dots,j_y,0,\dots)=(b_1-b_n, \cdots,b_{n-1}-b_n,0)\textrm{ and}\\
&(\lambda_\beta,\zeta)=(-a_1-b_1-r, \vert {\mathbf a}\vert-\vert {\mathbf b}\vert).\end{eqnarray}
\end{Def}
\smallskip
\begin{Rem}$\Sigma^{(r)}_n$ is the set $\Sigma$ of (\cite{k-v}).
\end{Rem}
\smallskip

Notice that $-a_1-b_1-r=-(a_1-a_n)-(b_1-b_n)-r-(a_n+b_n)$. If $(a_n+b_n)>0$, this parameter is less than the critical value.

\smallskip

\smallskip

We can reformulate Corollary~\ref{missing}:

\begin{Cor}\label{missing1}
All such modules $\tilde M(V_{(\Lambda,\zeta)^\xi})$ appear in the decomposition.
\end{Cor}

\medskip

For $\xi\in \Sigma^{(r)}_n$, let $\dim(\xi)$ denote the dimension. We can then state:

\begin{Thm}[\cite{k-v}] \label{k-v-thm}
\begin{equation}{\mathcal  F}_{2nr}=\oplus_{\xi\in \Sigma^{(r)}_n}\dim(\xi)\tilde M(V_{(\Lambda,\zeta)^\xi}).
\end{equation}This is a decomposition into irreducible unitary highest weight
representations of $u(n,n)^{\mathbb C}$. In case $r<2n$, if $(\Lambda,\zeta)^\xi$ appears, the $\zeta$ is unique, so the formula in Theorem~\ref{k-v-thm} also gives a decomposition into irreducible unitary highest weight
representations of $su(n,n)^{\mathbb C}$.
 In case $r\geq 2n$ assume that $\Phi:=-\lambda_\beta-i_1-j_1-r\geq0$. Furthermore, set $\vert {\mathbf i}\vert=i_1+\cdots+i_x$ and $\vert {\mathbf j}\vert=j_1+\cdots+j_y$, and let 
 $\zeta_i=\vert {\mathbf i}\vert-\vert {\mathbf j}\vert+ni$. Then $(\Lambda,\zeta_i)^\xi$ appears precisely for $i=-\, -\Phi+2,\dots,\Phi$.
 \end{Thm}
Phi
\pof This was already established in Theorem~\ref{mainth}  with only the case $r\geq 2n$ missing explicitly. However, this case follows as in (\ref{440}). \qed

\medskip

\begin{Prop}\label{minor-type}
The highest weight vectors are given by
\begin{equation}
\triangle({\mathbf x})^{\mathbf a}\tilde\triangle({\mathbf y})^{\mathbf b}v_{2nr}.
\end{equation}
If  $a_n\cdot b_n>0$, the highest weight vector 
\begin{equation}
\triangle({\mathbf x})^{{\mathbf a}-{\mathbf e}_n}\tilde\triangle({\mathbf y})^{{\mathbf b}-{\mathbf e}_n}v_{2nr}
\end{equation}
defines an equivalent $su(n,n)$ module.

\smallskip

In terms of the creation operators we have
\begin{equation}
\forall k: \triangle_k({\mathbf x})=\sum_{\sigma\in S_k}(-1)^{sgn(\sigma)}\prod_{i=1}^kA^\dagger_i(\sigma(i)),
\end{equation}and
\begin{equation}
\forall \ell: \tilde\triangle_\ell({\mathbf y})=\sum_{\sigma\in S_\ell}(-1)^{sgn(\sigma)}\prod_{i=1}^\ell A_{n+i}^\dagger(r-\ell+(\sigma(i))).
\end{equation}
\end{Prop}

\bigskip

\section{The $q$-harmonic representation}\label{5}

We introduce the quantized version of the Weyl algebra. As in
\cite{kilu}, we call this the Hayashi-Weyl Algebra since it
was studied by Hayashi in \cite{hi}. We stress, though, that the
algebra presented here is a modified, simpler version. Notably,
our $q$-is the $q^2$ of \cite{hi}.

\begin{Def}[The Hayashi-Weyl Algebra] Let $q$-be a non-zero
complex number such that $q^2 \neq 1$. The algebra ${\mathcal
H}{\mathcal W}_{s}$ is defined as an associative unital algebra
with generators $B_i,B_i^\dagger,L_i^{\pm 1}; i=1,2,\cdots,s$
and relations \begin{align} & L_i L_j=L_j L_i,  &&L_i
L_i^{-1}=L_i^{-1} L_i=1, &\\& B_iB_j=B_jB_i, &&
B^\dagger_iB^\dagger_j=B^\dagger_jB^\dagger_i, &\\& B_iB_j=B_jB_i
\quad i \neq j, &&\\& L_i B_j L_i^{-1}=q^{-\delta_{ij}} B_j, &&
L_i B^\dagger_j L_i^{-1}=q^{\delta_{ij}} B^\dagger_j, \label{25d}&\\&
B_iB^\dagger_i-qB^\dagger_iB_i=L_i^{-1}, &&
B_iB^\dagger_i-q^{-1}B^\dagger_iB_i=L_i.\label{25} \end{align}

The last pair of relations are equivalent to the following:
\begin{equation}
B_iB_i^\dagger=\frac{qL_i-q^{-1}L_i^{-1}}{q-q^{-1}}, \quad
B_i^\dagger B_i=\frac{L_i-L_i^{-1}}{q-q^{-1}}. \end{equation}

We denote by  ${\mathcal
H}{\mathcal W}^+_{s}$ and ${\mathcal
H}{\mathcal W}^-_{s}$ the subalgebras generated by the elements $B_i, i=1,\dots,s$ and the elements $B^\dagger_i, i=1,\dots,s$, respectively.
\end{Def}

\smallskip

 When $s=1$ we drop the subscripts.

\medskip

For later reference we list the following which is easily proved
by induction:

\begin{Lem}\label{5.2}For $N\in{\mathbb N}$, \begin{eqnarray}
B(B^\dagger)^N&=&q^{-N}(B^\dagger)^NB+[N]_q(B^\dagger)^{N-1}L\\
&=&q^{N}(B^\dagger)^NB+[N]_q(B^\dagger)^{N-1}L^{-1}.
\end{eqnarray} \end{Lem}

\medskip

\begin{Lem}[The $q$-Weyl-Serre Equations] \label{qws}\begin{eqnarray}
(B^\dagger)^2B-(q+q^{-1})(B^\dagger
BB^\dagger)+B(B^\dagger)^2=0,\\
B^2B^\dagger-(q+q^{-1})(BB^\dagger B)+B^\dagger B^2=0.
\end{eqnarray} \end{Lem}

\medskip

\begin{Rem}
More generally, any pair $(\hat B,\hat B^\dagger)=(\kappa L^a B,\tilde\kappa L^{-a} B^\dagger)$ satisfies the $q$-Weyl--Serre Equations, witk $a\in{\mathbb Z}$ and $\kappa,\tilde \kappa\in{\mathbb C}^*$.
\end{Rem}

\medskip

The introduced algebra has a natural - occasionally  to be called defining -
representation.

\begin{Def}The $q$-Fock Space ${\mathcal F}^{(q)}_{s}$ is an
infinite-dimensional space with basis $\{{\mathbf e}({\bf
m})\mid {\bf m} \in \mathbb N_0^{2n}\}$ and the $q$-Stone--von Neumann representation
$\pi^{qSvN}_{s}$
is a representation of ${\mathcal
H}{\mathcal W}_{s}$ in ${\mathcal F}^{(q)}_{s}$.
\smallskip
The actions of the
generators in this representation are given by \begin{equation*} L_i {\mathbf
e}(\mathbf{m})=q^{m_i}{\mathbf e}(\mathbf{m}), \quad B_i
{\mathbf e}(\mathbf{m})=[m_i]_{q}{\mathbf
e}(\mathbf{m}-\mathbf{e}_i),\quad B_i^\dagger {\mathbf
e}(\mathbf{m})={\mathbf e}(\mathbf{m}+\mathbf{e}_i),
\end{equation*} where $\mathbf{m}=\{m_1,\ldots,m_{s}\}$ and
$[m]_q=\frac{q^m-q^{-m}}{q-q^{-1}}$. Here, we set  ${\mathbf e}(\mathbf{m})=0$ if ${\bf m} \notin \mathbb N_0^{2n}$.

\smallskip

Notice that
\begin{equation}\pi^{qSvN}_{s}=\underbrace{\pi^{qSvN}_{1}
\otimes\cdots\otimes \pi^{qSvN}_{1}}_{s}.\end{equation} 
\end{Def}

\smallskip

It is obvious that the vector $v_{s}^{(q)}={\mathbf e}({\mathbf
0})$ is cyclic, satisfies $L_iv_{s}^{(q)}=v_{s}^{(q)}$, and is
annihilated by all $B_i$, $i=1,\dots, s$. 
\begin{Lem}$\pi^{qSvN}_{s}$ is irreducible. \end{Lem}

\proof It suffices to consider $\pi^{qSvN}_{1}$ for ${\mathcal
H}{\mathcal W}_{1}$. The operators in the representation are
shift operators which, since we assume $\forall n\in{\mathbb
N}:[n]_q\neq0$, omit no positions. \qed

\medskip

\begin{Lem} \begin{equation} \{(B^\dagger)^aL^b, B^cL^d\mid
a\in{\mathbb N}_0, c\in {\mathbb N}, b,d\in{\mathbb Z}\}
\end{equation} is a basis of ${\mathcal H}{\mathcal W}_{1}$.
\end{Lem}

\proof Due to (\ref{25}), it is clear that the set is spanning.
Suppose then that a finite linear combination \begin{equation}\label{528}p=\sum(
\alpha_{a}(B^\dagger)^ap_a+ \beta_{b}B^b\tilde p_b)=0\end{equation} for some Laurent polynomials $p_a,\tilde p_b$. Then $\forall N\in{\mathbb N}: L^NpL^{-N}=0$. By considering weights it follows that
$\forall a: (B^\dagger)^ap_a=0$ and   $\forall b: 
B^b\tilde p_b=0$. Suppose then that some $(B^\dagger)^ap_a=0$. One is easily brought into a situation where
\begin{equation} X_a=(B^\dagger)^a(a_0+a_1L+\cdots+a_r L^r)=0
\end{equation}
for some $r$ and some constants $a_0,\dots,a_r$.
Then $B^\dagger X_a-X_a B^\dagger=0$ and hence 
\begin{equation} (B^\dagger)^{a+1}(a_1(1-q)L+\cdots+a_r(1-q^r) L^r)=0.
\end{equation}
Continuing like that, one concludes that $ (B^\dagger)^{R}=0$ for some $R\in{\mathbb N}$. One may then apply Lemma~\ref{5.2} to conclude that (since $q$-generic), $(B^\dagger)^{R-1}=0$. Proceeding like that one gets the contradiction $1=0$. The case  $B^b\tilde p_b=0$ follows analogously.\qed

\medskip

\begin{Prop} $\forall s=1,2,\dots, {\mathcal H}{\mathcal W}_{s}$
is a simple algebra which is a domain. \end{Prop}

\proof It suffices to consider ${\mathcal H}{\mathcal W}_{1}$.
Suppose $p_1\cdot p_2=0$ for $p_1,p_2\in {\mathcal H}{\mathcal
W}_{1}$. One can easily shift this equation, if needed, to an
equation $p_3\cdot p_4=0$ where both elements $p_3,p_4$ are finite sums of
the general form $\sum_{x}(B^\dagger)^xp_x$ for Laurent polynomials $p_x$. By looking at
the highest powers of $B^\dagger$ one easily concludes that at
least one of the factors must be $0$.

Let us then consider a non-zero ideal $I$ in ${\mathcal
H}{\mathcal W}_{1}$, and let $p\in I$ be a non-zero element.  The proof now proceeds exactly as the previous proof. The only difference is that the equations $=0$ should be replaced by ``$\in I$''. After the manipulations, one then reaches the statement $1\in I$. This implies that $I$ is the full space. \qed

\bigskip

Equally obvious is:

\begin{Prop}\label{basis-prop}
The following is a basis of ${\mathcal F}^{(q)}_{s}$:
\begin{equation}
\{(B^\dagger_1)^{a_1}(B^\dagger_2)^{a_2}\cdots (B^\dagger_s)^{a_s}v_s^{(q)}\mid(a_1,a_2,\dots,a_s)\in{\mathbb N}_0^s\}.
\end{equation}
\end{Prop}

\begin{Cor}\label{inject}$\forall i: {\mathbb B}^\dagger_i$ acts injectively on ${\mathcal F}^{(q)}_{s}$. 
\end{Cor}

\medskip

\subsection{The Hayashi-Weyl algebra and ${\mathcal
U}_q(su(n,n))$}

We now proceed to give an explicit homomorphism $\psi_q^{(1)}$ of
${\mathcal U}_q(su(n,n)^\mathbb C)$ into ${\mathcal H}{\mathcal
W}_{2n}$. If $G_i, i=1, \dots, 8n-4 $, is a set of generators of ${\mathcal 
U}_q(su(n,n)^\mathbb C)$, we
set $\forall i: \psi_q^{(1)}(G_i)=G_i^{(1)}$  where the elements $G_i^{(1)}$ are:

\begin{Def} \label{hw2}\begin{align*} &
E^{(1)}_\beta=-B_1B_{n+1}, &
F^{(1)}_\beta=B^\dagger_1B^\dagger_{n+1}, \\ &
E^{(1)}_j=B^\dagger_jB_{j+1}, & F^{(1)}_j=B_jB_{j+1}^\dagger,
\quad 1\leq j<n, \\ & E^{(1)}_j=B^\dagger_jB_{j+1}, &
F^{(1)}_j=B_jB^\dagger_{j+1}, \quad 2n>j>n,\\
&K^{(1)}_\beta=q^{-1}L^{-1}_1L^{-1}_{n+1},&	 
K^{(1)}_j=L^{-1}_{j+1}L_j\textrm{ for }j<n\textrm{ or }j>n.
\end{align*}
\end{Def}

It is easy to see that it holds that
\begin{eqnarray*}
E^{(1)}_\beta
F^{(1)}_\beta-F^{(1)}_\beta E^{(1)}_\beta&=&\frac{K^{(1)}_\beta-{(K^{(1)}_\beta)}^{-1}}
{q-q^{-1}},\\
E^{(1)}_j
F^{(1)}_j-F^{(1)}_jE^{(1)}_j&=&\frac{K^{(1)}_j-{(K^{(1)}_j)}^{-1}}
{q-q^{-1}}.\end{eqnarray*}

\medskip

The crucial point is the following:
\begin{Lem}
The above operators satisfy the $q$-Serre relations for type $A_{2n-1}$.
\end{Lem}

\proof This follows easily from Lemma~\ref{qws}. \qed

\medskip

The following is quite obvious:

\begin{Def/Prop}
The element
\begin{equation}
\Xi_q:=L_1\cdot \cdots\cdot L_n \cdot L_{n+1}^{-1}\cdot\cdots\cdot L_{2n}^{-1}\label{Xi-q}
\end{equation}commutes with all the above generators.
\end{Def/Prop}

\medskip

The element $\Xi_q$ can be joined to ${\mathcal U}_q(su(n,n)^\mathbb C)$ so that we get ${\mathcal U}_q(u(n,n)^\mathbb C)$. This is useful just as ${\mathcal U}(u(n,n)^\mathbb C)$ was in the classical case.

\medskip

Having thus given $\psi_q^{(1)}$ on the generators, it is
determined on all expressions. We denote by ${\mathbb W}_{i,j}^{(1)}$ the
image of $W_{i,j}$ (c.f. \ref{26})) under the above map and note, in particular, that
$\psi_q^{(1)}(W_{1,1})=F^{(1)}_\beta$,
$\psi_q^{(1)}(W_{2,1})=(-q^{-1})(F^{(1)}_\beta F^{(1)}_1-
qF^{(1)}_1 F^{(1)}_\beta)= F^{(1)}_1 F^{(1)}_\beta-q^{-1}
F^{(1)}_\beta F^{(1)}_1= B^\dagger_2B^\dagger_{n+1}L_1$. 
We will henceforth change the notation as follows:
$B_i\rightarrow B_i(1)$, $B^\dagger_i\rightarrow
B^\dagger_i(1)$, $L^{\pm1}_i\rightarrow L^{\pm1}_i(1)$.

Define symbols

\begin{eqnarray} {L}_{k-1\downarrow}(1)=L_1(1)\cdots
L_{k-1}(1)\textrm{ for }k>1&,& {L}_0(1)=1,\\ {L}_{n+\ell-1\downarrow}(1)=L_{n+1}(1)\cdots L_{n+\ell-1}(1)\textrm{ for
}\ell>1&,& {L}_{n+0\downarrow}(1)=1. \end{eqnarray}

An easy computation reveals that \begin{equation}
{\mathbb W}_{k,\ell}^{(1)}= B^\dagger_k(1) B^\dagger_{n+\ell}(1){L}_{k-1\downarrow}
(1){L}_{n+\ell-1\downarrow}(1). \end{equation} On the matrix level,
we denote the image of the matrix ${\mathbb W}$ by ${\mathbb W}^{(1)}$. 
The entry of the latter at position $k,\ell$ is given by  ${\mathbb W}_{k,\ell}^{(1)}$. We also proceed
to define symbols ${\mathbb B}^\dagger_k(1)=B^\dagger_k(1){L}_{k-1\downarrow}(1)$ and notice that
\begin{eqnarray}{\mathbb B}^\dagger_b(1) {\mathbb
B}^\dagger_a(1)&=&q{\mathbb B}^\dagger_a(1){\mathbb
B}^\dagger_b(1)\textrm{ if }n\geq b>a\geq1 \textrm{ or }2n\geq
b>a\geq n+1,\\{\mathbb B}^\dagger_b(1) {\mathbb
B}^\dagger_a(1)&=&{\mathbb B}^\dagger_a(1){\mathbb
B}^\dagger_b(1)\textrm{ if }n\geq a\geq1 \textrm{ and }2n\geq
b\geq n+1 .\end{eqnarray}

We then have \begin{equation}{\mathbb
W}^{(1)}=\begin{pmatrix}{\mathbb B}_1^\dagger(1)\\\vdots\\
{\mathbb B}^\dagger_k(1)\\\vdots\\{\mathbb
B}^\dagger_n(1)\end{pmatrix} \begin{matrix}\\\cdot& ({\mathbb
B}^\dagger_{n+1}(1) & \hdots & {\mathbb B}^\dagger_{n+\ell}(1)
&\hdots & {\mathbb B}^\dagger_{2n}(1))\\&\\&\\&\\&\\
&\\&\\\end{matrix}. \end{equation}

\begin{Prop}\label{q2x2} The elements ${\mathbb W}_{i,j}^{(1)}$ satisfy the FRT($q^{-1}$) relations (\ref{frt}). In addition:  Let $a<c$ and $b<d$ be arbitrary. The $2\times2$ $q$-minor ${\mathbb W}^{(1)}_{a,b}{\mathbb W}^{(1)}_{c,d}-q^{-1}{\mathbb W}^{(1)}_{a,d}{\mathbb W}^{(1)}_{b,c}$ vanishes and, indeed, in this situation ${\mathbb W}^{(1)}_{a,b}{\mathbb W}^{(1)}_{c,d}=q^{-2} {\mathbb W}^{(1)}_{cd}{\mathbb W}^{(1)}_{a,b}$. \end{Prop}

\proof. It suffices to take $a=b=1$ and $c=d=2$. In this case, by inserting the definitions and using the relations, one gets
\begin{equation}
{\mathbb W}^{(1)}_{a,b}{\mathbb W}^{(1)}_{c,d}-q^{-1}{\mathbb W}^{(1)}_{a,d}{\mathbb W}^{(1)}_{b,c}=B^{\dagger}_{1}B^{\dagger}_{n+1}
B^{\dagger}_{2}B^{\dagger}_{n+2}L_1L_{n+1}-q^{-1}B^{\dagger}_{2}B^{\dagger}_{n+1}L_1B^{\dagger}_{1}B^{\dagger}_{n+1}L_{n+1}=0.
\end{equation}The other claims follow analogously. \qed

\medskip

In analogy with Proposition~\ref{3.11} and Theorem~\ref{3.12} we define ${\mathcal I}^{(q)}_{2\times 2}$ to be the ideal in ${\mathcal A}^-_q$ generated by all quantum $2\times2$ minors, and we define $({\mathcal A_q^-})^{(1)}$ to be the subalgebra of ${\mathcal H}{\mathcal W}_{2n}$ generated by the elements $w_{ij}^{(1)}$.

In the same way we get:

\begin{Prop}
\begin{equation}
({\mathcal A_q^-})/ {\mathcal I}^{(q)}_{2\times 2}\equiv ({\mathcal A_q^-})^{(1)}.
\end{equation}
\end{Prop}

\medskip

\begin{Thm}${\mathcal I}^{(q)}_{2\times 2}$ is prime.\label{q2-prime}
\end{Thm}

\bigskip

Analogously to Proposition~\ref{4.1} one gets the following special case of the of coming (\S6) general cases:

\begin{Prop}The cyclic module for ${\mathcal U}_q(su(n,n)^\mathbb C)$  defined by $v^{(q)}_{2n}$ through $\pi_{2n}^{qSvN}\circ \psi^{(1)}_q$, is an
irreducible highest weight representation of ${\mathcal
U}_q(su(n,n))$ of highest weight $(0,0,-1)$. \end{Prop}

\begin{Rem} This was studied from another perspective by D.
Shklyarov, S. Sinel'\-shchikov, A. Stolin, and L. Vaksman
(\cite{vaks}). \end{Rem}

\bigskip

\subsection{$q$-analogues of maps and tensor stuctures} 

We wish to generalize the maps and results from from \S\ref{s3}. For any
integer $r>1$ we identify \begin{equation} {\mathcal H}{\mathcal
W}_{2nr}=\underbrace{{\mathcal H}{\mathcal
W}_{2n}\otimes\cdots\otimes {\mathcal H}{\mathcal W}_{2n}}_{r\
copies}.\label{iden} \end{equation}

Specifically, we write the generators of ${\mathcal H}{\mathcal
W}_{2nr}$ as $B_s(k), B^\dagger_s(k)$, $s=1,\dots,n$ and
$k=1,\dots, r$. More generally, for a fixed $k=1,\dots,r$, we
write an element $G$ in the subalgebra generated by the elements
$B_s(k), B^\dagger_s(k), s=1,\dots,2n$ as $G(k)$. The
identification in ({\ref{iden}) is then generated by the
identifications
\begin{equation}G(k)\leftrightarrow1\otimes\cdots\otimes
1\otimes G(k)\otimes 1\otimes\cdots\otimes 1 \textrm{ (in the
$k$th position)}. \end{equation}

\medskip

\begin{Def} We denote by ${\mathcal H}{\mathcal W}^X_{nr}$ the subalgebra of ${\mathcal H}{\mathcal W}_{2nr}$ generated by elements $B_\ell(k), (B_\ell(k))^\dagger$ having a subscript $\ell=1,2,\dots,n$ and $1\leq k\leq r$. Likewise,   ${\mathcal H}{\mathcal W}^Y_{nr}$ denotes the subalgebra generated by elements $B_\ell(k), (B_\ell(k))^\dagger$  having a subscript $\ell=n+1,n+2,\dots,2n$ and $1\leq k\leq r$.
\end{Def}

\medskip

Clearly:

\begin{Lem}
\begin{equation}\label{hw-tensor}
{\mathcal H}{\mathcal W}^X_{nr}\otimes {\mathcal H}{\mathcal W}^Y_{nr}={\mathcal H}{\mathcal W}_{2nr}.
\end{equation}
\end{Lem}

\medskip

We first define a homomorpism $\psi_q^{(r+1)}$ of ${\mathcal
U}_q(su(n,n)^\mathbb C)$ into ${\mathcal H}{\mathcal W}_{2n(r+1)}$
for $r\in{\mathbb N}$. We do this by using the co-product
$\Delta^r$: \begin{equation}\label{r-times} \Delta^r: {\mathcal
U}_q({\mathfrak g}^{\mathbb C})\rightarrow \underbrace{{\mathcal
U}_q({\mathfrak g}^{\mathbb C})\otimes\cdots\otimes {\mathcal
U}_q({\mathfrak g}^{\mathbb C})}_{r+1}.\end{equation}
Then,\begin{equation}\label{95}\psi_q^{(r+1)}=(\underbrace{\psi_q^{(1)}
\otimes\cdot\otimes
\psi_q^{(1)}}_{r+1})\circ\Delta^r.\end{equation}

\medskip

The co-product we use is the one of  (\cite{jan}), given in its simplest 
${\mathcal U}_q(sl(2,{\mathbb C}))$ version for standard generators $e,f,K$ as 
\begin{eqnarray} \label{st1}\Delta e&=&e\otimes
1+K\otimes e,\\ \Delta f&=& f\otimes K^{-1}+1\otimes f,\\ \Delta
K^{\pm1}&=&K^{\pm1}\otimes K^{\pm1}.\label{st3} \end{eqnarray} \bigskip

To make the formulas more simple, we first set $L_{a,b}=L_a(b)$ in all cases. Then we define, for all
$x=1,\dots,n,n+1,\dots,2n$, and $b=1,\dots,r$ \begin{eqnarray}
L_{a,b^\uparrow}&=&L_a(b)L_a(b+1)\cdots L_a(r),\textrm{ and}\\
L_{a,b^\downarrow}&=&L_a(b)L_a(b-1)\cdots L_a(1)
.\end{eqnarray}
We extend the definition to any $b\in{\mathbb Z}$ such that  
$L_{a,b^\updownarrow}=1$ if $b\notin \{1,\dots,r\}$. In the following, $j\in\{1,\dots,n-1,n+1,\dots,2n-1\}$.

\begin{Def}
\begin{eqnarray}\label{e-b}
e_\beta^{(r)}&=&-\sum_{i=1}^rq^{-i+1}{B}_1(i){B}_{n+1}(i)L^{-1}_{1,
(i-1)^\downarrow}
L^{-1}_{n+1,(i-1)^\downarrow}\  ,\\
f_\beta^{(r)}&=&\sum_{i=1}^rq^{r-i}
{B}_1^\dagger(i){B}_{n+1}^\dagger(i)L_{1,(i+1)^\uparrow}
L_{n+1,(i+1)^\uparrow}\ ,\\ e_j^{(r)}&=&\sum_{i=1}^r
{B}_{j+1}(i){B}^\dagger_{j}(i)L^{-1}_{j+1,( i-1)^\downarrow}
L_{j,(i-1)^\downarrow}\ ,
\\ f_j^{(r)}&=&\sum_{i=1}^r
{B}_j(i){B}^\dagger_{j+1}(i) 
L_{j+1,(i+1)^\uparrow} L^{-1}_{j,(i+1)^\uparrow}
\label{f-j}\ ,\\
(K_\beta^{(r)})^{\pm1}&=&K^{\pm1}_\beta(1)\cdot\cdots\cdot K^{\pm1}_\beta(r), \\
(K_j^{(r)})^{\pm1}&=&K^{\pm1}_j(1)\cdot\cdots\cdot K^{\pm1}_j(r).
  \end{eqnarray}
\end{Def}

\medskip

Since $\triangle$ is a homomorphism, we get:

\begin{Prop}
The above operators satisfy the $q$-Serre relations.
\end{Prop}

Analogously to $r=1$ we get:

\begin{Def/Prop}The operator\begin{equation}
\Xi_q^{(r)}:=\Xi_q(1)\cdot\cdots\cdot \Xi_q(r)\label{q-center}\end{equation}
commutes with all of the above generators.
\end{Def/Prop}

\medskip

\begin{Def}
The representation 
$${\mathcal R}^{qharm}_r=\pi^{qSvN}_{2nr}\circ\psi_q^{(r)}$$is the ($r$th) $q$-Harmonic Representation of ${\mathcal U}_q(u(n,n)^{\mathbb C})$ in ${\mathcal F}^{(q)}_{2nr}$.
\end{Def}

\medskip

\subsection{The structure of $\psi_q^{(r+1)}({\mathcal A}_q^-)$} 

We first wish to compute ${\mathbb W}^{(r+1)}$ for $r\geq1$: \begin{Lem}In ${\mathcal
U}_q({\mathfrak g}^{\mathbb C})\otimes{\mathcal
U}_q({\mathfrak g}^{\mathbb C})$ it holds that
\begin{eqnarray}&\Delta(T_{\mu_{k}}T_{\mu_{k-1}}\cdots
T_{\mu_{1}}(F_\beta))=\\&\nonumber
T_{\mu_{k}}T_{\mu_{k-1}}\cdots T_{\mu_{1}}(F_\beta)\otimes
K^{-\beta-\mu_1-\cdots-\mu_{k}}+1\otimes
T_{\mu_{k}}T_{\mu_{k-1}}\cdots T_{\mu_{1}}(F_\beta)+\\&\nonumber
(q^{-1}-q)\sum_{1\leq a\leq k}T_{\mu_{k-a}}T_{\mu_{k-a-1}}\cdots
(F_\beta) \otimes (T_{\mu_{k}}\cdots
T_{\mu_{k+2-a}}(F_{\mu_{k+1-a}}))K^{-\beta-\mu_1-\cdots-\mu_{k-a}},
\end{eqnarray} or, equivalently,
\begin{eqnarray}&\Delta(W_{k+1,1})=\\&\nonumber W_{k+1,1}\otimes
K^{-\beta-\mu_1-\cdots-\mu_{k}}+1\otimes W_{k+1,1}+\\&\nonumber
(q^{-1}-q)\sum_{1\leq a\leq k}W_{k-a+1,1} \otimes
(-q^{-1})^a(T_{\mu_{k}}\cdots
T_{\mu_{k+2-a}}(F_{\mu_{k+1-a}}))K^{-\beta-\mu_1-\cdots-\mu_{k-a}}.
\end{eqnarray} \end{Lem}

\proof We use   Lemma~\ref{2.1} and (\ref{t-si}) repeatedly. We also observe that there will be expressions - with $E$ unspecified - of the form
$E(K^{-\alpha}F_\gamma-q F_\gamma K^{-\alpha})$ and 
$E(F_\gamma K^{-\alpha}-q K^{-\alpha}F_\gamma)$ where $\langle \alpha,\gamma\rangle=-1$. The first kind equals $E(q^{-1}-q) F_\gamma K^{-\alpha}$, the second is zero. The formulas then follow by induction. \qed

\medskip

We will generally set $W^{(r+1)}_{k,\ell}= \psi^{(r+1)}(W_{k,\ell})$ when there is no chance of confusing  the suberscript on  $W^{(r+1)}_{k,\ell}$ with a power.     More generally, using $\Delta^{r+1}=(\Delta^r\otimes 1)\Delta$,
we have:

\begin{Lem} \begin{eqnarray}&W^{(r+1)}_{k+1,1}=\\&\nonumber
W^{(r)}_{k+1,1}\otimes K^{-\beta-\mu_1-\cdots-\mu_{k}}+1\otimes
W^{(r)}_{k+1,1}+\\&\nonumber (q^{-1}-q)\sum_{1\leq a\leq
k}W^{(r)}_{k-a+1,1} \otimes
(-q^{-1})^a(T_{\mu_{k}}T_{\mu_{k-1}}\cdots
T_{\mu_{k+2-a}}(F_{\mu_{k+1-a}}))K^{-\beta-\mu_1-\cdots-\mu_{k-a}}.
\end{eqnarray}

Equivalently, using the notation $[A,B]_q=AB-q^{-1}BA$,
\begin{eqnarray*} W^{(r+1)}_{k+1,1}&=&W^{(r)}_{k+1,1}\otimes
K^{-\beta-\mu_1-\cdots
-\mu_{k}}\\&+&(q^{-1}-q)W^{(r)}_{k,1}\otimes
(-q^{-1})F_{\mu_k}K^{-\beta-\mu_1-\cdots -\mu_{k-1}}\\
&+&(q^{-1}-q)W^{(r)}_{k-1,1}\otimes (-q^{-1})
[F_{\mu_{k}},F_{\mu_{k-1}}]_qK^{-\beta-\mu_1-\cdots
-\mu_{k-2}}\\ &+&(q^{-1}-q)W^{(r)}_{k-2,1}\otimes (-q^{-1})
[F_{\mu_{k}},[F_{\mu_{k-1}},F_{\mu_{k-2}}]_q]_qK^{-\beta-\mu_1-\cdots
-\mu_{k-3}}\\ &+&\cdots\\ &+&(q^{-1}-q)W^{(r)}_{1,1}\otimes
(-q^{-1}) [[F_{\mu_k},\dots
F_{\mu_{3}},[F_{\mu_{2}},F_{\mu_1}]_q]_q\dots]_qK^{-\beta}\\
&+&1_R\otimes W^{(1)}_{k+1,1}. \end{eqnarray*} \end{Lem}

\medskip

\begin{Def}
$$\forall r=1,2,\dots, \ {\mathbb W}^{(r)}_{k,\ell}=\psi_q^{(r)}(W^{r}_{k,\ell}).$$
\end{Def}

From Definition~\ref{hw2}, and formula
(\ref{25}), we then get:

\medskip

\begin{Lem}\label{r+1} \begin{eqnarray*}
{\mathbb W}^{(r+1)}_{k+1,1}&=&{\mathbb W}^{(r)}_{k+1,1}\otimes q
\left(L_{k+1}L_{n+1}\right)(r+1)\\&+&(q^{-1}-q){\mathbb W}^{(r)}_{k,1}\otimes
(-1)\left(B^\dagger_{k+1}B_{k}L_{k}L_{n+1} \right)(r+1)\\
&+&(q^{-1}-q){\mathbb W}^{(r)}_{k-1,1}\otimes
(-1)\left(B^\dagger_{k+1}B_{k-1}L_{k}L_{k-1}L_{n+1}\right)(r+1)\\
&+&(q^{-1}-q){\mathbb W}^{(r)}_{k-2,1}\otimes
(-1)\left(B^\dagger_{k+1}B_{k-2}L_{k}
L_{k-1}L_{k-2}L_{n+1}\right)(r+1)\\ &+&\cdots\\
&+&(q^{-1}-q){\mathbb W}^{(r)}_{1,1}\otimes
(-1)\left(B^\dagger_{k+1}B_{1}L_{k}L_{k-1}\cdots
L_2L_1L_{n+1}\right)(r+1)\\ &+&1_R\otimes {\mathbb W}^{(1)}_{k+1,1}(r+1).
\end{eqnarray*} 
\end{Lem}

Recall that
${\mathbb W}^{(1)}_{k+1,1}(r+1)=B^\dagger_{k+1}B^\dagger_{n+1}
L_{k\downarrow}
$ (at the $(r+1)$th position).

\medskip

More generally,

\begin{eqnarray}\label{recu-first}
&&{\mathbb W}^{(r+1)}_{k+1,\ell+1}=\\&&{\mathbb W}^{(r)}_{k+1,\ell+1}\otimes q
\left(L_{k+1}L_{n+\ell+1}\right)(r+1)\\&+&(1-q^{-2})\sum_{x=1}^k q{\mathbb W}^{(r)}_{x,\ell+1}\otimes
\left(B^\dagger_{k+1}B_{x}L_{k}\cdots\nonumber L_xL_{n+\ell+1} \right)(r+1)\\\nonumber
&+&(1-q^{-2})\sum_{y=1}^\ell q{\mathbb W}^{(r)}_{k+1,y}\otimes
\left(B^\dagger_{n+\ell+1}B_{n+y}L_{n+\ell}\cdots L_{n+y}L_{k+1} \right)(r+1)\\  \nonumber
&+&(1-q^{-2})^2\sum_{x,y=1}^{k,\ell} q{\mathbb W}^{(r)}_{x,y}\otimes
\left(B^\dagger_{k+1}B^\dagger_{n+\ell+1}B_xB_{n+y}L_{n+\ell}\cdots L_{n+y}L_{k}\cdots L_x \right)(r+1)\\  \nonumber
    &+&1_R\otimes {\mathbb W}^{(1)}_{k+1,\ell+1}(r+1).
\end{eqnarray}

\medskip

\begin{Def} \label{elements}In ${\mathcal H}{\mathcal W}_{2nr}$,  for $k=1,\cdots, r-1$, set \begin{eqnarray}{\mathbb B}^\dagger_\ell(k,r)&=&B^\dagger_\ell(k) L_{(\ell-1)\downarrow}(k) L_{\ell}(k+1)\cdots L_{\ell}(r),\\ {\mathbb B}_\ell(k,r)&=&B_\ell(k) L^{-1}_{\ell\downarrow}(k)L^{-1}_{\ell}(k+1)\cdots L^{-1}_{\ell}(r),\\
{\mathbb B}^\dagger_{n+\ell}(k,r)&=&B^\dagger_{n+\ell}(k)
L_{n+\ell-1\downarrow (to\ n+1)}(k)L_{n+\ell}(k+1)\cdots L_{\ell}(r),\\
{\mathbb B}_{n+\ell}(k,r)&=&B_{n+\ell}(k) L^{-1}_{(n+\ell)\downarrow(to\ (n+1))}(k)L^{-1}_{n+\ell}(k+1)\cdots L^{-1}_{n+\ell}(r).
\end{eqnarray}\label{above}Moreover,  ${\mathbb B}_\ell^\dagger(r)={\mathbb B}_\ell^\dagger(r,r):={B}_\ell^\dagger(r)L_{(\ell-1)\downarrow}(k)$ and ${\mathbb B}_\ell(r)={\mathbb B}_\ell(r,r):={B}_\ell(r)L^{-1}_{\ell\downarrow}(k)$, for any $x,y\in\{1,\dots,2n\}$, respectively.
\end{Def}

\begin{Def}
Two operators $A,B$ quasi-commute if $AB-q^\alpha BA=0$ for some $q^\alpha$. If the generators of an algebra all quasi-commute we say that the algebra is a quasi-polynomial algebra. 
\end{Def}

\medskip
The following is straightforward:
\begin{Lem}\label{con}
The elements in Definition~\ref{elements}  quasi-commute  with the following notable exception:
\begin{equation}
\forall k,\ell:\ {\mathbb B}_\ell(k,r){\mathbb B}^\dagger_\ell(k,r)-q^{-2}{\mathbb B}^\dagger_\ell(k,r){\mathbb B}_\ell(k,r)=q^{-1}\cdot I.
\end{equation}
\end{Lem}

\medskip
Then, obviously:
\begin{Prop}
The elements in Definition~\ref{above} generate an algebra which has a PBW-like basis.
\end{Prop}

\medskip

\begin{Cor}\label{corwkl}
\begin{eqnarray}\label{w-k-l}
&&{\mathbb W}^{(r)}_{k+1,\ell+1}=\\\nonumber&&{\mathbb W}^{(r-1)}_{k+1,\ell+1}q
L_{k+1}(r)L_{n+\ell+1}(r)\\\nonumber&+&(1-q^{-2})\sum_{x=1}^k q{\mathbb W}^{(r-1)}_{x,\ell+1}L_{x}(r)L_{n+\ell+1}(r)
{\mathbb B}^\dagger_{k+1}(r,r){\mathbb B}_{x}(r,r)\\\nonumber
&+&(1-q^{-2})\sum_{y=1}^\ell q{\mathbb W}^{(r-1)}_{k+1,y}L_{k+1}(r)L_{n+y}(r)
{\mathbb B}^\dagger_{n+\ell+1}(r,r){\mathbb B}_{n+y}(r,r)\\  \nonumber
&+&(1-q^{-2})^2\sum_{x,y=1}^{k,\ell} q{\mathbb W}^{(r-1)}_{x,y}L_x(r)L_{n+y}(r)
{\mathbb B}^\dagger_{k+1}(r,r){\mathbb B}^\dagger_{n+\ell+1}(r,r){\mathbb B}_x(r,r){\mathbb B}_{n+y}(r,r)\\  
    &+&{\mathbb B}^\dagger_{k+1}(r,r){\mathbb B}^\dagger_{n+\ell+1}(r,r).\nonumber
\end{eqnarray} 
\end{Cor}
\medskip

\begin{Def} We denote by ${\mathcal H}{\mathcal W}^{\mathbb Z}_{2nr}$ the subalgebra generated by the elements in Definition~\ref{elements}. Furthermore, ${\mathcal H}{\mathcal W}^{\mathbb X}_{nr}$ denotes the subalgebra generated by those of the above generators having a subscript $\ell=1,2,\dots,n$ and  ${\mathcal H}{\mathcal W}^{\mathbb Y}_{nr}$ denotes the subalgebra generated by those of the above generators having a subscript $n+\ell=n+1,n+2,\dots, 2n$.
\end{Def}

\medskip

Clearly:

\begin{Lem}\label{lem-use1}
There is a PBW basis for ${\mathcal W}^{\mathbb Z}_{2nr}\cdot v_{2nr}^{(q)}$ of the form
\begin{equation}
\left({\mathbb B}_{1}^\dagger(1,r)\right)^{a_{1,1}}\left({\mathbb B}_{2}^\dagger(1,r)\right)^{a_{2,1}}\cdots \left({\mathbb B}_{2n}^\dagger(r,r)\right)^{a_{2n,r}}\cdot v_{2nr}^{(q)}.
\end{equation}
\end{Lem}

The following result is easy to see:

\begin{Lem}${\mathcal H}{\mathcal W}^{\mathbb X}_{nr}$ and ${\mathcal H}{\mathcal W}^{\mathbb Y}_{nr}$ commute and
\begin{equation}
{\mathcal H}{\mathcal W}^{\mathbb X}_{nr}\otimes{\mathcal H}{\mathcal W}^{\mathbb Y}_{nr}={\mathcal H}{\mathcal W}^{\mathbb Z}_{2nr}.
\end{equation}
\end{Lem}

\medskip

\begin{Def}
\begin{eqnarray}
{\mathbb A}(k,r)&:=&q^{1/2}\sum_{a\geq b} \left(\delta_{a,b}+(1-\delta_{a,b})(1-q^{-2}){\mathbb B}_{a}^\dagger(k,r){\mathbb B}_{b}(k,r)\right)E_{a,b},\\
{\mathbb C}(k,r)&:=&q^{1/2}\sum_{d\geq c} \left(\delta_{c,d}+(1-\delta_{c,d})(1-q^{-2}){\mathbb B}_{n+d}^\dagger(k,r){\mathbb B}_{n+c}(k,r)\right)E_{c,d},\quad\end{eqnarray}
\begin{eqnarray}
{\mathbb X}^{\mathbf 0}(k,r)&:=&\begin{pmatrix}{\mathbb B}_1^\dagger(k,r)\\\vdots\\
{\mathbb B}^\dagger_\ell(k,r)\\\vdots\\{\mathbb
B}^\dagger_n(k,r)\end{pmatrix}, 
{\mathbb Y}^{\mathbf 0}(k,r):=\begin{pmatrix}{\mathbb B}_{n+1}^\dagger(k,r)\\\vdots\\
{\mathbb B}^\dagger_{n+\ell}(k,r)\\\vdots\\{\mathbb
B}^\dagger_{2n}(k,r)\end{pmatrix},\\ {\mathbb W}^{\mathbf 0}(k,r)&:=& {\mathbb X}^{\mathbf 0}(k,r)\cdot {\mathbb Y}^{\mathbf 0}(k,r)^T.
\end{eqnarray}

\medskip

Analogously to Proposition~\ref{q2x2}, and with a similar proof, we have:

\begin{Prop}Let ${\mathbb W}_{i,j}^{\mathbf 0}(k,r), i,j=1,\dots,n$ denote the entries of  ${\mathbb W}^{\mathbf 0}(k,r)$, $k=1,\dots,r$. \label{qRxR} For a fixed, arbitrary $k$, these elements satisfy the FRT($q^{-1}$) relations (\ref{frt}). In addition:  Let $a<c$ and $b<d$ be arbitrary. The $2\times2$ $q$-minor ${\mathbb W}_{a,b}^{\mathbf 0}(k,r){\mathbb W}_{c,d}^{\mathbf 0}(k,r)-q^{-1}{\mathbb W}_{a,d}^{\mathbf 0}(k,r){\mathbb W}_{c,b}^{\mathbf 0}(k,r)$ vanishes and, indeed, in this situation ${\mathbb W}_{a,b}^{\mathbf 0}(k,r){\mathbb W}_{c,d}^{\mathbf 0}(k,r)=q^{-2} {\mathbb W}_{c,d}^{\mathbf 0}(k,r){\mathbb W}_{a,b}^{\mathbf 0}(k,r)$. \end{Prop}

\medskip

We let ${\mathcal H}{\mathcal W}^{{\mathbb Z},{\mathbf 0}}_{2nr}$ denote the subalgebra of ${\mathcal H}{\mathcal W}^{\mathbb Z}_{2nr}$ generated by the entries of ${\mathbb W}^{\mathbf 0}(k,r)$, $k=1,\dots,r$. Likewise, ${\mathcal H}{\mathcal  W}^{{\mathbb X},{\mathbf 0}}_{nr}$ denotes the subalgebra  generated by the entries of ${\mathbb X}^{\mathbf 0}(k,r)$ $k=1,\dots,r$, and ${\mathcal H}{\mathcal W}^{{\mathbb Y},{\mathbf 0}}_{nr}$ denotes the subalgebra  generated by the entries of ${\mathbb Y}^{\mathbf 0}(k,r)$, $k=1,\dots,r$.
\end{Def}

\medskip

The following is obvious:

\begin{Lem}
${\mathcal H}{ \mathcal W}^{{\mathbb Z},{\mathbf 0}}_{2nr}$ is a quasi-polynomial algebra. We have
\begin{equation}
{\mathcal H}{\mathcal W}^{{\mathbb Z},{\mathbf 0}}_{2nr}={\mathcal H}{\mathcal W}^{{\mathbb X},{\mathbf 0}}_{nr}\otimes {\mathcal H}{\mathcal W}^{{\mathbb Y},{\mathbf 0}}_{nr}.
\end{equation}
\end{Lem}

\medskip

\begin{Prop}
\begin{equation}\label{recu}
{\mathbb W}^{(r)}={\mathbb W}^{\mathbf 0}(r,r)+\sum_{k=2}^r {\mathbb A}(r,r)\cdots {\mathbb A}(k,r){\mathbb W}^{\mathbf 0}(k-1,r){\mathbb C}(k,r)\cdots {\mathbb C}(r,r).
\end{equation}
\end{Prop}

\pof Let us introduce two diagonal $n\times n$ matrices
\begin{equation}
{\mathbb L}^\downarrow(r)=Diag(L_{1}(r),\cdots, L_{n}(r))\textrm{ and }{\mathbb L}^\uparrow(r)=Diag(L_{n+1}(r), \cdots, L_{2n}(r)).
\end{equation}
Then, by (\ref{w-k-l}), it follows directly that \begin{equation}\label{recu1}
\boxed{{\mathbb W}^{(r)}={\mathbb W}^{\mathbf 0}(r,r)+{\mathbb A}(r,r){\mathbb L}^\downarrow(r){\mathbb W}^{(r-1)} {\mathbb L}^\uparrow(r){\mathbb C}(r,r)}.
\end{equation}
Proceeding recursively, one obtains the result if one makes the following two observations: 1) 
\begin{equation}
{\mathbb L}^\downarrow(r)\cdots {\mathbb L}^\downarrow(r-k){\mathbb A}(r-k-1,r-k-1)({\mathbb L}^\downarrow(r)\cdots {\mathbb L}^\downarrow(r-k))^{-1}={\mathbb A}(r-k-1,r),
\end{equation}with a similar statement for ${\mathbb C}(r-k-1,r)$.  

2) \begin{equation}
{\mathbb L}^\downarrow(r)\cdots {\mathbb L}^\downarrow(r-k){\mathbb W}(r-k-1,r-k-1){\mathbb L}^\uparrow(r-k)\cdots {\mathbb L}^\uparrow(r))={\mathbb W}(r-k-1,r).
\end{equation}
\qed

\medskip

\begin{Cor}\label{use-cor1}The entries of ${\mathbb W}^{(r)}$ are elements of ${\mathcal H}{\mathcal W}^{\mathbb Z}_{2nr}$.
\end{Cor}

\medskip

\begin{Def} \label{modif} Set ${\mathbb X}_r(r)={\mathbb X}(r,r)$, ${\mathbb Y}_r(r)={\mathbb Y}(r,r)$, and, for $k=1,\cdots,r-1$,
\begin{equation}
{\mathbb X}_r(k):={\mathbb A}(r,r)\cdots {\mathbb A}(k+1,r){\mathbb X}^{\mathbf 0}(k,r), {\mathbb Y}_r(k):={\mathbb C}^T(r,r)\cdots {\mathbb C}^T(k+1,r){\mathbb Y}^{\mathbf 0}(k,r).
\end{equation}
\end{Def}

We easily get the following two assertions (compare to (\ref{XYT})):

\begin{Prop}
\begin{equation}\label{540}{\mathbb W}^{(r)}={\mathbb X}_r(r){\mathbb Y}_r^T(r)\cdots + {\mathbb X}_r(k){\mathbb Y}_r^T(k)+\cdots + {\mathbb X}_r(1){\mathbb Y}_r^T(1).
\end{equation}
\end{Prop}

\begin{Cor}Let ${\mathbb X}_r$ be the $n\times r$ matrix whose $k$th column is ${\mathbb X}_r(k)$ and let ${\mathbb Y}_r$ be the $n\times r$ matrix whose $k$th column is ${\mathbb Y}_r(k)$. Then
\begin{equation}
{\mathbb W}^{(r)}={\mathbb X}_r{\mathbb Y}_r^T.\label{rank2}
\end{equation}\label{rank}
\end{Cor}

\medskip

Notice that the bases depend on $r$.

\medskip

Two more easy consequences are:

\begin{Cor}
\begin{equation}
{\mathcal H}{\mathcal W}^{\mathbb Z}_{2nr}\cdot v_{2nr}={\mathcal H}{ \mathcal W}^{{\mathbb Z},{\mathbf 0}}_{2nr}\cdot v_{2nr}.
\end{equation}
\end{Cor}

\smallskip

\begin{Cor}
The image ${\mathcal H}{\mathcal W}_{2nr}^{{\mathcal A}_q^-}$ of ${\mathcal A}_q^-$ in ${\mathcal H}{\mathcal W}_{2nr}$ by $\psi_{q}^{(r)}$ is generated by the entries of ${\mathbb W}^{(r)}$.
\end{Cor}

\bigskip

\subsection{Annihilators}

\begin{Def}\label{-1-def}We use the definition of a quantum minor in ${\mathcal A}_q^-$ as given in  (\ref{2}) with $a=q^{-1}$.  We let ${\mathcal I}^{(q)}_{k\times k}$ denote the ideal in ${\mathcal A}_q^-$ generated by all quantum $k\times k$ minors.
\end{Def}

\begin{Def}\label{id-def}
$Ann^{-,0}_q(2nr)$ denotes the set of elements in ${\mathcal A}_q^-$ that annihilate $v_{2nr}^{(q)}$ in the representation ${\mathcal R}^{qharm}_r$.
\end{Def}

It is easy to see that this is a ${\mathcal U}_q({\mathfrak k}^{\mathbb C})$-module which is a left ideal.

\medskip

Using the quantum Laplace expansions (see e.g. \cite{yam}), the following is not hard to prove:
\begin{Prop}\label{leftright}
The left ideal generated by $k\times k$ quantum minors is equal to the right ideal generated by $k\times k$ quantum minors.
\end{Prop}

We have:

\begin{Prop}\label{ann-det}
\begin{equation}
{\mathcal I}^{(q)}_{(r+1)\times (r+1)}\subseteq Ann_q^{-,0}(2nr).
\end{equation}
\end{Prop}

\pof This follows by the representation theory of ${\mathcal U}_q({\mathfrak k}^\mathbb C)$. The action of ${\mathcal U}_q({\mathfrak k}^\mathbb C)$ on ${\mathcal R}^{qharm}_r({\mathcal I}^{(q)}_{(r+1)\times (r+1)})\cdot v_{2nr}^{(q)}$ is just the restriction of the left action on ${\mathcal F}_{2nr}^{(q)}$. The latter is  the $r$th fold tensor product of the symmetric representation and hence  cannot accommodate more that $r\times r$ minors corresponding to anti-symmetric $r$ tensors. See also Proposition~\ref{q-most-gen}. \qed

\medskip

\begin{Def}
Let ${{\mathbb W}}^{(r,0)}$ denote the $n\times n$ matrix
\begin{equation}
{{\mathbb W}}^{(r,0)}=\sum_{k=1}^R {\mathbb W}^{0}(k,r).
\end{equation}
Specifically, the entries are given as
\begin{equation}
{{\mathbb W}}^{(r,0)}_{ij}=\sum_{k=1}^r{\mathbb B}^\dagger_i(k,r){\mathbb B}^\dagger_{n+j}(k,r)\in {\mathcal H}{\mathcal W}^{Z,{\mathbf 0}}_{2nr}.
\end{equation}
${{\mathbb W}}^{(r,0)}$ is {\bf the associated quasi-polynomial matrix at level $r$}.
\end{Def}

We shall see that for many purposes, the associated quasi-polynomial matrix at level $r$ contains enough information to reach conclusions about ${\mathbb W}^{(r)}$.

Similarly to Corollary~\ref{rank} we have:
\begin{Cor}Let ${\mathbb X}^{\mathbf 0}_r$ be the $n\times r$ matrix whose $k$th column is ${\mathbb X}^{\mathbf 0}(k,r)$ and let ${\mathbb Y}^{\mathbf 0}_r$ be the $n\times r$ matrix whose $k$th column is ${\mathbb Y}^{\mathbf 0}(k,r)$. Then
\begin{equation}
{{\mathbb W}}^{(r,0)}={\mathbb X}^{\mathbf 0}_r({\mathbb Y}^{\mathbf 0}_r)^T.
\end{equation}
\end{Cor}

\medskip

We now discuss $\ell\times\ell$ quantum determinants involving the entries of ${\mathbb W}^{(r,0)}$. The algebra ${\mathcal H}{ \mathcal W}^{{\mathbb Z},{\mathbf 0}}_{2nr}$ does not satisfy the FRT equations. Non-the-less, we apply (\ref{3}) to it. For simplicity we only give the top left ``determinant'', which we mark by ``TL''. (The formula (\ref{2}) will actually give the same result.) 
\begin{Lem}
\begin{eqnarray}\\\nonumber&(\qdet)^{TL}_{\ell\times\ell}({\mathbb W}^{(r,0)})=\\\nonumber&
\sum_{\sigma\in S_\ell} (-q)^{-\ell(\sigma)} \sum_{k_1,k_2,\dots,k_\ell}^{r,r,\cdots,r} {\mathbb B}^\dagger_{n+1}1(k_1){\mathbb B}^\dagger_{\sigma(1)}{\mathbb B}^\dagger_{n+2}(k_2){\mathbb B}^\dagger_{\sigma(2)}\cdots {\mathbb B}^\dagger_{n+\ell}(k_\ell) {\mathbb B}^\dagger_{\sigma(\ell)}=\\\label{k1k2}&
\sum_{k_1,k_2,\dots,k_\ell}^{r,r,\cdots,r} ({\mathbb B}^\dagger_{n+1}(k_1)\cdots {\mathbb B}^\dagger_{n+\ell}(k_\ell)) \sum_{\sigma\in S_\ell} (-q)^{-\ell(\sigma)} ({\mathbb B}^\dagger_{\sigma(1)}(k_1)\cdots {\mathbb B}^\dagger_{\sigma(\ell)}(k_\ell))\label{pw}.\nonumber
\end{eqnarray}
\end{Lem}

\medskip

If we then consider the case $\ell>r$ (still, $\ell\leq n$) then, in the expressions 
\begin{equation}
\sum_{\sigma\in S_\ell} (-q)^{-\ell(\sigma)} ({\mathbb B}^\dagger_{\sigma(1)}(k_1){\mathbb B}^\dagger_{\sigma(2)}(k_2)\cdots {\mathbb B}^\dagger_{\sigma(\ell)}(k_\ell)),
\end{equation}one must have $k_i=k_j$ for at least one pair $(i,j)$ with $i\neq j$.

\begin{Lem}\label{top R+1}
Suppose that $k_{i_0}=k_{j_0}$ with $i_0<j_0$. We may assume that no $k_i$, $i_0< i <j_0$ is equal to $k_{i_0}$. Then 
\begin{equation}
\sum_{\sigma\in S_\ell} (-q)^{-\ell(\sigma)} {\mathbb B}^\dagger_{\sigma(1)}(k_1){\mathbb B}^\dagger_{\sigma(2)}(k_2)\cdots {\mathbb B}^\dagger_{\sigma(\ell)}(k_\ell))=0.
\end{equation}

\end{Lem}

\pof Let $k_{i_0}=k_{j_0}=a\in {\mathbb N}$.  It is easy to see that  ${\mathbb B}^\dagger_i(k)$ commutes with ${\mathbb B}^\dagger_j(\ell)$ if $(i-j)(k-\ell)\neq 0$. We focus on the terms  ${\mathbb B}^\dagger_{\sigma(i_0)}(a) {\mathbb B}^\dagger_{\sigma(j_0)}(b)$ as $\sigma$ varies through $S_\ell$. For any choice of values of $\sigma$ on all positions $i\neq i_0,j_0$ there are exactly two possible choices of $(\sigma(i_0),\sigma(j_0))$, which we may denote as $(c,d)$ and $(d,c)$ with $c<d$.  The last one is a ``cross-over'' and has a length which is one more than the length of the other. But ${\mathbb B}^\dagger_d(a){\mathbb B}^\dagger_c(a)=q{\mathbb B}^\dagger_c(a){\mathbb B}^\dagger_d(a)$ (Definition~\ref{elements} and (\ref{25d})). Thus, the two expressions cancel. \qed

\medskip

We then conclude:

\begin{Cor}\label{cor-det}
Any $(r+1)\times (r+1)$ minor of ${\mathbb W}^{(r,0)}$ is zero. ($r+1\leq n$.)
\end{Cor}

\medskip

We now offer a technical computation to be used below:

\begin{Lem}Assume $i>x$ and $j>y$. Then
\begin{eqnarray*}
&\left({\mathbb B}^\dagger_{i}(r){\mathbb B}_{x}(r){\mathbb  W}^{(r-1)}_{x,j}+{\mathbb B}^\dagger_{n+j}(r){\mathbb B}_{n+y}(r){\mathbb  W}^{(r-1)}_{i,y}+{\mathbb B}^\dagger_{i}(r){\mathbb B}^\dagger_{n+j}(r){\mathbb B}_{x}(r){\mathbb B}_{n+y}(r){\mathbb  W}^{(r-1)}_{x,y}\right)\cdot \\&\left({\mathbb B}^\dagger_{a}(r){\mathbb B}^\dagger_{n+b}(r)+ {\mathbb B}^\dagger_{c}(r){\mathbb B}^\dagger_{n+d}(r){\mathbb  W}^{(r-1)}_{f,g}\right)\cdot v_{2nr}^{(q)}={\big[}q^{-1}\delta_{a,x}{\mathbb B}^\dagger_{i}(r){\mathbb B}^\dagger_{n+b}(r){\mathbb  W}^{(r-1)}_{a,j}+\\&q^{-1}\delta_{b,y}{\mathbb B}^\dagger_{a}(r){\mathbb B}^\dagger_{n+j}(r){\mathbb  W}^{(r-1)}_{i,b}+q^{-2}\delta_{a,x}\delta_{b,y}{\mathbb B}^\dagger_{i}(r){\mathbb B}^\dagger_{n+j}(r){\mathbb  W}^{(r-1)}_{a,b}+\\&q^{-1}\delta_{c,x}{\mathbb B}^\dagger_{i}(r){\mathbb B}^\dagger_{n+d}(r){\mathbb  W}^{(r-1)}_{c,j}{\mathbb  W}^{(r-1)}_{f,g}+ q^{-1}\delta_{d,y}{\mathbb B}^\dagger_{c}(r){\mathbb B}^\dagger_{n+j}(r){\mathbb  W}^{(r-1)}_{i,d}{\mathbb  W}^{(r-1)}_{f,g}+\\& q^{-2}\delta_{c,x}\delta_{d,y}{\mathbb B}^\dagger_{i}(r){\mathbb B}^\dagger_{n+j}(r){\mathbb  W}^{(r-1)}_{c,d}{\mathbb  W}^{(r-1)}_{f,g}{\big]}\cdot  v_{2nr}^{(q)}.
\end{eqnarray*}
\end{Lem}

\pof Using Lemma~\ref{con} and the fact that all ${\mathbb B}_a(r)$ commute with all ${\mathbb W}^{(r-1)}_{b,c}$, this is a straightforward computation. \qed

\medskip

Now recall Corollary~\ref{corwkl}. By paying attention to degrees and observing everywhere that 
\begin{equation}
{\mathbb B}^\dagger_{i}(r){\mathbb B}^\dagger_{n+j}(r)= {\mathbb W}^{(r,0)}_{i,j}- {\mathbb W}^{(r-1,0)}_{i,j}
\end{equation}
we obtain, by applying the last lemma repeatedly:

\smallskip

\begin{Lem}\label{filter}
There exist, depending on the choices, homogeneous polynomials $p_s^{(r,0)}$ of degree $s$ in the entries of ${\mathbb W}^{(r,0)}$ and polynomials $\hat p^{(r-1)}_{t}$ in the entries of ${\mathbb W}^{(r-1,0)}$ and ${\mathbb W}^{(r-1)}$  such that, when $k\geq2$ (in the case $k=1$ the second term is $0$)
\begin{eqnarray*}
{\mathbb W}^{(r)}_{i_1,j_1}\cdots {\mathbb W}^{(r)}_{i_k,j_k} v^{(q)}_{2nr}&=&{\mathbb W}^{(r,0)}_{i_1,j_1}\cdots {\mathbb W}^{(r,0)}_{i_k,j_k} v_{2nr}\\&+&
\sum_{s=1}^{k-1} p_s^{(r,0)}\cdot\hat p^{(r-1)}_{s}v^{(q)}_{2nr}.
\end{eqnarray*}
\end{Lem}

\bigskip

\begin{Rem} Using Corollary~\ref{cor-det} the above Lemma offers an alternative proof of Proposition~\ref{ann-det}. On e.g. the level of ${\mathcal U}_q({\mathfrak k}^\mathbb C)$ modules, we may filter the $R$th fold tensor product - construed as a series of $R$ tensor products from, say, the left, by  the previous $(r-1)$th tensor product. Indeed, we may (compare to the discussions about tensor products in Section~\ref{4})  pass to the quotient by the left ideal generated by the elements  $({\mathbb W}^{(r-1)}_{i,j})v^{(q)}_{2nr}$ - i.e we can set them equal  to zero. Finally, we can apply Corollary~\ref{cor-det}.\end{Rem}

\medskip

Let $M_{top}(r)$ denote the top leftmost  $r\times r$ minor of 
${\mathbb W}^{(r)}$.
      
\begin{Cor}
$(M_{top}(r)) v^{(q)}_{2nr}$ is non-zero. 
\end{Cor}

\pof It suffices to prove that the analogous element obtained from ${\mathbb W}^{(r,0)}$ is non-zero. Consider the, obviously non-zero, element $\prod_{i=1}^r {\mathbb B}_i^\dagger (i,r){\mathbb B}_{n+i}^\dagger (i,r)$. This is a summand in  (\ref{pw}) precisely when $\forall i:k_i=i$ and $\sigma=Id$. \qed

\begin{Cor}$(M_{top}(r))^k v^{(q)}_{2nr}$ is non-zero for any $k\in{\mathbb N}$.
\end{Cor}

\pof This follows similarly.  \qed

More generally, let $X_i(r)$ to be the $i\times i$ minor of ${\mathbb W}^{(r)}$ consisting of the first $i$ rows and columns.

\begin{Cor}
The element $\prod_{i=1}^{r}X_{i}^{N_{i}}v^{(q)}_{2nr}$ is non-zero for any $(N_1,\cdots, N_{r})\in{\mathbb N}_0^{r}$.
\end{Cor}

\pof If any such vector were to belong to the left ideal $Ann^{-,0}_q(2nr)$ a simple tensor product argument reveals that $(M_{top}^r)^k\in Ann^{-,0}_q(2nr)$ for some $k\in{\mathbb N}$ which contradicts the previous corollary. \qed

\medskip

The above elements are highest weight vectors for ${\mathcal U}_q(u(n)\times u(n))$ (c.f Proposition~\ref{q-minor-type}). In the language of Young diagrams (here: double diagrams), we get exactly those that have at most $r$ rows. In the ideal ${\mathcal
I}^{(q)}_{(r+1)\times (r+1)}$ all ${\mathcal U}_q(u(n)\times u(n))$ types will necessarily have Young diagrams with at least $(r+1)$ rows. Thus we conclude:

\medskip

\begin{Thm}\label{539}
$$Ann^{-,0}_q(2nr)={\mathcal
I}^{(q)}_{(r+1)\times (r+1)}.$$
\end{Thm} 

\bigskip

\subsection{$q$-prime}

We now turn our attention to {\em prime ideals}.

\medskip

\begin{Lem}\label{weight}
Let $p\in {\mathcal H}{\mathcal W}^0_{2nr}$, let $L=L_k(\ell)$ for some fixed $k,\ell$, and let 
\begin{equation}
p=\sum_{i=1}^N p_i,\textrm{ where }\forall i: Lp_iL^{-1}=q^{n_i}p_i.
\end{equation}
Suppose the elements $n_1,\cdots,n_N\in{\mathbb N}_0$ are pairwise different. Suppose that $pv^{(q)}_{2nr}=0$. Then
\begin{equation}\forall i=1,\cdots,N:\ p_iv^{(q)}_{2nr}=0.\end{equation}
\end{Lem}

\pof We have that for all $s\in {\mathbb N}_0$: $L^spv^{(q)}_{2nr}=0$. Hence
\begin{equation}
\forall s\in {\mathbb N}_0:\  \sum_{i=1}^N q^{sn_i}p_i v^{(q)}_{2nr}=0.
\end{equation}
The claim follows easily by subtractions and multiplications by $q$. See also Artin's Theorem on linear independence of characters (\cite{artin}). \qed

\medskip

\begin{Thm}[Q-Prime]\label{q-prime}
The $q$-determinantal ideals are prime.
\end{Thm}

\proof

Set $W=W^{(r)}$ (recall that they  satisfy the FRT($q^{-1}$) equations). Further, set 
\begin{eqnarray}
{\mathbf a}&=&(a_{n,1},a_{n,2},\cdots, a_{n,n}),\\
{\mathbf b}&=&(b_{1,1},\cdots,b_{i,j},\cdots, b_{n-1,n}),\ i\leq  n-1,\\
	p&=&\sum \alpha_{{\mathbf a},{\mathbf b}}W^{\mathbf b}W^{\mathbf a},\textrm{ and}\\\hat p&=&\sum_{{\mathbf c},{\mathbf d}}\beta_{{\mathbf c},{\mathbf d}}W^{\mathbf d}W^{\mathbf c}\ ({\mathbf c},{\mathbf d}\textrm{ defined analogously}).
\end{eqnarray}

Assume $(p\cdot\hat p)v_{2nr}^{(q)}$=0. Then either $pv_{2nr}=0$ or  $\hat pv_{2nr}^{(q)}=0$.

\proof

The elements $p,\hat p\in{\mathcal A}^-_q$ are only determined modulo ${\mathcal
I}^{(q)}_{(r+1)\times (r+1)}$. We will use this fact to make some assumptions about them.

 We order the elements ${\mathbf a},{\mathbf c}$ lexicographical so that $(n,n)$ is the biggest index and $W_{n,n}$ is placed furthest to the right.  Let ${\mathbf a_0}$ be the biggest power in ${\mathbf a}$, and ${\mathbf c_0}$ defined analogously. The elements ${\mathbf b}, {\mathbf d}$ are ordered similarly, row by row, ending with the first row at the left. We use a PBW basis based on this order.
\begin{eqnarray}
p&=&\sum_{\mathbb b}\alpha_{{\mathbf a_0},{\mathbf b}}W^{\mathbf b}W^{\mathbf a_0} +\ L.o.t.s\textrm{ and, similarly}\\
\hat p&=&\sum_{\mathbb b}\beta_{{\mathbf c_0},{\mathbf d}}W^{\mathbf d}W^{\mathbf c_0}  +\ L.o.t.s\ .
\end{eqnarray}
The symbol $L.o.t.s$ in both equations stand for lower order terms.

Observe that $W^{\mathbf a}W^{\mathbf b}=W^{\mathbf b}W^{\mathbf a}$ modulo lower order terms.

Consider then

\begin{eqnarray}
&(\sum_{\mathbb b}\alpha_{{\mathbf a_0},{\mathbf b}}W^{\mathbf b})\cdot(\sum_{\mathbb b}\beta_{{\mathbf c_0},{\mathbf d}}W^{\mathbf d})W^{\mathbf a_0}W^{\mathbf c_0}v_{2nr}^{(q)}=\\&p_0\cdot\hat p\cdot W^{\mathbf a_0}W^{\mathbf c_0}\cdot  v_{2nr}^{(q)},
\end{eqnarray}
where $p_0=\sum_{\mathbb b}\alpha_{{\mathbf a_0},{\mathbf b}}W^{\mathbf b}$ and $\hat p_0=\sum_{\mathbb b}\beta_{{\mathbf c_0},{\mathbf d}}W^{\mathbf d}$. 

\medskip

{\bf The assumptions we make is that $p_0\cdot v_{2nr}^{(q)}\neq 0$ and 
$\hat p_0\cdot v_{2nr}^{(q)}\neq 0$}. Indeed, if e.g. $p_0\cdot v_{2nr}^{(q)}=0$ then $p_0\in {\mathcal I}^{(q)}_{R+1}$ and the latter is a 2-sided ideal by Proposition~\ref{leftright}, hence $p_0{\mathbf W}^{{\mathbf a}_0}$ can be left out of our considerations.

\medskip

By construction, ${\mathbb B}^\dagger_{n}(k,r)$ and ${\mathbb B}^\dagger_{2n}(k,r)$ do not take part in $p_0,\hat p_0$ for any $k=1,\dots,r$. Similarly, ${\mathbb B}_{n}(k,r)$ and ${\mathbb B}_{2n}(k,r)$ do not take part in $p_0,\hat p_0$ for any $k=1,\dots,r$. Indeed, by Corollary~\ref{corwkl}, not even in $p,\hat p$.

\medskip

Let us first look at $({\mathbb W}^{(r)}_{nn})^N v_{2nr}^{(q)}$. The key ingredient is  (\ref{w-k-l}), or its equivalent form (\ref{recu}). Observe that no operators ${\mathbb B}_n(k,r)$ or  ${\mathbb B}_{2n}(k,r)$ take part in $W_{nn}^{(r)}$, and that if a term ${\mathbb B}_{i}(k,r)$ is present, then $i\neq n,2n$ and hence may be moved to the right where it eventually will annihilate $v_{2nr}^{(q)}$.   We then get that 
\begin{equation}
(W_{nn})^N v_{2nr}^{(q)}=\left({\mathbb B}^\dagger_n(1,r){\mathbb B}^\dagger_{2n}(1,r)+\cdots +{\mathbb B}^\dagger_n(r,r){\mathbb B}^\dagger_{2n}(r,r)\right)^Nv_{2nr}^{(q)}.
\end{equation}
More generally, in the $n$th row at column $j$ we only get annihilation operators ${\mathbb B}_y(k,r)$ with $y<j$, and ${\mathbb B}_x(k,r)$ with $x<n$, hence, by our choice of ordering,  we  get

\medskip

{\small
\begin{eqnarray}&
W_{n1}^{a_{n,1}}W_{n2}^{a_{n,2}}\cdots W_{nn}^{a_{n,n}}
 v_{2nr}^{(q)}=\\&
 \left({\mathbb B}^\dagger_n(1,r){\mathbb B}^\dagger_{n+1}(1,r)+\cdots +{\mathbb B}^\dagger_n(r,r){\mathbb B}^\dagger_{n+1}(r,r)\right)^{a_{n,1}}\left({\mathbb B}^\dagger_n(1,r){\mathbb B}^\dagger_{n+2}(1,r)+\cdots +{\mathbb B}^\dagger_n(r,r){\mathbb B}^\dagger_{n+2}(r,r)\right)^{a_{n,2}}\cdot\nonumber\\&\cdots  \left({\mathbb B}^\dagger_n(1,r){\mathbb B}^\dagger_{2n}(1,r)+\cdots +{\mathbb B}^\dagger_n(r,r){\mathbb B}^\dagger_{2n}(r,r)\right)^{a_{n,n}}v_{2nr}^{(q)}=
 \\&\left({\mathbb W}^{(r,0)}_{n1}\right)^{a_{n,1}}\left({\mathbb W}^{(r,0)}_{n2}\right)^{a_{n,2}}\cdots \left({\mathbb W}^{(r,0)}_{nn}\right)^{a_{n,n}}
 v_{2nr}^{(q)}.
\end{eqnarray}
}
\medskip

Let us now look at  $L_n(1)$ and $L_{2n}(1)$ weights:

\smallskip

We have that 
\begin{equation}
L_n(1)^k(p\cdot \hat p)L_n(1)^{-k}v_{2nr}^{(q)}=0,
\end{equation}
hence, by Lemma~\ref{weight}, any weight space is zero.

It follows that $p_0\cdot \hat p_0$ annihilates for instance 
 \begin{equation}({\mathbb B}^\dagger_n(1,r))^{a_{n,1}+a_{n,2}+\cdots+a_{n,n}}({\mathbb B}^\dagger_{n+1}(1,r))^{a_{n,1}}({\mathbb B}^\dagger_{n+2}(1,r))^{a_{n,2}}\cdot\cdots\cdot  ({\mathbb B}^\dagger_{2n}(1,r))^{a_{n,n}}\ v_{2nr}^{(q)}\label{591}.\end{equation} 

  We rewrite  $p_0\cdot \hat p_0$ in such a way that the terms ${\mathbb B}_x(k,r)$ move to the right. At the same time we now write it in terms of the variables ${\mathbb B}^\dagger_x(k,r)$ and ${\mathbb B}_y(k,r)$, $x,y=1,\dots,2n$ but with the aforementioned exceptions.

 Observe, by e.g. proposition~\ref{540},  that $p_0\cdot \hat p_0$ contains no elements of the form ${\mathbb B}_x(1,r)$, $x=1,\dots,2n$. Furthermore, any ${\mathbb B}_x(k,r)$, $x=1,\dots,2n$, and $k=2,\dots, R$ annihilate the element in (\ref{591}) automatically.  Suppose then that $p_0\cdot \hat p_0$ contains a biggest non-trivial summand $h\in{\mathcal H}{\mathcal W}^{\mathbb Z}_{2nr}$ and that $h\cdot v_{2nr}^0\neq0$. C.f.  Corollary~\ref{use-cor1}. Clearly then, by Lemma~\ref{lem-use1}, also 
 \begin{equation}
 h\cdot ({\mathbb B}^\dagger_n(1,r))^{a_{n,1}+a_{n,2}+\cdots+a_{n,n}}({\mathbb B}^\dagger_{n+1}(1,r))^{a_{n,1}}({\mathbb B}^\dagger_{n+2}(1,r))^{a_{n,2}}\cdot\cdots\cdot  ({\mathbb B}^\dagger_{2n}(1,r))^{a_{n,n}}\cdot v_{2nr}^{(q)}\neq0.
 \end{equation}
 
This is a contradiction, so, in conclusion, it follows that $p_0\cdot \hat p_0 v_{2nr}^{(q)}=0$. As mentioned, $p_0 v_{2nr}^{(q)}\neq 0$ and $\hat p_0 v_{2nr}^{(q)}\neq 0$. 
 
 \smallskip
 
We thus arrive at a similar situation, but where the $n$th row of ${\mathbb W}^{(r)}$ has been removed.

\medskip

Now remove the last column $(i,j)=(i,n)$ by a similar argument. 

\medskip

Suppose first that $n=r+1$ so that we are looking at the quotient module by the ideal generated by the $(r+1)\times (r+1)$ quantum determinant. Here we observe that a non-zero polynomial in the variables $W_{ij}$ with $\max\{i,j\}\leq n-1$ can never be a product of a polynomial and the $n\times n$ quantum determinant. Neither can the  product of two such because the mentioned condition defines a sub-algebra. So if $n=r+1$ we reach a contradiction.

\smallskip

Of course, we always have that $n\geq r+1$. If $n=r+k$, with $k>1$ we can by the previous arguments make the reduction $(p,\hat p)\rightarrow (p_0,\hat p_0)\rightarrow \cdots$ until we reach a pair $p_{k-1},\hat p_{k-1}$ which depend only on the top $r=n-k$ variables. But we cannot get an $r+1$ determinant from such variables, so  by Theorem~\ref{539} we cannot get zero. This contradiction then proves the general theorem. 

 \qed

\bigskip

\section{The full decomposition of the $q$-Fock spaces}\label{6}

The $q$-Stone--von Neumann module ${\mathcal
F}^{(q)}_{2nr}=\pi^{(q)}_{2nr}({\mathcal H}{\mathcal
W}_{2nr})v^{(q)}_{2nr}$ becomes a module for ${\mathcal
U}_q(u(n,n))$ through the homomorphism $\psi_q^{(r)}$
(\ref{95}) and by adding the element $\Xi_q^{(r)}$ (\ref{q-center}). We need to investigate it as a ${\mathcal
U}^+_q(su(n,n))$ module. We will not distinguish between the elements 
(\ref{e-b}-\ref{f-j}) and their actions in this module.

We aim to prove an analogue of Theorem~\ref{k-v-thm}. One interesting aspect is that while we will be working with the unital abelian algebra  $({\mathcal H}{\mathcal
W}_{2nr})_{2nr}^+$ generated by the elements $({\mathbb B}_x(y))^\dagger$,   
we, non-the-less, arrive in a natural way at quantum minors.
We will, with obvious modifications, use the same notation as in \S4. Now, as usual, a highest weight $\Lambda$ and a highest weight vector $v_\Lambda$ of a representation  of ${\mathcal U}_q({\mathfrak g})$ satisfies
\begin{equation}\label{hwv} K^\xi v_\Lambda=q^{\langle \Lambda,\xi\rangle}v_\Lambda
\end{equation} 
for any $\xi$ in the root lattice. We will further often assume that 
\begin{equation}\Xi_q^{(r)}v_\Lambda=\zeta\cdot v_\Lambda
\end{equation}in which case we replace $(\Lambda,v_\Lambda)$ by  $((\Lambda,\zeta),v_{(\Lambda,\zeta)})$.

\medskip

In the spirit of (\ref{hwv}), roots and weights will be identified with their ``classical'' counterparts ($q=1$). Highest weight representations are defined analogously and, due to M. Rosso (\cite{rosso}) and G. Lusztig (\cite{lu0}), in the generic case, the dimensions of finite-dimensional highest weight modules are independent of $q$.

\medskip

Proceeding as in \S4, $\tilde M_q(V_{\tau_\Lambda})$ now  denotes the highest weight module of ${\mathcal U}_q(su(n,n))$ defined by the same highest weight data as  $\tilde M(V_{\tau_\Lambda})$ and it is a quotient by the analogous invariant sub spaces.

\medskip

What is not so clear, and must be established, is that there is a representation of ${\mathcal U}_q(gl(r))$ on ${\mathcal F}^{(q)}_{2nr}$ which commutes with the action of  ${\mathcal U}_q(su(n,n))$ as well as  with $\Xi_q^{(r)}$. Furthermore, it is not clear what should replace the highest weight vectors $\Delta$  though the correct answer, namely $\Delta_q$, seems so obvious. After all, the vector space ${\mathcal F}^{(q)}_{2nr}$ has a basis built up from monomials in commuting operators acting on the ``vacuum'' - just like in the case of ${\mathcal F}_{2nr}$, and hence, these two spaces are linearly isomorphic. As we shall see, this observation is not important; it is the actions that dictate what the highest weight vectors should be. Naturally they will be given in terms of the operators $B^\dagger_i(k)$.

\medskip
 
It follows immediately that ${\mathcal
F}^{(q)}_{2nr}=\pi^{(q)}_{2nr}({\mathcal H}{\mathcal
W}_{2nr})v^{(q)}_{2nr}$ becomes a module for ${\mathcal
U}_q(u(n)\times u(n))$ through the homomorphism $\psi_q^{(r)}$ and the element $\Xi_q^{(r)}$. The vectors in  ${\mathcal
F}^{(q)}_{2nr}$ transform under symmetrized tensor products of the defining representation, so it is clear that we have the following analogue of Proposition~\ref{most-gen} (c.f. Corollary~\ref{minor-type}):

\smallskip

\begin{Prop}\label{q-most-gen}If $r<n$, the most general ${\mathcal U}_q({\mathfrak k}^{\mathbb C})$-type in ${\mathcal F}_{2nr}$ has a $q$-highest weight  of the form
\begin{equation}
{(a_1,\cdots,a_r,0,\cdots),(b_1,\cdots,b_r,0,\cdots),-a_1-b_1-r)}.
\end{equation}
If $r\geq n$, the most general ${\mathcal U}_q({\mathfrak k}^{\mathbb C})$-type in ${\mathcal F}^{(q)}_{2nr}$ has a $q$-highest weight of the form
\begin{eqnarray}
{(a_1,\cdots,a_{n-1},0),(b_1,\cdots,b_{n-1},0,),-a_1-b_1-\tilde r)},\textrm{ with }a_{n-1},b_{n-1}\geq0,\\\textrm{ and } \tilde r=r, r+1,r+2,\dots.\qquad\nonumber
\end{eqnarray}
The multiplicities are the same as in the classical case.
\end{Prop}

\medskip

With these provisos, we can now formulate the main results of this case.  The explanations and proofs will take up this last section of the article.

\medskip            

\begin{Thm}\label{q-k-v-thm}
\begin{equation}{\mathcal  F}^{(q)}_{2nr}=\oplus_{\xi\in \Sigma^{(r)}_n}\dim(\xi)\tilde M_q(V_{(\Lambda,\zeta)^\xi}). 
\end{equation}
When $q$-is real and positive, the representations $M_q(V_{\tau_\Lambda})$ are
unitary.
\end{Thm}

\medskip

The analogue of Proposition~\ref{minor-type} is (compare also to Definition~\ref{asDef} and Proposition~\ref{asDef2}).

\begin{Prop}\label{q-minor-type}
If ${\mathbf a}=(a_1,\cdots,a_s)$, set $\ell(\mathbf a)=s$. Let
\begin{equation}
\forall k\leq\min\{n,r\}: \triangle^q_k({\mathbf x})=\sum_{\sigma\in S_k}(-q)^{sgn(\sigma)}\prod_{i=1}^kB_i^\dagger(\sigma(i)),
\end{equation}
\begin{equation}
\forall \ell\leq\min\{n,r\}: \tilde\triangle^q_\ell({\mathbf y})=\sum_{\sigma\in S_\ell}(-q)^{sgn(\sigma)}\prod_{i=1}^\ell B_{n+i}^\dagger(r-\sigma(i)).
\end{equation}Further, let 
\begin{equation}\triangle^q({\mathbf x})^{\mathbf
a}= \prod_{k=1}^{\ell(\mathbf a)}\label{6.5}
\left(\triangle^q_k({\mathbf x})\right)^{a_k}, \end{equation}
\begin{equation}\tilde\triangle^q({\mathbf y})^{\mathbf
b}= \prod_{k=1}^{\ell(\mathbf b)}
\tilde\triangle^q_k({\mathbf y})^{b_k} .\end{equation}

\smallskip

If ${\ell(\mathbf a)}+{\ell(\mathbf b)}\leq r$, the vectors

\begin{equation}
\triangle^q({\mathbf x})^{\mathbf a}\tilde\triangle^q({\mathbf y})^{\mathbf b}v_{2nr}^{(q)} \end{equation}
are highest weight vectors both for ${\mathcal U}_q(su(n,n))$ and ${\mathcal U}_q(gl(r))$. Furthermore, any joint highest weight vector is of this form.
\end{Prop}

\bigskip

\subsection{The details; NYM}

We wish to describe in detail the modules involved and not just 
appeal to the ``well-known'' fact that the representation theory at $q$-generic 
is the same as for $q=1$. Recall the important result of de Consini, Eisenbud, and Procesi (\cite{EPdC}) in the classical case. Our main tool now is the article by M. Noumi, H. Yamada, and K. Mimachi 
(henceforth NYM; \cite{yam}),  which will give us the analogue for the quantum case. Their work relies on the following result by M. Rosso (quoting (\cite{yam})
\begin{Thm}[\cite{rosso}] If $q$-is not a root of unity, any finite dimensional 
${\mathcal U}_q(sl(n,{\mathbb C}))$ module is completely reducible.
\end{Thm}

\smallskip

All our modules will be weight modules by construction.

\medskip

In analogy with the classical case we set:

\begin{Def} A {\bf top $q$-pluri-harmonic polynomial of weight $(\Lambda,\zeta)$} (and of order $r$)  is an
element $p \in {\mathcal F}^{(q)}_{2nr}$ which is annihilated by
${\mathcal U}_q^{>0}$ and which satisfies
\begin{eqnarray}
\forall\alpha\in\Pi: K_\alpha
p&=&q^{\langle\alpha,\Lambda\rangle}p\textrm{ for some weight }\Lambda\\
\Xi_q^{(r)}p&=&q^\zeta p.
\end{eqnarray}
\end{Def}

\medskip

Parallel to the classical case we now wish to introduce an extra module structure 
on ${\mathcal
F}^{(q)}_{2nr}$. 

\medskip

At first we will study two sub-spaces of ${\mathcal
F}^{(q)}_{2nr}$, namely one built up  solely by applying operators $B^\dagger_i(j)$ with $i\leq n$, and one solely by  $B^\dagger_{n+i}(j)$ with $i\leq n$, respectively, to the vacuum vector $v_{2nr}$. 

\medskip

This is where (\cite{yam}) can be utilized. In it a bi-module structure on a certain algebra, 
to be defined right below, is introduced. We will set up a {\bf vector space} isomorphism 
between this algebra and each of the two mentioned sub-spaces of ${\mathcal F}^{(q)}_{2nr}$, thus furnishing these subspaces
with bi-module structures too. 

\medskip

Here are the notations and structures from (\cite{yam}):

\smallskip

The coordinate ring ${\mathcal A}(r,s,{\mathbb C})$ is the
${\mathbb C}$ algebra generated by variables $w_{i,j}$ with
$1\leq i\leq r$ and $1\leq j\leq s$ satisfying the relations
\begin{eqnarray}\label{130} &w_{ij}w_{ik}=qw_{ik}w_{ij}\ ,
w_{ji}w_{ki}=qw_{ki}w_{ji}\ (j<k),\\
&w_{i\ell}w_{jk}=w_{jk}w_{i\ell}\ (i<j,\ k<\ell), \\
&w_{ik}w_{j\ell}-w_{j\ell}w_{ik}=(q-q^{-1})w_{i\ell}w_{jk}\
(i<j,\ k<\ell). \label{132}\end{eqnarray}

\smallskip

  In $su(N)$    we identify 
$\mu_i\leftrightarrow \varepsilon_i-\varepsilon_{i+1}$ for any simple root 
$\mu_i, i=1,\dots,N-1$. Here, $\{\varepsilon_1,\dots,\varepsilon_N\}$ is the 
standard orthonormal basis of ${\mathbb R}^N$. ${\mathcal U}_q(su(N))$ is then 
given in its usual guise, compatible with this.

Slightly adapted and expanded to our notation we now cite:
\begin{Thm}{\cite[(1.35.1)-(1.35.c)]{yam}} 
There is a  left action by ${\mathcal U}_q(u(s))$ and a right action by 
${\mathcal U}_q(u(r))$  on the vector space spanned by the elements $w_{i,j}; 
i=1,\dots, s; j=1,\dots, r$  given as follows;

\begin{eqnarray}\label{133-bis1}L_k\star w_{ij}&=&q^{\langle\varepsilon_k,\varepsilon_j\rangle}w_{ij},
\\\label{133} K_{\mu_k} \star
w_{ij}&=&q^{\langle\mu_k,\varepsilon_j\rangle}w_{ij}\\
&=&L_k\star L_{k+1}^{-1}\star w_{ij},\\
E_{\mu_k}\star w_{ij}&=&\delta_{j,k+1}w_{i,j-1},\\ F_{\mu_k}\star
w_{ij}&=&\delta_{j,k}w_{i,j+1}\label{134},
\\\label{135-bis1} w_{ij}\star L_{k}
&=&q^{\langle{\varepsilon_k},\varepsilon_i\rangle}w_{ij},
\\\label{135} w_{ij}\star K_{\mu_k}
&=&q^{\langle{\mu_k},\varepsilon_i\rangle}w_{ij}\\\label{135-bis2}&=&w_{ij}\star L_k\star L_{k+1}^{-1},\\ w_{ij}\star
E_{\mu_k}&=&\delta_{i,k}w_{i+1,j},\\ w_{ij}\star
F_{\mu_k}&=&\delta_{i,k+1}w_{i-1,j}.\label{136} \end{eqnarray}
\end{Thm}

The stated actions on the elements $w_{ij}$ do indeed commute, and from this  one gets the actions from the left and right on ${\mathcal 
A}(s,r,{\mathbb C})$: If
$\triangle a=\sum_i a^1_i\otimes a^2_i$ then \begin{eqnarray}\label{139}
a\star(\phi\psi)&=&\sum_i(a^1_i\star \phi)(a^2_i\star \psi),\\\label{140}
(\phi\psi)\star a&=&\sum_i(\phi\star a^1_i)(\psi\star a^2_i).
\end{eqnarray}
For reference, we will call this the left NYM action, and the right NYM action, 
respectively. The resulting algebras of operators will be denoted {\bf the right NYM operators} and {\bf the left NYM operators}. In the following, whenever the symbol $\star$ appears, it refers to one of these actions. It will be clear from its position if it is a right or a left action.

\medskip

There is in (\cite{yam}) even an action on a localization by the quantum 
determinant of this algebra, but we shall not need this level of 
sophistication.

\medskip

We will give explicitly the actions below. The co-product used in (\cite{yam}) 
is different from ours. It is the one introduced by Drinfeld and Jimbo (\cite{drin}, \cite{jim}). Given in its simplest ${\mathcal U}_q(sl(2,{\mathbb C}))$ version 
for generators $E,F,K^{\pm1}$ it is \begin{eqnarray}\label{d1}\Delta_{DJ}
E&=&E\otimes
K^{-1/2}+K^{1/2}\otimes E,\\ \Delta_{DJ} F&=&F\otimes
K^{-1/2}+K^{1/2}\otimes F,\\ \Delta_{DJ}
K^{\pm1}&=&K^{\pm1}\otimes K^{\pm1} \label{d3}.\end{eqnarray}
 
\medskip
 
\begin{Rem} The above necessitates the introduction of a fixed choice of $q^{1/2}$.
After that one can introduce operators $K^{\pm\frac12}$. Using this, we will show that one can adapt their results to ours. The factors  $K^{\pm\frac12}$ and $q^{\pm\frac12}$ are inherent to NYM, and hence cannot be avoided when we use their results. Once we know that our results are true, we could then reformulate them without these factors, but we will refrain from doing this cumbersome exercise, as it adds little insight. \end{Rem}

\medskip

In NYM =\cite{yam}[p. 36] the notion of a $q$-minor $\triangle_q$ is introduced. It is the one given in  (\ref{triangle}) with $a=q$.

\medskip

We will state an analogue to (\ref{6.5}): Set, for ${\mathbf m}=(m_1,\dots,m_r)\in{\mathbb 
N}_0^r$,
\begin{equation}\triangle_{q}(w)^{\mathbf m}=\prod_{i=1}^r\triangle_{q,i}(w)^{m_i}.
\end{equation}

\medskip

The following is proved in (\cite{yam}) (Combine Theorem~2.12 and Theorem~2.4 
ibid. Also see the argument at the end of Section~2.3 ibid.)

\medskip

\begin{Thm}{\cite{yam}} As a bi-module,\label{yam-bi}
\begin{equation}
{\mathcal A}(n,n,{\mathbb C})=\oplus_{{\mathbf m}\in{\mathbb 
N}_0^n}W({\mathbf m}),
\end{equation}
where $W({\mathbf m})$ is the irreducible bi-module with highest weight vector  
$\triangle_{q}(w)^{\mathbf m}$. 
\end{Thm}

\medskip

\begin{Rem} We stress that the minors considered in Section~\ref{5} use $a=q^{-1}$. They are computed in algebras generated by the variables ${\mathbb W}^{(r)}_{i,j}$ whose  terms contain at least one  ${\mathbb B}^\dagger_i(k,r) $ and one ${\mathbb B}^\dagger_{n+j}(k,r)$. See (\ref{rank2}) for a more precise statement. The variables $w_{ij}$ we now consider are hence {\bf not at all} like the variables ${\mathbb W}^{(r)}_{i,j}$.

There is a further point that should be noted: In the bi-module structure (to be defined presently) the right action has an interchange of $E\rightarrow F$ so that a highest weight vector from the right action is annihilated by elements $F_a$.

\smallskip
 After this, Theorem~\ref{yam-bi}  is seen as being a $q$-analogue of the Peter-Weyl Theorem. As we shall see later in Proposition~\ref{left32}, this is precisely what we shall need.
\end{Rem}

\bigskip

\subsection{The left and right NYM actions}

\smallskip

We here make a general definition of some quantum multiplication, quantum differentiation, and quantum scale operators in an algebra ${\mathcal A}$ (or even just a vector space) in which there is a (PBW) basis defined in terms of some $N$ ``variables''  $x_i$ which are already ordered so that $i<j\Leftrightarrow x_i<x_j$ and $x_1$ is placed furthest to the left.
\begin{equation}
{\mathcal B}=\{x^{\mathbf b}=x_1^{b_1}\cdots x_N^{b_N}\in {\mathcal A}\mid {\mathbf b}\in{\mathbb N}^N_0\}.
\end{equation}
\begin{Def}$\forall  i=1,\dots, N$:
\begin{eqnarray} M_{i}x^{\mathbf
b}&=&x^{{\mathbf b}'} \textrm{ where }{{\mathbf b}'}={\mathbf
b}+e_{i},\\ D_{i}x^{\mathbf b}&=&[b_{i}]_q x^{{\mathbf b}''}
\textrm{ where }{{\mathbf b}''}={\mathbf b}-e_,\\
H^{\pm1}_{i}x^{\mathbf b}&=&q^{\pm b_{i}}x^{\mathbf 
b}.\end{eqnarray} 
\end{Def}

\smallskip

The operators $M_i,D_i$, and $H_i^{\pm}$ of course depend on the basis. Below we shall encounter bases for some special choices of $\sigma\in S_N$,
\begin{equation}
{\mathcal B}_\sigma=\{x^{\mathbf b}=x_{\sigma(1)}^{b_1}\cdots x_{\sigma(N)}^{b_N}\mid {\mathbf b}\in{\mathbb N}^N_0\}, 
\end{equation}
resulting from different orderings of the variables. In at least one instance it will hold, for some $\sigma_0\in S_N$ that 
\begin{equation}\forall {\mathbf b}\in{\mathbb N}^N_0:\ x_{\sigma_0(1)}^{b_{\sigma_0(1)}}\cdots x_{\sigma_0(N)}^{b_{\sigma_0(N)}}=x_1^{b_1}\cdots x_N^{b_N}.
\end{equation}
In this case we say that the bases are equivalent. Clearly, in such a case the resulting operators $M,D$, and $H^\pm$ will be the same. 

\medskip

We can use any such basis ${\mathcal B}$ to set up a vector space isomorphism ${\mathcal I}^{\mathcal B}_{{\mathcal A},{\mathcal P}_N}$ between ${\mathcal A}$ and the space ${\mathcal P}^N={\mathcal P}(z_1,\cdots,z_n)$ of complex polynomials in $N$ variables
\begin{equation}\label{ident}
{\mathcal I}^{\mathcal B}_{{\mathcal A},{\mathcal P}_N}: x_1^{b_1}\cdots x_N^{b_N}=x^{\mathbf b}\rightarrow z^{\mathbf b}=z_1^{b_1}\cdots z_N^{b_N}.
\end{equation}
In this way, any operator on ${\mathcal A}$ that can be expressed in terms of the operators $M,D$, and $H^{\pm}$ can be transported to ${\mathcal P}^N$, and vice versa.

\medskip

Any two PBW bases of ${\mathcal P}^N$ of the above kind yield the same operators $M,D,H^{\pm}$.

\medskip

We consider an order $>_R$ of the pairs $(i,j)$ as follows:
\begin{equation}
(i,j)>_R(s,t)\Leftrightarrow\left\{\begin{array}{l}j<t\textrm{ or}\\j=t, s<i.\end{array}\right.
\end{equation}
We write a variable $w_{ij}$ to the left of a variable $w_{st}$ if $(i,j)>_R(s,t)$. This way we get a PBW basis ${\mathcal B}^R$ consisting of monomials in the variables $w_{ij}$. Using (\ref{135}-\ref{136}) repeatedly we get
\begin{eqnarray}
\left(\sum_\beta\left(\prod_{(i,j)>_R(a,\beta)}w_{ij}^{b_{ij}}\right)w_{a,\beta}^{b_{a,\beta}}\left(\prod_{(a,\beta)>_R(s,t)}w_{st}^{b_{st}}\right)\right)\star E_a=\\
\sum_\beta\left((\prod_{(i,j)>_R(a,\beta)}w_{ij}^{b_{ij}})\star K_a^{1/2}\right)(w_{a,\beta}^{b_{a,\beta}}\star E_a)\left((\prod_{(a,\beta)>_R(s,t)}w_{st}^{b_{st}})\star K_a^{-1/2}\right).
\end{eqnarray}
Here, the expressions in the brackets can easily be computed. This, together with an issue of a change $q\leftrightarrow q^{-1}$ between NYM and our setup, is actually the reason for chosing this particular basis. 

The right action of $F_a$ can be computed analogously. We denote by $E^R_a({\mathcal B^R})$ and
$F^R_a({\mathcal B^R})$, respectively, the { right NYM operators} on ${\mathcal P}_{s,r}$ computed in the basis ${\mathcal B^R}$. Finally the right actions $L^R_a({\mathcal B}^r)$, $K^R_a({\mathcal B}^r)$ of $K_a$ are similarly determined. To state the result, we define the symbols $H^\pm_{a,b\uparrow}$, $H^\pm_{a\downarrow,b}$, etc. in analogy with the symbols $L^\pm_{a,b\uparrow}$, $L^\pm_{a\downarrow,b}$, etc. of \S5. We can then state the result:
\begin{Prop}[Right NYM operators]
\begin{eqnarray}
E^R_a({\mathcal B}^r)&=&q^{1/2}\sum_{\ell=1}^s\label{43}
D_{a,\ell}M_{a+1,\ell}
H^{-1/2}_{a,\ell+1\uparrow}H^{1/2}_{a+1,\ell+1\uparrow}H^{1/2}
_{a,\ell\downarrow}H^{-1/2}_{a+1,\ell\downarrow}(K_a^r)^{-1}\\
&=&q^{1/2}\sum_{\ell=1}^s\label{43}
D_{a,\ell}M_{a+1,\ell}
H^{-1}_{a,\ell+1\uparrow}H_{a+1,\ell+1\uparrow}(K_a^r)^{-1/2}.\nonumber\\\textrm{Similarly,}\\
\label{42}F^R_a({\mathcal B}^r)&=&q^{1/2}\sum_{\ell=1}^s
 D_{a+1,\ell}M_{a,\ell}H^{-1/2}_{a,\ell\uparrow}H^{1/2}_{a+1,\ell\uparrow}H^{
1/2 }_{a,\ell-1\downarrow}H^{-1/2}_{a+1,\ell-1\downarrow}(K_a^r),\\
L^R_a({\mathcal B}^r)&=&\prod_j H_{a,j},\\
K^R_{\mu_k}({\mathcal B}^r)&=&\prod_j H_{k,j}H_{k+1,j}^{-1}.
\end{eqnarray}
\end{Prop}

\medskip

We now introduce a linear isometry ${\mathcal I}_{s,r}^{\mathcal B}: {\mathcal
P}_{s,r} \rightarrow {\mathcal F}^{(q)}_{sr}$ given on a basis $z^{\mathbf
b}=\prod_{k,\ell}(z_{k,\ell})^{b_{k,\ell}}$ by

\begin{equation} {\mathcal I}_{s,r}^{\mathcal B}(z^{\mathbf
b})=\prod_{k,\ell}(B^\dagger_{k}(\ell))^{b_{k,\ell}}\cdot
v^{(q)}_{sr}. \end{equation}

\smallskip

 The operators $D_{i,j}, H^\pm_{i,j}$, and $M_{i,j}$ are  transported to ${\mathcal 
F}^{(q)}_{sr}$ by setting
 \begin{equation}D_{i,j}v^{(q)}_{sr}=0,\  H^\pm_{i,j}v^{(q)}_{sr}=v^{(q)}_{sr},\textrm{ and }
M_{i,j}v^{(q)}_{sr}=B^\dagger_{i}(j)v^{(q)}_{sr}.
 \end{equation}

\medskip

Using the linear isomorphism ${\mathcal I}^{\mathcal B}_{{\mathcal A},{\mathcal P}_N}$ with ${\mathcal A}={\mathcal A}(s,r,{\mathbb C})$ and $N=sr$, as well as the linear isomorphism ${\mathcal I}^{\mathcal B}_{s,r}$, it follows that ${\mathcal F}^{(q)}_{sr}$ as well as ${\mathcal P}_{s,r}$ become bi-modules. 

\medskip

In correspondence with (\ref{hw-tensor}) we have
\begin{equation}
{\mathcal
F}^{(q)}_{2nr}={\mathcal
F}^{X,(q)}_{nr}\otimes {\mathcal
F}^{Y,(q)}_{nr},
\end{equation}
and we have
\begin{eqnarray}
{\mathcal
P}_{2nr}&=&{\mathcal
P}^X_{nr}\otimes{\mathcal
P}^Y_{nr},\\
{\mathcal A}(r,2n,{\mathbb C})&=&{\mathcal A}^X(r,n,{\mathbb C})\otimes{\mathcal A}^Y(r,n,{\mathbb C}).
\end{eqnarray}

Here, ${\mathcal
P}^X_{nr}$ has variables $z_{ij}$ with $1\leq i\leq n$ and $1\leq j\leq r$ while ${\mathcal
P}^Y_{nr}$ has variables $z_{ij}$ with $n+1\leq i\leq 2n$ and $1\leq j\leq r$. The spaces ${\mathcal A}^X(r,n,{\mathbb C})$ and ${\mathcal A}^Y(r,n,{\mathbb C})$ are defined analogously.

In the same spirit,

\begin{eqnarray}{\mathcal
F}^{X,(q)}_{nr}&=&{\mathcal H}{\mathcal W}^X_{nr}\cdot v_{2nr}^{(q)},\\
{\mathcal
F}^{Y,(q)}_{nr}&=&{\mathcal H}{\mathcal W}^Y_{nr}\cdot v_{2nr}^{(q)}
.\end{eqnarray}

Under linear isomorphisms, the spaces labeled by an $X$ are equivalent, as are the spaces labeled by $Y$. A (say, left) operator $G$ on a space labeled by, say, $X$ may occationally get a label $G(X)$, and similarly for $Y$.

\medskip

In conclusion, we now have a bi-module structure on the $q$-Fock module 
${\mathcal
F}^{(q)}_{sr}$, or, equivalently, on ${\mathcal
P}_{sr}$.

\medskip

It will become important later that we do not just use the bi-module structure on ${\mathcal
F}^{(q)}_{2nr}$ but use bi-module structures on ${\mathcal
F}^{X,(q)}_{nr}$ and ${\mathcal
F}^{Y,(q)}_{nr}$ where the left action on the latter is  not the NYM action, but a modified version thereof.

\medskip

Returning to the formulas  (\ref{e-b}-\ref{f-j}), it is clear that they also 
can be formulated in ${\mathcal
F}^{(q)}_{2nr}$, or, equivalently, on ${\mathcal
P}_{2nr}$. The operators $x^{(r)}_j$, $x=e,f$ can even be viewed as operators on ${\mathcal
P}^X_{nr}$ when $1\leq j\leq n$ and on
${\mathcal
P}^Y_{nr}$ when $n+1\leq j\leq 2n$.

\medskip

We list these formulas for comparison (Notice: We compute this in the basis ${\mathcal B}^R$.):

\smallskip

\begin{Prop}
\begin{eqnarray}\label{e-b-n}
e_\beta^{(r)}&=&\sum_{i=1}^rq^{-i+1}{D}_{1,i}{D}_{n+1,i}H^{-1}_{1,
(i-1)^\downarrow}
H^{-1}_{n+1,(i-1)^\downarrow}\\
f_\beta^{(r)}&=&\sum_{i=1}^rq^{r-i}
{M}_{1,i}{M}_{n+1,i}H_{1,(i+1)^\uparrow}
H_{n+1,(i+1)^\uparrow}\\ e_j^{(r)}&=&\sum_{i=1}^r\label{e-j-n}
{D}_{j+1,i}{M}_{j,i}H^{-1}_{j+1,( i-1)^\downarrow}
H_{j,(i-1)^\downarrow}, 
\\ f_j^{(r)}&=&\sum_{i=1}^r
{D}_{j,i}{M}_{j+1,i} H^{-1}_{j,(i+1)^\uparrow}
H_{j+1,(i+1)^\uparrow}
\label{f-j-n}. \end{eqnarray}
\end{Prop}

\medskip

To compare with our expressions we use generally
$E=q^{-1/2}eK^{1/2}$ and $F=q^{-1/2}fK^{-1/2}$ but with $e,f$
computed from the NYM expressions. These are also NYM operators.

Comparing these formulas to those in  (\ref{e-j-n}-\ref{f-j-n}) we get (notice that $E\leftrightarrow f$ and $F\leftrightarrow e$):

\begin{Prop}\label{left32} 
\begin{eqnarray}
E^R_a({\mathcal B}^r)&=&q^{1/2}f_a^{(r)}(K_a^r)^{-1/2},\\
F^R_a({\mathcal B}^r)&=&q^{1/2}e_a^{(r)}(K_a^r)^{1/2}.
\end{eqnarray}
\end{Prop}

\begin{Cor}\label{left3}
The operators $e_a^{(r)}, f_a^{(r)}$ are right NYM operators. In particular, they commute with all operators acting from the left.
\end{Cor}

\pof (of Corollary~\ref{left3}). This follows immediately since the operators $(K_a^r)$ are right operators. \qed

\medskip

We can compute the left NYM operators analogously, but here we chose an order $>_L$ defined as follows:

\begin{equation}
(i,j)>_L(s,t)\Leftrightarrow\left\{\begin{array}{l}i<s\textrm{ or}\\i=s, t<j.\end{array}\right.
\end{equation}

\smallskip

Corresponding to this we get a PBW basis ${\mathcal B}^L$. The computation of the { left NYM operators} $E^L_a({\mathcal B}^L)$ and $F^L_a({\mathcal B}^L)$ in this basis follows analogously. In fact, one can simply interchange lower indices while making the switch $E\leftrightarrow F$. One gets:

\begin{Prop}[Left NYM operators]
\begin{eqnarray}
E^L_a({\mathcal B}^L)&=&q^{-1/2}\sum_{\ell=1}^s
D_{\ell,a+1}M_{\ell,a}H^{-1/2}_{\ell-1\downarrow,a+1}
H^{1/2}_{\ell-1\downarrow,a}H^{-1/2}_{
\ell\uparrow,a}H^{1/2}_{\ell\uparrow,a+1},\\ F^L_a({\mathcal B}^L)&=&q^{-1/2}\sum_{\ell=1}^s
D_{\ell,a}M_{\ell,a+1}H^{-1/2}_{\ell\downarrow,a+1}
H^{1/2}_{\ell\downarrow,a}H^{-1/2}_{
(\ell+1)\uparrow,a}H^{1/2}_{(\ell+1) \uparrow,a+1},\\
L^L_a({\mathcal B}^r)&=&\prod_j H_{j,a},\\
K^L_{\mu_k}({\mathcal B}^r)&=&\prod_j H_{j,k}H_{j,k+1}^{-1}.
\end{eqnarray} 
\end{Prop}

\medskip

\begin{Lem}
Set $K_\beta=\prod_\ell H_{1,\ell}H_{n+1,\ell}$. Then $K_\beta$ commutes with all left NYM operators.
\end{Lem}

\medskip

We now perform some computations with the aim of finding some left NYM operators that commute with $e_\beta$ and $f_\beta$. Recalling the homomorphism $\Gamma^{(r)}_3$ of Proposition~\ref{res33}, our strategy will be to look for some left operators on ${\mathcal A}^X(r,n,{\mathbb C})$ and ${\mathcal A}^Y(r,n,{\mathbb C})$, respectively. Such operators will be labeled by $X$,$Y$, respectively. We restrict our search to a small class of promising operators:

\medskip

Set
\begin{eqnarray}
&A_a:=H_{1,a}^{\varepsilon_1}
H_{1,a+1}^{-\varepsilon_2}
H_{n+1,a}^{\beta_1}
H_{n+1,a+1}^{-\beta_2},\\\nonumber
&B_a:=H_{1,a}^{\varepsilon_3}
H_{1,a+1}^{-\varepsilon_4}
H_{n+1,a}^{\beta_3}
H_{n+1,a+1}^{-\beta_4}.\end{eqnarray}

Then,

\begin{eqnarray}
&D_{1,a+1}M_{1,a}A_ae_\beta-e_\beta
D_{1,a+1}M_{1,a}A_a=\\\nonumber
&q^{-a+1}\left(q^{-\varepsilon_1-\beta_1}M_{1,a}D_{1,a}-
D_{1,a}M_{1,a}\right)D_{1,a+1}D_{n+1,a}A_a
H^{-1}_{1,a-1\downarrow}
H^{-1}_{n+1,a-1\downarrow}+\\\nonumber
&q^{-a}(q^{\varepsilon_2+\beta_2+1}-1)D_{1,a+1}D_{n+1,a+1}
H^{-1}_{1,a\downarrow}
H^{-1}_{n+1,a\downarrow}D_{1,a+1}M_{1,a}A_a,
\end{eqnarray}

\begin{eqnarray}
&D_{n+1,a}M_{n+1,a+1}B_ae_\beta-e_\beta
D_{n+1,a}M_{n+1,a+1}B_a=\\\nonumber
&q^{-a}\left(q^{\varepsilon_4+\beta_4}M_{n+1,a+1}
D_{n+1,a+1}-qD_{n+1,a+1}M_{n+1,a+1}\right)D_{n+1,a}
D_{1,a+1}B_a 
H^{-1}_{1,a\downarrow}
H^{-1}_{n+1,a\downarrow}+\\\nonumber
&q^{-a+1}(q^{-\varepsilon_3-\beta_3}-1)D_{1,a}D_{n+1,a}
H^{-1}_{1,a-1\downarrow}
H^{-1}_{n+1,a-1\downarrow}D_{n+1,a}M_{n+1,a+1}B_a,
\end{eqnarray}

\begin{eqnarray}
&D_{1,a+1}M_{1,a}A_af_\beta-f_\beta
D_{1,a+1}M_{1,a}A_a=\\\nonumber
&q^{r-a-1}\left(q^{-\varepsilon_2-\beta_2}D_{1,a+1}M_{1,a+1}
-M_{1,a+1}D_{1,a+1}\right)
M_{1,a}M_{n+1,a+1}A_a
H^{}_{1,a+2\uparrow}
H_{n+1,a+2\uparrow}+\\\nonumber
&q^{r-a}\left(q^{\varepsilon_1+\beta_1+1 }-1\right)
M_{1,a}M_{n+1,a}
H_{1,a+1\uparrow}
H_{n+1,a+1\uparrow}D_{1,a+1}M_{1,a}B_a,
\end{eqnarray}
and
\begin{eqnarray}
&D_{n+1,a}M_{n+1,a+1}B_af_\beta-f_\beta
	D_{n+1,a}M_{n+1,a+1}B_a=\\\nonumber
&q^{r-a}\left(q^{\varepsilon_3+\beta_3}D_{n+1,a}
M_{n+1,a}-qM_{n+1,a}D_{n+1,a}\right)M_{1,a}
M_{n+1,a+1}B_a
H_{1,a+1\uparrow}
H_{n+1,a+1\uparrow}+\\\nonumber
&q^{r-a-1}\left(q^{-\varepsilon_4-\beta_4}-1\right)M_{1,a+1}
M_{n+1,a+1}
H_{1,a+2\uparrow}
H_{n+1,a+2\uparrow}D_{n+1,a}M_{n+1,a+1}B_a.
\end{eqnarray}

\medskip

The operators $A_a,B_a$ of course depend upon the four variables $\varepsilon_i,\beta_i; i=1,2,3,4$. 

\medskip

From now on we express the relevant operators in terms of the basis ${\mathcal B}^L$.

 \medskip 

Using the above computations we easily reach:
\smallskip

\begin{Prop}\label{commu}The operator
\begin{equation}D_{1,a+1}M_{1,a}A_a-\alpha D_{n+1,a}M_{n+1,a+1}B_a
\end{equation}commutes with $e_\beta$ and $f_\beta$ if and only if $\alpha=1$, $\varepsilon_1+\beta_1+1=0$,  $\varepsilon_2+\beta_2+1=0$, $\varepsilon_3+\beta_3=0$, $\varepsilon_4+\beta_4=0$, $\varepsilon_1=\varepsilon_3$, and
$\varepsilon_2=\varepsilon_4$. \end{Prop}

\medskip

Because of this result we now set:

\begin{Def}
\begin{eqnarray}
E_a(X)&=&\sum_{\ell=1}^nD_{\ell,a+1}
M_{\ell,a} H_{\ell-1\downarrow,a}
H^{-1}_{\ell-1\downarrow,a+1} H^{-1}_{n+1,a}H_{n+1,a+1}\\
&=&q^{1/2}E_a^L(X)({\mathcal B}^L)(K^L_a(X))^{1/2}(K^L_a(Y))^{-1}.\\
&&\textrm{Similarly,}\\
F_a(X)&=&\sum_{\ell=1}^nD_{\ell,a}
M_{\ell,a+1} H^{-1}_{\ell+1\uparrow,a}
H_{\ell+1\uparrow,a+1} \\
&=&q^{1/2} F_a^L(X)({\mathcal B}^L)(K^L_a(X))^{-1/2}.
\end{eqnarray}

In the same way, we define elements $E_a(Y), F_a(Y)$:
\begin{eqnarray}E_a(Y)&=&q^{1/2}E_a^L(Y)({\mathcal B}^L)(K^L_a(Y))^{1/2}(K^L_a(X))^{-1},\textrm{ and}\\
F_a(Y)&=&q^{1/2} F_a^L(Y)({\mathcal B}^L)(K^L_a(Y))^{-1/2}.
\end{eqnarray}
\end{Def}
\medskip

\begin{Cor}The elements 
$(E_a(X)-F_a(Y))$  and $(E_a(Y)-F_a(X))$ commute with $e_\beta,f_\beta$.
\end{Cor}

\pof  The result follows follows from  Proposition~\ref{commu}
as far as the first term is concerned since the term has been constructed so that the full commutativity when $\ell\neq1$ is covered. The second equation follows by symmetry. \qed

\medskip

\begin{Prop}
\begin{equation}
E_a^L(X)({\mathcal B}^L)(K^L_a(X))^{1/2}(K^L_a(Y))^{-1}- F_a^L(Y)({\mathcal B}^L)(K^L_a(Y))^{-1/2}
\end{equation}

and

\begin{equation}
E_a^L(Y)({\mathcal B}^L)(K^L_a(Y))^{1/2}(K^L_a(X))^{-1}- F_a^L(X)({\mathcal B}^L)(K^L_a(X))^{-1/2}
\end{equation}
commute with $e_\beta,f_\beta$. 
\end{Prop}
\medskip

\pof  The two expressions are $q^{-1/2}(E_a(X)-F_a(Y))$  and $q^{-1/2}(E_a(Y)-F_a(X))$, respectively. \qed

\bigskip

Let us write  $k_a(Z)=K^L_a(Z)$,   $e_a(Z)=E_a^L(Z)({\mathcal B}^L)$, and $f_a(Z)= F_a^L(Z)({\mathcal B}^L)$; $Z=X, Y$. Notice that the operators $e_a(Z),f_a(Z), k_a(Z)^{\pm1}$ by NYM (\cite{yam}) satisfy the $q$-Serre relations. In particular,   $\left[e_a(Z),f_a(Z)\right]=\frac{k_a(Z)-k_a(Z)^{-1}}{q-q^{-1}}$;  $Z=X, Y$.

\medskip

Notice that \begin{equation}(k_a(X)k_a(Y)^{-1})=L_a(X)L_{a}^{-1}(Y)L_{a+1}^{-1}(X)L_{a+1}(Y),\end{equation} and any power of  $L_a(X)L_{a}^{-1}(Y)$ commutes with $e_\beta,f_\beta$.

\medskip

From this we conclude:

\begin{Prop} The following left NYM operators 
\begin{eqnarray}
{\mathbf e}_a&=& e_a(X)k_a(Y)^{-1/2}-f_a(Y)k_a(X)^{-1/2},\textrm{ and}\\
{\mathbf f}_a&=& f_a(X)k_a(Y)^{-1/2}-e_a(Y)k_a(X)^{-1/2}
\end{eqnarray}
commute,  for any $a=1,\cdots,r$, with $e_\beta,f_\beta$ as well as with the operators $e_j^{(r)}, f_j^{(r)}$ for any $j\in\{1,\dots,n-1,n+1,\dots,2n-1\}$.
\end{Prop}

\medskip

A simple computation gives:

\begin{Prop}
\begin{equation}
[{\mathbf e}_a,{\mathbf f}_a]=\frac{k_a(X)k_a(Y)^{-1}-k_a(Y)k_a(X)^{-1}}{q-q^{-1}}.
\end{equation}
\end{Prop}

Set ${\mathbf k}_a=k_a(X)k_a(Y)^{-1}$.

\begin{Prop}[Drinfeld Double]
The elements ${\mathbf e}_a,{\mathbf f}_a,{\mathbf k}_a^{\pm1}$, $a=1,\dots,r$ satisfy the quantum Serre relations for $su_q(r)$.
\end{Prop}

\pof First of all, it is easy to see that $[{\mathbf e}_a,{\mathbf f}_b]=0$ if $a\neq b$. The relations involving $k_a^{\pm1}$ are also straightforward. To address the equations for the quantum adjoint, consider two different elements ${\mathbf e}_a,{\mathbf e}_b$. Write, in straightforward notation, ${\mathbf e}_a=\alpha-\beta$, and 
${\mathbf e}_b=\gamma-\delta$. We must then prove that
\begin{equation}
(\alpha-\beta)(\alpha-\beta)(\gamma-\delta)-[2]_q(\alpha-\beta)(\gamma-\delta)(\alpha-\beta)+(\gamma-\delta)(\alpha-\beta)(\alpha-\beta)=0.
\end{equation}
Observing that $\alpha\delta=q^{-1}\delta\alpha, \beta\gamma=q\gamma\beta, \alpha\beta=q^2\beta\alpha$, and $\gamma\delta=q^{2}\delta\gamma$, the equations involving ``cross-terms'' are seen to vanish. Since the elements  $e_a(Z), f_a(Z)$ already satisfy the $q$-Serre, the claim follows. \qed

\medskip

See e.g. J. Xiao (\cite{double}) for a study of the Drinfeld Double.

\medskip

\begin{Def}
We denote by ${\mathcal U}^D_q(u(r))$ the particular presentation of ${\mathcal U}_q(u(r))$ thus constructed.
\end{Def}

\medskip

\begin{Prop}
There is a representation of  ${\mathcal U}^D_q(u(r))$ 
on ${\mathcal F}_{2nr}^{(q)}$.
\end{Prop}

\pof We can use the generators ${\mathbf k}^{\pm1}_a$, ${\mathbf e}_a$, and ${\mathbf f}_a$ defined above. \qed 

}
\medskip

We then obtain:

\smallskip

\begin{Thm}
The representations of ${\mathcal U}_q(u(n,n))$ and ${\mathcal U}^D_q(u(r))$ 
on ${\mathcal F}_{2nr}^{(q)}$ commute.
\end{Thm} 

\medskip

There is also an analogue of Proposition~\ref{commut}:

\begin{Prop}
An element $h\in{\mathcal H}{\mathcal W}_{2nr}$ commutes with $\psi_q^{(r)}({\mathcal U}_q({\mathfrak u}(n,n)^{\mathbb C})$ if and only $h\in {\mathcal U}^D_q({\mathfrak u}(r)^{\mathbb C})$.
\end{Prop}

\noindent{\bf Proof:}(Sketched) We have that $h$ is a polynomial in operators ${\mathbb D}_*, {\mathbb M}_*$, and ${\mathbb K}^{\pm1}_*$. Focusing on, say ${\mathcal H}{\mathcal W}^X_{nr}$, we have that $h$ commutes with $\psi_q^{(r)}({\mathcal U}_q({\mathfrak u}(n))^{\mathbb C})^X$. Via the linear isomorphisms, we can view these as acting on ${\mathcal A}^X(r,n,{\mathbb C})$, and these are right NYM operators. Since irreducible representations of $\psi_q^{(r)}({\mathcal U}_q({\mathfrak u}(n))^{\mathbb C})^X$ of a specific type in ${\mathcal A}^X(r,n,{\mathbb C})$ all occur in the same degree, it follows that $h$ must preserve these degrees. We can then, crucially, employ the $q$-commutator result of R. B. Zhang (\cite{r-zhang}). It is clear that we have all of
${\mathcal U}_q(gl(n))^{\mathbb C})^X$ acting from the right. Thus, this piece of $h$ must be a left NYM operator.

By similar reasoning involving ${\mathcal A}^Y(r,n,{\mathbb C})$ we eventually conclude, similarly to the classical case,  that 
\begin{equation}
h\in ({\mathcal U}_q({\mathfrak u}(r))^{\mathbb C})^X({\mathcal U}_q({\mathfrak u}(r))^{\mathbb C})^Y.
\end{equation}
There is now the proviso that we need operators $k_a(X)^{\pm 1/2}$ and $k_a(Y)^{\pm 1/2}$ acting as scaling operators in the representations. This involves a simple choosing of a $q^{1/2}$.

 Parallel to the classical case, the center of  ${\mathcal H}{\mathcal W}_{2nr}$ is ${\mathbb C}$ as follows by considering commutators of the elements $K_*^{\pm1}$. We here use that $q$-is not a root of unity. From then on, the proof proceeds in the same way. \qed

\bigskip

\noindent{\bf Proof of Theorem~\ref{q-k-v-thm} and Proposition~\ref{q-minor-type}:} It follows from Theorem~\ref{yam-bi} that, as a ${\mathcal U}_q(u(n))\times {\mathcal U}_q(u(n))\times{\mathcal 
U}_q(u(r))\times {\mathcal 
U}_q(u(r))^{opposite}$ module, the highest weight vectors in ${\mathcal F}^{(q)}_{2nr}$ are given as
\begin{equation}\forall {\mathbf m_1}, {\mathbf m_1}\in{\mathbb N}_0^r:\ \triangle_{q}^{\mathbf m_1}(x)\widetilde{\triangle_{q}^{\mathbf 
m_2}}(y) \cdot v^{(q)}_{2nr}.
\end{equation}

\medskip
We can then see, by direct evaluation of $e_\beta$ on these,  that the element $\triangle_{q}^{\mathbf m}(x)\widetilde{\triangle_{q}^{\mathbf 
n}}(y)$ is a  top $q$-pluri-harmonic polynomial if ($\ell({\mathbf m})$ DEF?)
\begin{equation}\ell({\mathbf m})+\ell({\mathbf n})\leq r.\end{equation}In this 
case, this vector is the highest weight vector for an irreducible ${\mathcal 
U}_q(u(r))$ module of highest weight $(m_1,\dots, m_{\ell({\mathbf 
m})},0,\dots,0,-n_{\ell({\mathbf n})},\dots,-n_1)$. The non-zero vectors in 
this module are all top $q$-pluri-harmonic polynomials of highest weight
\begin{equation}
(\Lambda,\zeta)=(({\mathbf m},{\mathbf n},-r-m_1-n_1), \vert{\mathbf m}\vert - \vert{\mathbf n}\vert).
\end{equation}
By weight comparisons (c.f. (\cite{rosso}), there can be no more top $q$-pluri-harmonic polynomials than the ones listed.

\medskip

After this, it is clear that we have the same correspondence between a special class of irreducible representations of ${\mathcal 
U}_q(u(r))$, though with a new $q$-weight interpretation of $(\Lambda,\zeta)^\xi)$, as in Definition~\ref{extra}.

\medskip

With this, we also have a {\bf Proof of Theorem~\ref{q-k-v-thm}. \qed

\medskip

\begin{Rem}
There remains now perhaps only to discuss what ``quantized covariant differential operators'' should be. In line with Remark~\ref{cov-dif} everything in the beginning carries nicely over, indeed all the way to (\ref{cov-dif-0}). After that one can take the path of using operators ${\mathbb D}_*, {\mathbb M}_*$, and ${\mathbb K}^{\pm1}_*$ defined in specific bases. This line will be pursued elsewhere.
\end{Rem}

\end{document}